\documentclass[reprint,
 superscriptaddress,
 amsmath,amssymb,
]{revtex4-2}

\usepackage{graphicx}
\usepackage{dcolumn}
\usepackage{bm}
\usepackage{physics}
\usepackage{amsmath}
\usepackage{amsthm}
\usepackage{xcolor}
\usepackage{wrapfig}
\usepackage{ulem}
\usepackage{import}
\usepackage[caption=false]{subfig}
\usepackage{float}

\begin{document}

\preprint{APS/123-QED}

\title{High-fidelity two-qubit gates between fluxonium qubits with a resonator coupler}

\author{Emma L. Rosenfeld}\thanks{Current affiliation: Google Research}
\affiliation{AWS Center for Quantum Computing, Pasadena, CA 91125, USA}

\author{Connor T. Hann}
\affiliation{AWS Center for Quantum Computing, Pasadena, CA 91125, USA}

\author{David I. Schuster}
\affiliation{AWS Center for Quantum Computing, Pasadena, CA 91125, USA}
\affiliation{Department
of Applied Physics, Stanford University, Stanford, CA 94305, USA}

\author{Matthew H. Matheny}\thanks{mmathen@amazon.com}
\affiliation{AWS Center for Quantum Computing, Pasadena, CA 91125, USA}

\author{Aashish A. Clerk}
\affiliation{AWS Center for Quantum Computing, Pasadena, CA 91125, USA}
\affiliation{Pritzker School of Molecular Engineering, University of Chicago, Chicago IL 60637, USA}

\date{\today}

\begin{abstract}
We take a bottom-up, first-principles approach to design a two-qubit gate between fluxonium qubits for minimal error, speed, and control simplicity. Our proposed architecture consists of two fluxoniums coupled via a resonator. Using a simple linear coupler has many practical benefits, including the possibility of material optimization for suppressing loss, reducing fabrication complexity, and increasing yield by circumventing the need for Josephson junctions. Crucially, a resonator-as-coupler approach also suggests a clear path to increased connectivity between fluxonium qubits, by reducing capacitive loading when the coupler has a high impedance. After performing analytic and numeric analyses of the circuit Hamiltonian and gate dynamics, we tune circuit parameters to destructively interfere sources of coherent error, revealing an efficient, fourth-order scaling of coherent error with gate duration. For component properties from the literature, we predict an open-system average CZ gate infidelity of $1.86 \times 10^{-4}$ in 70ns.
\end{abstract}

\maketitle

\section{Introduction}

A formidable prerequisite for low-overhead quantum error correction is demonstrating entangling gate errors well below the threshold of an error correcting code. The fluxonium circuit, consisting of a small Josephson junction with shunt inductance as well as capacitance, has recently re-emerged as a long-lived qubit with the potential for entangling gate fidelities surpassing the current state-of-the-art transmon fidelities \cite{koch_charging_2009,manucharyan_fluxonium_2009, masluk_microwave_2012, manucharyan_evidence_2012, nguyen_high-coherence_2019, somoroff_millisecond_2021, kapit_small_2022, bao_fluxonium_2022, ficheux_fast_2021, xiong_arbitrary_2021, nesterov_cnot_2022, dogan_demonstration_2022, nesterov_proposal_2021, chen_fast_2022, moskalenko_high_2022, simakov_coupler_2023, setiawan_fast_2023, weiss_fast_2022, ciani_microwave-activated_2022, moskalenko_tunable_2021, nesterov_microwave-activated_2018, simakov_high-fidelity_2023, ma_native_2023, nguyen_scalable_2022, zhang_tunable_2023}. While recent fluxonium gate demonstrations are remarkable \cite{ding_high-fidelity_2023, zhang_tunable_2023, dogan_demonstration_2022}, further work is required to develop gate and control architectures that are optimal for large-scale devices consisting of many fluxonium qubits with a high degree of connectivity \cite{nguyen_scalable_2022}. Specifically, fluxonium exhibits qualitatively different matrix elements, spectra, and robustness to noise \cite{koch_charging_2009, manucharyan_fluxonium_2009, masluk_microwave_2012, manucharyan_evidence_2012, nguyen_high-coherence_2019, somoroff_millisecond_2021, mizel_right-sizing_2020, zhu_asymptotic_2013, herrera-marti_tradeoff_2013, hassani_superconducting_2022, gyenis_moving_2021, lin_protecting_2018} compared to the transmon. This precludes a simple translation of best-practices for transmon gates to fluxonium. Instead, first-principles studies are required to understand the opportunities and risks of new fluxonium-based gate schemes. 

\begin{figure}[ht]
\includegraphics[width=0.45\textwidth]{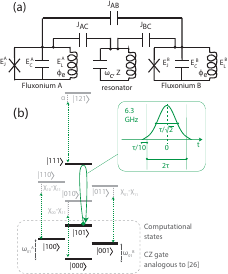}
\caption{Proposed circuit architecture and spectrum. (a) The fluxonium-resonator-fluxonium architecture includes two fluxonium qubits (left and right) capacitively coupled to a resonator (center). For an analysis of potential grounding configurations, see Appendix A. (b) Energy spectrum and states involved in the gate. Solid black and grey lines represent energy levels. Populated states are shown in black. Arrows represent the effect of the control tone, with the resonant transition in solid green, and detuned driving of other coupler excitations, at detunings $\chi_{ij}$ and $\alpha$, shown with green dashed arrows. Inset: control pulse of the gate. The coupler charge is modulated at its resonance frequency specifically when the fluxoniums' state is $\ket{11}$. For our simulations, the pulse envelope is Gaussian and truncated by linear ramps in the amplitude up and down from zero, alternatively, an offset or cosine pulse may equally be used. The total gate duration $t_g = 2.2\tau$.
}\label{fig:fig1}
\end{figure}

In this work, we present a comprehensive analysis of a promising candidate circuit and control scheme for fast, high-fidelity gates between fluxonium qubits. As we discuss, this approach incorporates many best-practices that give it strong practical advantages, especially when one considers scaling to multi-qubit circuits. Fig.~\ref{fig:fig1}(a) depicts our proposed gate architecture: 
a pair of fluxonium qubits connected through a coupler \cite{sung_realization_2021, yan_tunable_2018} formed by a simple linear resonator. 
One might expect that the lack of any intrinsic coupler nonlinearity would severely restrict the choice of possible gates schemes (e.g.~to resonator-induced-phase gates \cite{paik_experimental_2016} that use linear bosonic dynamics, or to approaches based on virtual coupler excitations \cite{puri_high-fidelity_2016}). 
Here, we show this is not the case, and in fact one can use approaches typically associated with highly nonlinear coupling elements.  This is achieved by strategically coupling the resonator to the highly anharmonic fluxonium qubits, in such a way that the coupling resonator inherits a significant nonlinearity. This in turn enables fast, single-photon excitation of the coupler that are conditional on the fluxoniums' states. 

In particular, we model a microwave-activated CZ gate achieved via a control scheme similar to that used in the recent work of Ding et al.~in Ref.~\cite{ding_high-fidelity_2023}, which entangled two fluxonium qubits through selective microwave excitations of a transmon qubit coupler using two linear charge drives, one on each fluxonium qubit. In this work, we use a single drive tone, optimized only over its frequency and amplitude, to drive a resonator coupler from its ground state to first excited state and back, selective on the qubits being in the computational $\ket{11}$ state and yielding a state-selecting phase shift of $\pi$. 

Our work extends Ref.~\cite{ding_high-fidelity_2023} in several additional directions, all related to our choice of a linear coupler and developed for pragmatic integration at scale. Perhaps most importantly, the resonator-as-coupler approach could facilitate increased connectivity between fluxonium qubits in a large circuit. Achieving large connectivity is both required for quantum error correction and is an open problem. Maintaining the fluxonium's small self-capacitance for a typical $E_C \sim 1$ GHz becomes challenging as the connectivity increases, because of capacitive loading and extra parasitic capacitances to the ground plane in a practical design. For integration in a realistic circuit capable of implementing a surface code, a high impedance coupler may be chosen to connect nearby fluxoniums with minimal coupling capacitance. In particular, for ultrahigh impedance resonators with impedance over $1 \text{ k}\Omega$, e.g. fabricated from low-loss, high kinetic inductance materials \cite{shearrow_atomic_2018, gao_physics_2008, amin_loss_2022} or junction arrays \cite{masluk_microwave_2012}, the loading efficiency would be improved compared to a transmon coupler. 

Pending experimental tests, there may also be performance advantages in our choice of a coupler that does not contain Josephson junctions. The resonator quality factor can approach $\sim 5 \times 10^6$ or higher by fabricating it out of high-quality materials such as tantalum in a single fabrication layer \cite{shi_tantalum_2022}, suppressing coupler-induced gate errors compared to more conventional couplers in a case when aluminum-based junction loss channels limit the transmon lifetime. 

Beyond loss mitigation, using a resonator coupler may also improve frequency targeting and device yield compared to setups with junction-based couplers. Achieving a desired resonance frequency for microwave driving the gate can be achieved to the $10^{-3}$ level in resonators \cite{kollar_hyperbolic_2019} (with additional study needed for the high impedance case). Conversely, junction energies are exponentially sensitive to the thickness of the junction barrier and drift over time due to aging processes within thin film dielectrics \cite{schafer_annealing_1991}. One might anticipate that using a resonator coupler removes the ability to in situ tune the coupler frequency.  This could reduce robustness to spectrum variations due to mis-targeting of the junctions in the fluxonium circuits. However, in principle, junction mis-targeting could be mitigated even when using resonator couplers.  One could achieve frequency tunability with the addition of a SQUID \cite{svensson_microwave_2018}, by using current-tunable kinetic inductance resonators \cite{vissers_frequency-tunable_2015} or by employing frequency trimming \cite{valles-sanclemente_post-fabrication_2023}. These techniques are not required in our study, in which we assume junction energy targeting at $2\%$ accuracy. Finally, the energy participation of a resonator is more diffuse compared to a transmon, potentially reducing the likelihood of coherent coupling between two-level systems (TLS) and the driven transition in our gate. However, additional study of the spatial distribution of TLS is needed to better understand this potential advantage. 

\begin{table}[t!]
\centering
\begin{tabular}{||c | c | c | c | c | c | c||}
 \hline
 $E_C^{A(B)}$ & $E_J^{A(B)}$ & $E_L^{A(B)}$ & $J_{A(B), c}$ & $J_{A,B}$ & $\omega_c / 2\pi$ & $Z_c$ 
 \\ [0.5ex] 
 \hline
 2(2) & 7.1(7.2) & 0.3(0.3) & 0.33(0.33) & 0.1 & 7.08 & 190 $\Omega$ \\ 
 \hline

 \hline
\end{tabular}
\caption{Representative circuit parameters analyzed in this work. $E_C^{i}, E_J^i, E_L^i$ (GHz) are capacitive, junction, and inductive energies for fluxonium $i$, while the resonator parameters are frequency $\omega_c / 2\pi$ (GHz) and impedance $Z_c$ ($\Omega$). Two-body coupling strengths between elements $i$ and $j$ are $J_{ij}$. For additional results using different circuit parameters with qualitatively similar conclusions, in particular the ultrahigh-impedance regime, see Appendix \ref{alt_params}.}
\label{table:1}
\end{table}

In what follows, we present both analytic and numerical studies of our two-fluxonium setup, demonstrating that our analytic estimates of coherent error are in qualitative agreement with numerics. We describe a fortuitous destructive interference of coherent error sources that can be engineered over a wide range of circuit parameters. This is made possible by the large induced nonlinearity of the resonator, inherited from the fluxonium qubits, and allows a coherent error that scales with gate time as $\epsilon_c \propto (1/t_g)^{p}$, where $p \approx 4$. We identify an exemplary set of realistic device and control parameters (see Fig.~\ref{fig:fig1}a and Table \ref{table:1}, with additional examples in Appendix \ref{alt_params}) chosen for low gate error, duration, and practical implementation, while maintaining favorable circuit parameters for long single-qubit lifetimes and coherence. Our results suggest the possibility of high fidelity operation, robust to small changes in flux and Josephson junction energies. For example, we predict that coherent errors of $2 \times 10^{-6}$ could be achieved using a 100ns gate time. Adding state-of-the-art levels of loss to our simulations, we predict a total (coherent and incoherent) error of $1.86 \times 10^{-4}$ for a 70 ns gate.


\section{Setup and basic gate mechanism}

\subsection{Identifying optimal, scalable gate strategies}

The approach to two qubit gates analyzed in this work was chosen after a careful consideration of a multitude of strategies. 
In this search, we stipulated several conditions for enhancing scalability and practicality. To start, we focused on microwave-activated techniques \cite{simakov_high-fidelity_2023}.  While the fluxoniums must be connected for an entangling gate, their interaction strength should be zero when the gate is not running to reduce parasitic entanglement during a quantum algorithm. Such spurious, ``always-on" ZZ-interactions are suppressed in our circuit, so we do not need baseband flux control \cite{bao_fluxonium_2022} or compensating Stark shift tones to zero them when the gate is off \cite{dogan_demonstration_2022, ficheux_fast_2021, ciani_microwave-activated_2022, nesterov_cnot_2022, nesterov_microwave-activated_2018}. This is achieved with our coupler-based architecture \cite{sung_realization_2021, yan_tunable_2018}, following the work of Ding et al. \cite{ding_high-fidelity_2023}. In particular, two coupling pathways, each contributing to the ZZ interaction, cancel each other: virtual excitation exchange through the resonator, and virtual excitation exchange directly between fluxoniums via a capacitor that connects them (see Fig. \ref{fig:fig1}(a)). We tune the circuit parameters such that the total effect is to suppress the always-on ZZ-interaction strength to the $\sim 1$ kHz level.  

We also considered cross-talk in designing the control scheme. By ensuring the appropriate hierarchy of charge matrix elements, we reduce spurious driving of fluxonium transitions when charge modulating the resonator. Since the coupler charge matrix element ($\sim 1$) is larger than any of the fluxoniums' (which have a maximum charge matrix element of $\sim 0.5$), the microwave power needed for fast excitation is reduced, and its effect on any nearby fluxonium transitions is already slightly suppressed. Simultaneously, the fluxonium's computational state transition charge matrix element may remain small ($< 0.05$) to protect from dielectric loss rates \cite{somoroff_millisecond_2021, koch_charging_2009}, as we do not drive fluxonium computational state transitions directly \cite{nesterov_cnot_2022, nesterov_microwave-activated_2018, ficheux_fast_2021}. Further, driving the coupler specifically may reduce nearest-neighbor cross-talk; approaches requiring only a single control tone acting specifically on the coupler, which is connected to two fluxoniums, were deemed preferable compared to driving the fluxoniums, which would each be connected to four couplers when running a surface code. Specifically, in the presence of cross-talk, driving the resonator may reduce calibration time because of the smaller cross-talk matrix with nearest neighbors ($3\times 3$ vs. $5 \times 5$). However, experimental testing is needed to determine if this minor benefit can be realized, or if more optimized control schemes are needed. Finally, while recent demonstrations of parametric flux gates are impressive \cite{zhang_tunable_2023, moskalenko_high_2022}, we focused our study on microwave-driven gates, avoiding a complex analysis of the sidebands associated with parametric modulation in a larger circuit. 

Additionally, we sought a gate that would tolerate disorder and noise within the circuit. We required robustness to the sizeable loss rates of non-computational, higher excited states of the fluxonium. This narrowed the design space to gates which drive state-selective excitations of the resonator coupler, which should boast a long lifetime for high gate fidelity when it is fabricated out of a high quality material. Robustness to junction and flux variation is also required at typical experimental levels \cite{zhang_universal_2021, sun_characterization_2023}.
To maintain flux-noise protection of the fluxonium coherence, we use low $E_L \lesssim 1$ GHz. Geometric mutual inductive coupling may inefficiently compete for coupling strength against fluxoniums' large inductive shunt, while galvanic coupling may introduce cross-talk. Therefore, we use capacitive rather than inductive couplings \cite{ma_native_2023, zhang_tunable_2023}. Robustness to junction mistargeting is numerically evident, potentially because the key interactions involved in our gate are near-resonant, leading to small shifts in gate speeds when the spectrum is perturbed.

Finally, for quantum error correction, our architecture must suggest a path to feasible capacitive loading in a larger circuit with increased coordination number. This is a key risk for fluxonium-based architectures \cite{zhang_tunable_2023, ding_high-fidelity_2023, nguyen_scalable_2022} which, to our knowledge, have not yet been coupled at high connectivity. As discussed above, the wide range of impedances available to electromagnetic resonators should enable increased connectivity beyond the impressive two-qubit demonstrations shown to date.
 
\subsection{Hamiltonian}

Having provided a motivation for our general circuit architecture, we now turn to a detailed description of the setup.
Our proposed circuit is described by a Hamiltonian $\mathcal{\hat{H}} = \mathcal{\hat{H}}_0 + \mathcal{\hat{H}}_{int}$, where the uncoupled Hamiltonian is 
\begin{equation}\label{Hamiltonian_0}
\mathcal{\hat{H}}_0/h = \sum_{i = A, B} 4 E_{C, i} \hat{n}_i^2 + \left( \frac{E_{L, i}}{2} \right) \hat{\phi}_i^2 + E_{J, i} \cos \left(\hat{\phi}_i \right) + \frac{\omega_c}{2\pi} \hat{c}^\dagger \hat{c},
\end{equation}
and the interaction Hamiltonian is 
\begin{equation}\label{H_int}
\mathcal{\hat{H}}_{int}/h =
 J_{A, c} \hat{n}_A \hat{n}_c + J_{B, c} \hat{n}_B \hat{n}_c + J_{A, B} \hat{n}_A \hat{n}_B.
\end{equation}
The bare Hamiltonian $\mathcal{\hat{H}}_0$ contains the fluxoniums' uncoupled Hamiltonian (charge and phase operators $\hat{n}_i$, $\hat{\phi}_i$, junction energies $E_{J, i}$, inductive energy $E_{L, i}$, and capacitive energy $E_{C, i}$), along with the coupler resonator (photon annihilation operator $\hat{c}$, charge number operator $\hat{n}_c \equiv \sqrt{\frac{R_K}{16\pi Z_c}}i(\hat{c}^\dagger - \hat{c})$, where $Z_c$ is its impedance and $R_K \equiv h/e^2$ is the resistance quantum). For both fluxoniums, the external flux values are $\phi_e^{A(B)} = \pi$, corresponding to the ``sweet spot" point of maximum coherence time. The interactions $\mathcal{\hat{H}}_{int}$ in Eq. \eqref{H_int} are all pairwise charge couplings. We diagonalize the static Hamiltonian $\mathcal{H}$, neglecting the drive, to extract the dressed states under the two-body couplings. In our circuit, the computational states are $\ket{i, 0, j}, i, j \in \{0, 1\}$, where state indices from left to right are \textit{fluxonium A, coupler, fluxonium B}. The notation $\ket{i,k,j}_{b}$ denotes an eigenstate of $\mathcal{\hat{H}}_0$, and $\ket{i,k,j}$ denotes an eigenstate of $\mathcal{\hat{H}}$. An eigenstate $\ket{i, k, j}$ corresponds to the state of maximum overlap with the bare excitations $\ket{i, k, j}_b$. 

For representative circuit parameters, see Table \ref{table:1}, with candidate capacitance matrices and additional exemplary parameter sets in Appendix \ref{appendix_a_circuit}. For example, see Appendix \ref{appendix_a_circuit} for candidate capacitance matrices, trading off capacitive loading on the fluxoniums with increased resonator impedance for implementation in a surface code. We choose parameters and coupling strengths such that dominant static hybridizations between the coupler and fluxoniums do not exceed $\sim$15\%, although smaller hybridization at the level of $\sim$10\% may be possible for increased resonator impedance (something that, depending on material choice, could be achieved while still retaining a high quality factor, although further research is needed \cite{shearrow_atomic_2018, chang_improved_2013, vissers_low_2010, amin_loss_2022, shi_tantalum_2022, place_new_2021}). This constraint promotes feasibility of our gate when extending it to a degree-four connected lattice, as required for surface code quantum error correction. While the hybridization is relatively large, it is specific to the fluxonium $1 \rightarrow 2$ plasmon transitions, while other transitions are minimally hybridized and therefore may be leveraged for other functions in a quantum processor, e.g., readout. In particular, these hybridizations are between coupler one photon states and fluxonium plasmon excitations, and are defined as:
\begin{eqnarray*}
 h_A & \equiv |\bra{2,0,1}_{b}\ket{1,1,1}|^2, \,\,\,\,
 h_B \equiv |\bra{1,0,2}_{b}\ket{1,1,1}|^2.
\end{eqnarray*}
These are chosen for a combination of optimal gate performance, as detailed in this work, and feasibility of integration into a larger circuit. 

\subsection{Basic gate mechanism}
As shown in Fig. \ref{fig:fig1}(b) and mentioned above, our CZ gate is realized through an excitation of a single photon in the coupler that occurs only when the fluxoniums are both in the first excited state, i.e., the computational $\ket{11}$ state. Specifically, a geometric phase of $\pi$ is accumulated as the coupler is driven to its first excited state and back (solid green arrow, Fig \ref{fig:fig1}(b)) \cite{ding_high-fidelity_2023}. To achieve the state-selective nature of the coupler excitation, and therefore a state-selective phase on the computational states as required for a CZ gate, we couple the circuit elements such that the resonator frequency strongly depends on the state of the fluxonium qubits. Specifically, we selectively couple the fluxonium plasmon (1-2) transition and coupler, and their virtual energy exchange provides the key dispersive coupling strengths $\chi_{ij}$ and nonlinearity $\alpha$ required for the gate. All other virtual transitions are ideally irrelevant to the gate dynamics and are constructed to have minimal hybridization. 

During the gate, we charge modulate the resonator, driving it at a frequency corresponding to its resonance when the fluxoniums are in the joint $\ket{11}$ state \cite{ding_high-fidelity_2023}. If the fluxoniums' state is $\ket{11}$ at the start of our gate, the resonator is excited from its ground state, to a single photon excitation and back, creating a geometric phase of $\pi$. The next resonator excitation is detuned (green dashed arrow with $\alpha$ label, Fig. \ref{fig:fig1}(b)) and the second photon state remains unpopulated, due to a nonlinearity $\alpha$ inherited from the fluxoniums further described below. If the fluxoniums are in any other state, the drive is again detuned (Fig. \ref{fig:fig1}(b), green dashed arrows with $\chi_{ij}$ labeling), the resonator is not excited, and no phase is accumulated, thus completing the CZ unitary. 

In our simulations, the charge drive on the resonator is a pulse with a Gaussian envelope of width $\tau / \sqrt{2}$. (Alternate pulses such as a cosine shape may equally be used). The pulse amplitude is $\Omega(t) \equiv \Omega_0 e^{-4t^2 / \tau^2}$, where $\Omega_0 \approx 2 /\sqrt{\pi} n_{c, 01} \tau$, as required for a full $2\pi$ rotation. Here, $n_{i, jk}$ denotes the matrix element of the $\ket{j}_{bare}-\ket{k}_{bare}$ transition of circuit element $i \in \{\text{fluxonium A, fluxonium B, coupler} \}$. To smoothly truncate the Gaussian pulse envelope to a finite gate time, we also add a linear ramp up from zero to $\Omega(-\tau)$ and down from $\Omega(\tau)$ to zero, over a time $\tau/10$ leading to a full gate duration $t_g = 2.2 \tau$ (see Fig.~\ref{fig:fig1}(b), inset). Following \cite{ding_high-fidelity_2023}, the pulse resonantly excites $\ket{1, 0, 1}$ to $\ket{1, 1, 1}$ and back (see green solid arrow in 
Fig.~\ref{fig:fig1}(b)), where the relevant states are shown in black horizontal bars. We always drive the state-selective coupler excitation via a charge drive, as its large matrix element facilitates efficient driving with minimal induced Stark shifts or population leakage. The drive frequency $h\omega_d / (2\pi) \approx \langle \mathcal{H} \rangle_{111} - \langle \mathcal{H} \rangle_{101}$ excites the coupler conditional on the fluxoniums' $\ket{11}$ computational state. 

Focusing on the relevant states in the dressed basis, the interaction Hamiltonian $\mathcal{H}_{\rm int}^d$ includes the terms:
\begin{equation}\label{Hamiltonian_diagonal}
 \begin{aligned}
 \mathcal{H}_{\rm int}^d /h = \chi \ket{1, 1, 1}\bra{1, 1, 1} + (\alpha + \chi) \ket{1, 2, 1} \bra{1, 2, 1} \\ +
 \sum_{i, j \neq \{1, 1\}} \chi_{ij} \ket{i, 1, j}\bra{i, 1, j} + ...
 \end{aligned}
\end{equation}

The crucial interaction terms are in the first line: the effective frequency of the coupler is shifted by a dispersive coupling strength $\chi$ when the fluxoniums are in computational state $\ket{11}$, which enables selective driving of the $\ket{1, 0, 1}$ to $\ket{1, 1, 1}$ transition that is at the heart of our gate. The inherited coupler nonlinearity $\alpha$ prevents the drive from adding more than just a single photon to the coupler. Here, $\alpha$ is induced by hybridization of the coupler (a linear resonator) and the highly nonlinear fluxoniums. It sets a natural speed limit on our gate; for gate times shorter than $\sim 1/\alpha$, this nonlinearity is not sufficient to prevent driving of the $\ket{1,2,1}$ state from influencing and corrupting the gate dynamics, primarily through unwanted Stark shifts of the desired $\ket{1, 0, 1}-\ket{1, 1, 1}$ transition. 

The terms in the second line of Eq.~(\ref{Hamiltonian_diagonal}) set another speed limit on the gate, i.e.~$t_{\text{min}} \sim |1/\delta\chi_{ij}|$. For gates faster than this scale, we expect detuned driving of the coupler even if the qubits are not in the $|11 \rangle$ state, leading to Stark shifts and spurious phase accumulation on the computational states 
(see Fig.~\ref{fig:fig1}, green dashed arrows). Both sources of spurious phase accumulation may be compensated to some degree with single-qubit $Z$ rotations following the gate. However, for general $\chi_{ij}$ and $\alpha$, there will be some state-selective nature to the spurious phases, which cannot be mitigated with single qubit rotations. 

As described below, we find circuit parameters such that $\chi_{ij} -\chi \sim \alpha$. We show that in this case, the spurious phases accumulated on all four computational states are almost equivalent to single-qubit $Z$ rotations (and hence can be trivially eliminated via virtual $Z$ gates). This fortuitous behavior is the result of a destructive interference effect, and results in high CZ fidelities.
With additional fine-tuning of the drive, this destructive interference effect persists even when $t_{g} \sim 1/|\chi_{ij}-\chi| \sim 1/\alpha$, leading to efficient, fourth-order scaling of coherent error with gate duration. Throughout this work, we simulate our system using scqubits \cite{chitta_computer-aided_2022, groszkowski_scqubits_2021}.

Finally, there is one additional problematic interaction in $\mathcal{H}_{\rm int}^d$: fluxonium-fluxonium $Z_A Z_B$-interactions in the computational manifold of the form $(\ket{1_A}\bra{1_A} - \ket{0_A} \bra{0_A}) (\ket{1_B}\bra{1_B} - \ket{0_B} \bra{0_B})$, where the interaction strength is given by $h \eta\equiv \langle \mathcal{H}\rangle_{101} - \langle \mathcal{H}\rangle_{100} - \langle \mathcal{H}\rangle_{001} + \langle \mathcal{H}\rangle_{000}$. This interaction generates parasitic entanglement between qubits when the gate is off, and reduces the microwave-activated nature of the gate. We impose a requirement in our circuit design that $|\eta| < 2$ kHz in order to reduce the associated error of single-qubit gates faster than $100$ns to below $\sim 10^{-4}$. To do so, inspired by previous work \cite{ding_high-fidelity_2023}, we engineer the capacitance matrix to realize small $\eta$ (Appendix \ref{appendix_a_circuit}). For the circuit parameters studied here, $\eta \approx -1$ kHz. We note that small $\eta$ can be achieved independently of our choice of frequency allocations and coupling strengths (Appendix \ref{appendix_c_hamiltonian}). 

\subsection{\label{sec:level2}Designing circuit parameters for low error}

As discussed above, the nonlinear couplings $\chi_{ij}, \alpha$ in Eq.~\ref{Hamiltonian_diagonal} set speed limits on our gate time. Here we analytically estimate these parameters and directly relate them to circuit parameters. 
We then leverage this understanding to choose frequency allocations and coupling strengths to maximize $|\chi-\chi_{ij}|$ and $\alpha$ values. We favor small hybridization to minimize inherited dissipation rates from non-computational, lossy states of the fluxoniums (see Table \ref{table:1}). We next estimate the resulting coherent errors as a function of gate duration, with qualitative agreement to numerical simulations. These estimates motivate our circuit parameter and control choices: by fine-tuning the drive frequency to eliminate any remaining spurious phases, both sources of unwanted Stark shifts destructively interfere to enable fast gates with low coherent errors.

Our goal is to maximize $\chi_{ij}-\chi \sim \alpha$, with a fixed, tolerable amount of qubit-resonator hybridization. Our approach is to strongly couple the resonator one-photon state specifically with the fluxoniums' highly nonlinear plasmon transition to ensure sufficient $\chi-\chi_{ij}, \alpha$. We first employ perturbation theory in the fluxonium-resonator couplings to understand the dependence of the dispersive couplings $\chi_{ij}$ in Eq.~(\ref{Hamiltonian_diagonal}) on circuit parameters. 
For our choice of circuit parameters (which are typical, see Appendix \ref{appendix_c_hamiltonian}), the dominant virtual transitions contributing to the $\chi_{ij}$ values involve the fluxonium 1-2 and 0-3 transitions. For example, to second order in perturbation theory, $\chi$ can be estimated as
\begin{equation}\label{chi_11_estimate}
 \begin{aligned}
 \chi^{(2)} \approx -\frac{J_{A, c}^2 |n_{A, 12} n_{c, 01} |^2}{\Delta_A} - \frac{J_{B, c}^2 |n_{B, 12} n_{c, 01}|^2}{\Delta_B} \\
 \end{aligned}
\end{equation}
where we use $\chi^{(n)}$ to denote the $n$th order correction to $\chi$. The charge matrix elements are described with notation $n_{i, kl}$ for circuit element $i$, on the $\ket{k}-\ket{l}$ transition, and $\omega_{kl}^{(i)}$ is its transition angular frequency. We define $(2\pi) \Delta_i \equiv \omega_{12}^{(i)}-\omega_c$, where $\omega_c$ is the bare coupler angular frequency and $i \in \{A, B\}$ (assumed throughout the rest of the text). 

To ensure large $\delta \chi_{ij} \equiv \chi-\chi_{ij}$, we choose $(2\pi)\Delta_i \ll \omega_c, \omega_{12}^{(i)}$ and $|J_{i, c} n_{i, 12} n_{c, 01} / \Delta_i| \lesssim 1$. In this regime, 
the only significant virtual transitions correspond to the processes in Eq.~\eqref{chi_11_estimate}. The coupler-plasmon exchange is no longer strictly perturbative; nonetheless, we can pick parameters to yield both large $\delta \chi_{ij}$ and only moderate hybridization $h_{A(B)} \approx 0.15$ (while also keeping other virtual-transition pathways strongly detuned, see Appendix \ref{appendix_c_hamiltonian} for details). 
As the fluxoniums' plasmon transitions are highly anharmonic, for a given hybridization between elements, they impart a strong nonlinear character into the system. 
By setting $\Delta_i \sim 0.5 $ GHz, we then tune the hybridization between elements, by design almost entirely due to to the plasmon-coupler exchange, ensuring satisfactory hybridization for scaling and maintaining long coupler lifetimes.

\begin{figure}[h]
\includegraphics[width=0.48\textwidth]{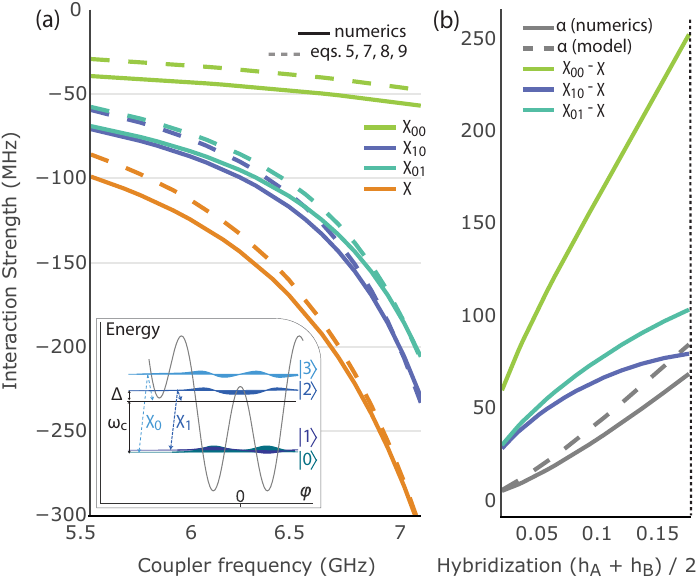}
\caption{ Key interaction parameters. 
(a) The dispersive couplings $\chi_{ij}$ between the resonator and fluxoniums for the fluxonium state $\ket{i_A, j_B}$. Solid lines are results extracted by numerically diagonizing $\mathcal{\hat{H}}$, while the dashed lines are our estimates in e.g.,
Eqs.~\eqref{chi_11_estimate}, \eqref{chi_10_estimate}. Inset: representative spectrum for fluxonium A \cite{groszkowski_scqubits_2021}. Grey solid line is the potential, while the solid black line represents the energy of a single photon in a resonator. Predominant perturbative pathways setting the dispersive interaction between fluxonium A and the coupler are shown. Crucially, $\chi_1^A \gg \chi_0^A$ since $\omega_{23}^{A(B)} / (2\pi) \gg \Delta_{A(B)}$ (see text for details). This enables large $\chi_{ij} - \chi$, a key requirement for reduced coherent errors. (b) Inherited coupler anharmonicity and $\chi_{ij}-\chi$ values, sweeping coupler frequency over the same values in (a), but instead plotting against the resulting hybridization. Solid lines are numerically exact results, while the dashed line follows our model diagonalizing within the plasmon-coupler exchange subspace. We choose a regime where $\alpha \sim \chi_{ij}-\chi$ (vertical black dashed line) for small coherent errors. }\label{fig:fig2}
\end{figure}

In this regime, the energy shifts arising from plasmon-coupler virtual exchange go beyond leading-order perturbation theory, but can be accurately captured by a simple model where the plasmon transition of each fluxonium is treated as an effective two-level qubit. 
Then, $\chi$ is obtained by diagonalizing $\mathcal{H}$ within the plasmon-coupler exchange subspace:
\begin{equation}\label{cqed_chi_11}
\chi \approx \frac{\Delta}{2} - \frac{1}{2}\sqrt{8 g^2 + \Delta^2},
\end{equation}
where we have treated both fluxoniums plasmon transitions as identical, approximating $\Delta_A = \Delta_B = \Delta $ and $J_{A, c}n_{A, 12} n_{c, 01} = J_{B, c}n_{B, 12} n_{c, 01} = g$ for simplicity.
The agreement between our model (Eq.~\eqref{cqed_chi_11}, 
Fig.~\ref{fig:fig2}(a), dashed orange line) and numerics 
(Fig.~\ref{fig:fig2}(a), solid orange line) persists even beyond the perturbative regime as $J_{i, c} n_{i, 12} n_{c, 01}/\Delta_i \rightarrow 1$. 

To selectively drive the coupler excitation conditional on the $\ket{1, 0, 1}$ state, we require $|\delta\chi_{ij}| > 1/t_g$, to avoid exorbitant Stark shifts of the $\ket{i, 0, j} \neq \ket{1, 0, 1}$ computational states. Our approach is to reduce $\chi_{ij}$ by suppressing the coupler frequency shift $\chi_0^i$ when fluxonium $i$ is in the ground state, the strongest contribution to which is virtual excitation exchange of a coupler photon with the fluxonium's 0-3 transition. This is achieved by increasing the 0-3 transition frequency well beyond $\omega_{12}^{i}$, and choosing a plasmon-above-coupler frequency allocation (Fig. \ref{fig:fig2}(a), inset), thereby maintaining large $|\delta \chi_{ij}|$.  For example, consider the unwanted dispersive shift $\chi_{10}$ in Eq.~\ref{Hamiltonian_diagonal}. Perturbation theory yields
\begin{equation}\label{chi_10_estimate}
 \begin{aligned}
 \chi_{10}^{(2)} \approx -\frac{J_{A, c}^2 |n_{A, 12} n_{c, 01} |^2}{\Delta_A} - \frac{ (2\pi) J_{B, c}^2 |n_{B, 03} n_{c, 01}|^2}{(\omega_{03}^B - \omega_c)} \\
 \end{aligned}
\end{equation}
The second term corresponds to virtual exchange between the coupler and the fluxonium 0-3 transition; it 
increases $|\chi_{10}|$ due to the large charge matrix element $n_{i, 03} \approx n_{i, 12} \approx 0.5$ (light blue dashed arrows, Fig. \ref{fig:fig2}(a), inset). 
This is deleterious for our gate, as it effectively reduces the state-selectivity of the coupler excitation $\delta\chi_{10}$, since $\chi_{10} \rightarrow \chi$ as $\omega_{12}^B \rightarrow \omega_{03}^B$. To compensate, we increase $E_C/E_J$ to about 0.3,
which effectively pushes the fluxonium second and third excited states out of well formed by the miminima of the cosine potential. This increases $\omega_{23}^{i} / (2\pi)$ by a significant amount to $\sim 2$ GHz.
(Alternatively, to facilitate capacitance budgeting in a larger architecture, $E_L/E_J$ may be increased instead to e.g. $E_L \sim 0.7$ GHz and $E_C \sim 1.3$ GHz). Finally, using a plasmon-above-coupler frequency allocation ensures that the plasmon transition will be closest to the coupler excitation compared to the fluxoniums' 0-3 excitation, as desired. 

We estimate $\chi_{ij}$ by again diagonalizing within the plasmon-coupler excitation exchange subspace. 
To treat the $\ket{0}_i$ states, we simply add the relevant second order shift from the 0-3 contributions, which by design are small. The $\chi_{ij}$ values are approximately
\begin{align}
\chi_{10} = \frac{\Delta_A - \sqrt{4 g_A^2 + \Delta_A^2}}{2} - \frac{ (2\pi) J_{B,c} |n_{B, 03} n_{c, 01}|^2 }{\omega_{03}^B - \omega_c} \\
\chi_{01} = \frac{\Delta_B - \sqrt{4 g_B^2 + \Delta_B^2}}{2} - \frac{ (2\pi)J_{A,c} |n_{A, 03} n_{c, 01}|^2 }{\omega_{03}^A - \omega_c} \\ 
\chi_{00} = - \frac{(2\pi)J_{A,c} |n_{A, 03} n_{c, 01}|^2 }{\omega_{03}^A - \omega_c} - \frac{ (2\pi)J_{B,c} |n_{B, 03} n_{c, 01}|^2 }{\omega_{03}^B - \omega_c}
\end{align}
where $g_{A(B)} = J_{A(B), c} |n_{A(B), 12} n_{c, 01}|$. These estimates for $\chi_{01}, \chi_{10}$ and $\chi_{00}$ (Fig. \ref{fig:fig2}(a), dashed green and blue lines) show strong agreement with the numerically exact results (solid lines), confirming our understanding of the key coupling mechanisms in our system. 
Throughout the coupler frequency sweep, we achieve frequency-selectivity of the coupler excitations of $|\delta\chi_{ij}|\approx 100$ MHz. 

In addition to coherent errors arising from a finite 
$|\delta\chi_{ij}|$, spurious non-resonant driving of the coupler two photon state must also be addressed. Recall that the inherited nonlinearity $\alpha$ of the coupler is given by the shift of the $\ket{1, 2, 1}$ state. In our chosen parameter regime, $\alpha$ is largely due to virtual transitions involving fluxonium plasmon excitations, and can thus be obtained to good approximation from our effective model (where each plasmon transition is treated as an effective two-level qubit coupled to the coupler resonator). 
The value of $\alpha$ obtained from full numerics (grey solid line in Fig. \ref{fig:fig2}(b)) approximately follows the predictions of our effective model (grey dashed line, Fig. \ref{fig:fig2}(b)).  

Finally, coherent errors can be further reduced by employing destructive interference between the unwanted phases generated by the two mechanisms discussed above (finite $\alpha$, finite $\delta \chi_{ij}$). For this interference to be possible, we require approximately $\alpha \sim \delta \chi_{ij}$. Although the $|\delta \chi_{ij}|$ values are similar for a wide range of hybridization values, we must increase the hybridization for sufficient $\alpha$. For the resonator impedance in Table \ref{table:1}, this is around a hybridization of 20\% and 14\% for $h_A$ and $h_B$, respectively (black dashed vertical line, Fig. \ref{fig:fig2}(b)). 
Indeed, we find that by requiring coupler anharmonicities of at least $\sim 70$ MHz $\sim \delta\chi_{ij}$, we can achieve low coherent error rates. 
For the remainder of this paper, we fix the coupler anharmonicity at $\alpha \sim 70$ MHz, resulting in circuit parameters shown in Table \ref{table:1}, with the coupler frequency at 7.08 GHz. By exploiting fluxoniums' strong nonlinearity, we have maximized $|\delta\chi_{ij}|$ and $\alpha$, with feasible hybridization, by choosing small $\Delta$, large $\omega_{03}^i$, and a plasmon-above-coupler frequency allocation. 
\section{Coherent error}

Having identified circuit parameters that are both realistic and optimized for our chosen gate, we now focus on understanding and mitigating sources of coherent gate error, using the standard metric of state-averaged coherent gate fidelity \cite{pedersen_fidelity_2007}:
\begin{equation}
F_c = \frac{\text{Tr}(\hat{U}^\dagger \hat{U}) + |\text{Tr}(\hat{U}_{\text{CZ}}^\dagger \hat{U})|^2}{20},
\end{equation}
where $\hat{U}$ is the propagator of the simulation including coherent errors, and $\hat{U}_{\text{CZ}}$ is the target ideal CZ unitary. We apply our understanding of the Hamiltonian shown in the previous section to study the time-domain behavior of our circuit under the gate control sequence. For the parameters of interest and smooth pulse envelopes, we find that the dominant coherent errors are unwanted Stark shifts arising from detuned driving of the coupler. This includes spurious driving of the coupler single-photon state when the qubits are in computational states other than $\ket{11}$, and spurious driving of the second coupler excited state when the qubits are in the $\ket{11}$ state. Population leakage errors can be suppressed by using gate durations $t_g \gtrsim 1/\alpha$ (where $\alpha$ is the induced coupler nonlinearity, c.f.~Eq.~\eqref{Hamiltonian_diagonal}) \cite{ribeiro_systematic_2017, setiawan_analytic_2021}. Here, we estimate each effect on coherent error and find qualitative agreement with time-domain simulations of our circuit and gate control, then use our model to suggest further improvements to fidelity. In particular, we identify destructive interference that leads to strong suppression of the Stark-shift induced errors. 

\subsection{\label{sec:level3} Stark shifts due to unwanted driving of coupler $\ket{1}$ state}
Applying charge driving to the coupler includes modulation on all transitions with a nonzero charge matrix element. Specifically, other coupler excitations, such as $\ket{i, 0, j}-\ket{i, 1, j}$ with $(i, j) \neq (1, 1)$, have about the same matrix element as the desired transition and experience equally strong modulation. The finite state-selectivity of the coupler frequency results in detuned driving of these transitions, leading to additional phase shifts $\varphi_{ij}$ on the computational states (see Fig. \ref{fig:fig1}(b), green dashed arrows for an illustration). We approximate the drive envelope as a Gaussian for all time, without linear truncation (see Fig \ref{fig:fig1}(a)), to estimate the Stark shifts. Integrating the Stark shift from the slowly-varying drive amplitude $\Omega(t) \equiv 2 e^{-4 t^2/\tau^2} / \sqrt{\pi} \tau |\bra{1, 1, 1} n_c \ket{1, 0, 1}|$ with detuning $\delta\chi_{ij}$ on the $\ket{i, 0, j}-\ket{i, 1, j}$ transition, we estimate $\varphi_{ij}$ for $ij \neq 11$, where $\ket{i, 0, j} \rightarrow e^{-i \varphi_{ij}} \ket{i, 0, j}$ under the gate, as
\begin{equation}\label{stark_shift}
\varphi_{ij} = \sqrt{\frac{\pi}{2}} \frac{|m_{ij}|^2}{\delta \chi_{ij}\tau},
\end{equation}
where $m_{ij}= \bra{i, 1, j} \hat{n}_c \ket{i, 0, j} / \bra{1, 1, 1} \hat{n}_c \ket{1, 0, 1}$ is the ratio of the matrix elements of the drive acting on the driven transition compared to the transition being Stark shifted ($\ket{i, 0, j}-\ket{i, 1, j}$. For small hybridization, $m_{ij} \approx 1$. In Fig.~\ref{fig:fig3}(a), we plot Eq.~\eqref{stark_shift} for $\varphi_{10}$ (dashed navy line) and $\varphi_{01}$ (dashed teal line) for a range of $t_g$ values, which reveals $\varphi_{ij}$ values of more than 10 degrees for $t_g < 100$ ns, a potentially significant source of error. Simulating our circuit and gate control scheme in the time domain, we plot numerics results for $\varphi_{01} - \varphi_{00}$ and $\varphi_{10} - \varphi_{00}$ as accumulated on the superposition $\psi_{ij}(0) = (\ket{0,0,0} + \ket{i, 0, j})/\sqrt{2}$ under our gate (Fig. \ref{fig:fig3}(a), dots). All simulations in this work are performed in the dressed basis, with the drive operator $\Omega(t) \hat{n}_{c}\cos{\omega_d t}$ proportional to the bare coupler operator $\hat{n}_c$, also written in the dressed basis. The simulations occur in the lab frame, without a rotating wave approximation. We include the full cosine potential of the fluxoniums, and truncate the system Hilbert space to include the 45 lowest-energy dressed eigenstates. We find $0.01\%$ relative variation in coherent error for multiple gate times, sweeping the simulation dimension from 40 to 50 (Appendix \ref{appendix_f_methods}).

\subsection{\label{sec:level} Stark shifts due to unwanted driving of coupler $\ket{2}$ state}
Similarly, an additional phase shift $ \varphi_{11}$ from spurious detuned driving of the coupler $\ket{2}$ excitation leads to deviation from the ideal gate by imparting an extra phase $\varphi_{11}$ on the $\ket{1, 0, 1}$ computational state, causing an overshoot or undershoot of the ideal phase of $\pi$. We plot the relative phase for $\varphi_{11}$ as found under a time-domain simulation the gate dynamics (green dots, Fig.~\ref{fig:fig3}a), and find that it agrees well with our model of phase accumulated from detuned driving of the next coupler excitation (green line, Fig.~\ref{fig:fig3}a). Following the approach from \cite{ribeiro_systematic_2017}, our estimate for $\varphi_{11}$ is given by a second-order Magnus expansion in the drive terms on the second coupler excitation, in a rotating frame and in an interaction picture as described subsequently. We consider a Hamiltonian $H = H_0 + V$, where the ``desired" evolution is $H_0 / h = \frac{\Omega(t) n_{c, 01}}{2} \ket{1, 0, 1} \bra{1, 1, 1} + \text{H.c.}$, in a frame rotating at the drive frequency close to resonance with the $\ket{1,0, 1}-\ket{1, 1, 1}$ transition and under a rotating-wave approximation. We account for the second order Magnus expansion evolution $\hat{\theta}_2(t)$ contributions to the evolution
\begin{equation}\label{theta_2_t}
\hat{\theta}_2(t) = -\frac{1}{2\hbar^2}\int_{-\infty}^{t_1}\int_{-\infty}^{t_2} [\hat{V}(t_1), \hat{V}(t_2)] dt_1 dt_2,
\end{equation}
where the perturbation 
\begin{equation}\label{V}
\hat{V}/h = \frac{\Omega(t)}{2} n_{c, 12}e^{2\pi i \alpha t} \ket{1, 2, 1}\bra{1, 1, 1} + \text{H.c.}
\end{equation}accounts for the drive on the next coupler excitation. Solving for $\hat{\theta}_2(t)$, we neglect terms proportional to $\dot{\Omega}(t)$, approximate the pulse envelope as Gaussian and assume $\Omega(-\infty) = \Omega(\infty) = 0$ to find $\hat{\theta}_2$ of 
\begin{equation}\label{theta_2_t_solved}
\begin{aligned}
\hat{\theta}_2(t) = + 2\pi i \int_{-\infty}^t dt_1 \frac{\Omega(t_1)^2 |n_{c, 12}|^2}{2\alpha} \Big(\cos^2{\theta_0(t_1)} |111\rangle \langle 111| \\ + \sin^2\theta_0(t_1) |101\rangle\langle 101| \\ + \sin\theta_0(t_1)\cos\theta_0(t_1) (-i |101\rangle \langle 111| + i\ket{111}\bra{101}) \\ - |121\rangle \langle 121|\Big),
\end{aligned}
\end{equation}
where $\theta_0(t) /2\pi \equiv \frac{1}{4} \left( 1 + \text{Erf}(2t / \tau ) \right)$ represents the population transfer on the $\ket{1, 0, 1}-\ket{1, 1, 1}$ transition.

We are most interested in phase accumulated on the $\ket{1, 0, 1}$ state, such that under the gate $\ket{1, 0, 1} \rightarrow - e^{-i\varphi_{11}} \ket{1, 0, 1}$, where $\varphi_{11}$ should be close to zero. Once again, we assume a first-order Magnus expansion of the dynamics so that each term can be treated separately, and focus on the dispersive effects. Integrating the coefficient to the $\ket{1, 0, 1}\bra{1, 0, 1}$ to extract $\varphi_{11}$, we arrive at $\varphi_{11}$ of
\begin{align}\label{varphi_11}
 \varphi_{11} \approx \frac{-\pi |n_{c, 12}|^2}{\alpha} \int_{-\infty}^{\infty} \Omega(t)^2 \sin^2{\theta_0(t)} dt \nonumber \\ 
 = \frac{-1.58}{\tau \alpha}.
\end{align}
We plot our estimate Eq.~\eqref{varphi_11} in Fig.~\ref{fig:fig3}(a), green dashed line, along with the numerically exact values (green dots) and find agreement within $\sim$1 degree, breaking down at short gate durations, which we attribute to the breakdown of the dispersive approximation, since the population leakage becomes significant. 
\subsection{Interference between Stark shift phase errors}

The dominant effect of nonzero $\varphi_{ij}$ values is to add unwanted single-qubit phase accumulation. We mitigate these with the addition of virtual gates, by adding optimized single-qubit $Z$ rotations onto the qubits at the end of our simulations. In an experiment, single-qubit $Z$ gates may be trivially performed in software. However, a linear combination of the $\varphi_{ij}$ values, corresponding to the joint $Z_A Z_B$ rotations, remains. In particular, the entangling power of our gate is set by the relative phase $\varphi$
\begin{equation}\label{delta_varphi}
\varphi = \varphi_{11} + \varphi_{00} - \varphi_{01} - \varphi_{10}
\end{equation}
which measures how close our gate is to a maximally entangling $Z_A Z_B$ gate modulo single qubit operations; $\varphi = 0$ is the ideal case. A non-zero $\varphi$ results in non-zero coherent error, which for small $\varphi$ is given by $ \epsilon_{ \varphi} \equiv 3 \varphi^2 / 20$ (Appendix \ref{appendix_f_methods}), \cite{ding_high-fidelity_2023}. 

\begin{figure}
 \centering \includegraphics[width=0.45\textwidth]{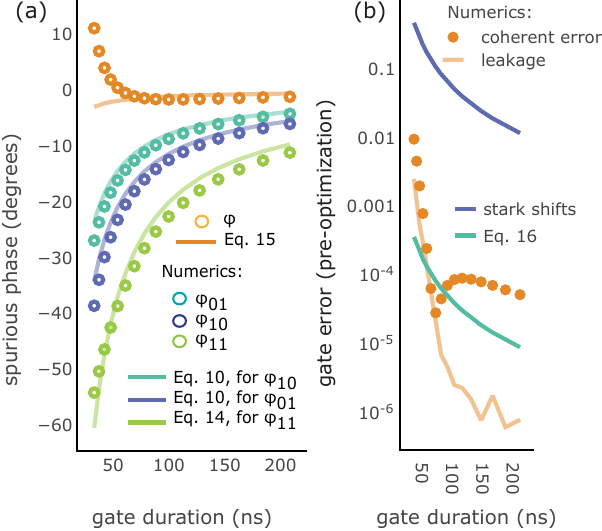}
 \caption{Coherent dynamics and error, driving the $\ket{1,0,1 }-\ket{1, 1, 1}$ transition on resonance. (a) Phase accumulated on the computational states as a function of gate duration. Numerically exact results are shown with dots, while Eqs. \eqref{stark_shift}, \eqref{varphi_11}, and \eqref{delta_varphi} are given by the transparent lines. (b) Resulting coherent error $\epsilon_\varphi$ as a function of gate duration. Numerically extracted values for coherent error are brown dots, while the numerically extracted values for population leakage are shown as the transparent line. Our model for coherent error given by Eq. \eqref{approx_coherent_error} is shown as the green line. The similarities between our model and the numerically exact results implies that the majority of the coherent error is dispersive in nature, prompting additional control optimization described below. The $\sim$10x higher coherent error from the Stark shifts on the $\ket{i, 0, j}, i, j \neq 1, 1$ states is shown by the blue line, demonstrating the destructive interference between the two coherent error sources occurs.}
 \label{fig:fig3}
\end{figure}

For our system and parameters, we find that, remarkably, the two mechanisms for unwanted Stark shifts (spurious driving of coupler $\ket{2}$ excitation versus spurious driving of its $\ket{1}$) excitation largely cancel in their contribution to the parameter $\varphi$. Following Eqs. \eqref{stark_shift} and \eqref{varphi_11}, and \eqref{delta_varphi}, we find that the gate infidelity for long times scales as:
\begin{align} \label{approx_coherent_error}
 \epsilon_\varphi & \simeq
 \frac{1}{\tau^2} \left[
 \frac{c_1}{\alpha } 
 + c_2 \left(
 \frac{|m_{01}|^2}{\delta \chi_{01}} + \frac{|m_{10}|^2}{\delta \chi_{10}}
 - \frac{|m_{00}|^2}{\delta \chi_{00}} \right)
 \right]^2
\end{align}
where $c_1 = 0.612, c_2 = 0.485$. 


In Eq.~\eqref{approx_coherent_error}, destructive interference minimizes the dominant sources of coherent error; the first term is positive, and the second term is negative (since $\delta\chi_{ij} < 0$, $\alpha > 0$, and $|\delta\chi_{00}| > |\delta\chi_{10}| \sim |\delta\chi_{01}|$). Combining the $\varphi_{ij}$ values derived above, the resulting $\varphi$ as a function of gate duration is shown in Fig. \ref{fig:fig3}(a). It shows qualitative agreement between the numerically exact results (orange dots) and our estimate from Eqs. \eqref{stark_shift}, \eqref{varphi_11}, and \eqref{delta_varphi} (orange line), in the regime where population leakage is negligible above $t_g \sim 50$. 

In Fig.~\ref{fig:fig3}(b), we plot the resulting estimate for gate error due to Stark shifts (c.f.~Eq.~\eqref{approx_coherent_error}), and compare against the errors found from a full time-domain numerical simulation. Note these results do not include any additional optimization of the drive (which can dramatically decrease the error, as we describe below). 
For gate times $t_g \gtrsim 50 $ns where leakage is minimal, there is a rough order-of-magnitude agreement between the coherent errors found in the full numerical simulation (orange dots) and the analytic prediction based on phase errors (green line). This suggests that for these gate times, Stark shifts are a dominant coherent error mechanism. Deviations from the analytic prediction likely arise as the destructive interference effect in Eq.~\eqref{approx_coherent_error} (which largely cancels the leading order Stark-shift errors) leads to increased sensitivity to higher order terms. 

Fig.~\ref{fig:fig3}(b) also illustrates the strength of this destructive interference effect, by showing results for a case where we only consider Stark shifts on computational states other than $\ket{11}$ (i.e.~for the case $\alpha \rightarrow \infty$). In this case there is no possibility for interference, and we see that the coherent error would increase by a factor of $\sim 10$ (navy blue line). 


\subsection{Additional coherent error supression via drive detuning}\label{sec:drive_optimization}

We achieve significant improvements in the phase error of our gate, beyond Eq.~\eqref{approx_coherent_error}, by shifting the frequency of our state-selective control tone to recover $\varphi \approx 0$. 
We estimate the optimal detuning $\delta\omega /2\pi \equiv \omega_{\text{drive}} / 2\pi - (\langle \mathcal{\hat{H}}\rangle_{111}/h - \langle \mathcal{\hat{H}}\rangle_{101}/h) $ by adding a term $-\frac{\delta \omega}{2\pi} |1, 1, 1\rangle \langle 1, 1, 1|$ to the perturbation $\hat{V}$ in equation \eqref{theta_2_t}, resulting effectively in $\frac{\Omega(t_1)^2}{2\alpha} \rightarrow \frac{\Omega(t_1)^2}{2\alpha} + \delta\omega$ in equation \eqref{theta_2_t_solved}. 
Again integrating the phase on $\ket{1, 0, 1}$, we arrive at an estimated phase incurred on the $\ket{1, 0, 1}$ state of 
\begin{equation}
\varphi_{11}(\delta \omega) = \frac{-1.58}{ \tau\alpha} - 0.515 (\delta\omega) \tau,
\end{equation}
such that the optimal detuning $\delta\omega$ is roughly

\begin{align} \label{optimal_delta}
 \frac{\delta\omega}{2\pi} & \simeq
 \frac{1}{\tau^2} \left[
 \frac{d_1}{\alpha } 
 + d_2 \left(
 \frac{|m_{01}|^2}{\delta\chi_{01}} + \frac{|m_{10}|^2}{\delta\chi_{10}}
 - \frac{|m_{00}|^2}{\delta\chi_{00}} \right)
 \right]
\end{align}
where $d_1 = -0.488, d_2 = -0.387$. 

\begin{figure}[h]
\includegraphics[width=0.45\textwidth]{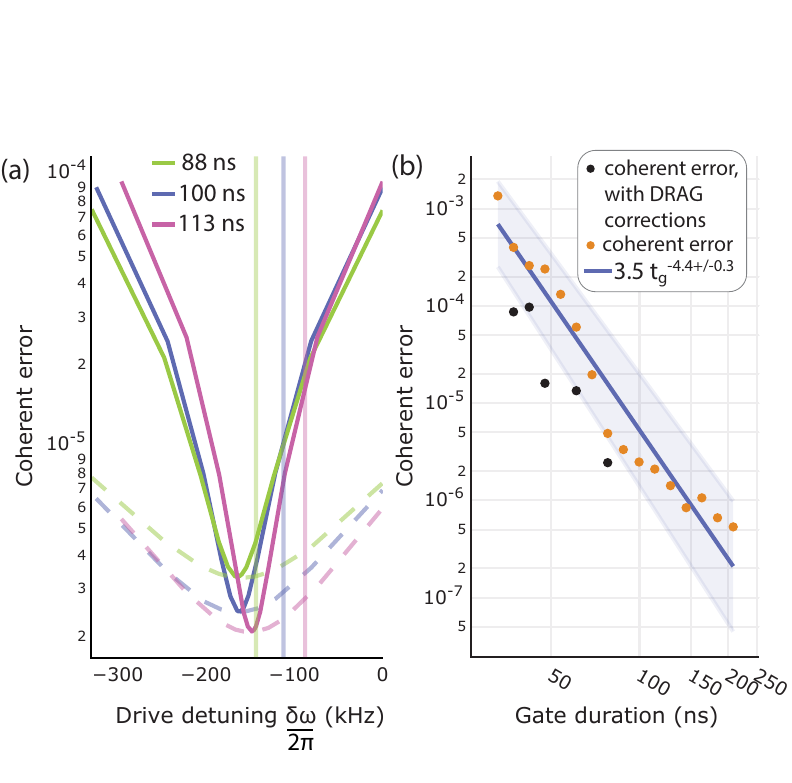}
\caption{Optimizing the drive frequency for minimal coherent error. 
(a) Solid lines: coherent gate error as a function of drive detuning, for three different gate durations. In each case, there is a dramatic reduction of coherent error at an optimal detuning. Vertical lines indicate the analytic prediction for the optimal drive detuning, Eq.~\eqref{optimal_delta}. Dashed lines: error due only to leakage. At the optimal detuning, the total error is almost entirely due to leakage. 
(b) Scaling of coherent error, for optimized drive detuning as a function of gate duration $t_g$. We fit to a power law to recover an efficient scaling $\sim t_g^{-4.4(3)}$, where error bars are $1\sigma$ residuals from the fit. All the coherent error is from population leakage.}\label{fig:fig4}
\end{figure}
As an example, we plot the coherent gate error \cite{pedersen_fidelity_2007} for a total gate duration $t_{g} \equiv 2.2 \tau$ of 88 ns, 100 ns, and 113 ns in Fig. \ref{fig:fig4}(a), (solid lines), sweeping the drive frequency in the time domain simulation through the estimated optimal values from equation \eqref{optimal_delta} (vertical lines). 
For each gate duration $t_g$, there is a clear optimal $\delta \omega$, resulting in over an order of magnitude reduction in coherent error compared to the resonant driving case. We also plot the population leakage as a function of drive detuning as dashed transparent lines in Fig. \ref{fig:fig4}(a); at the numerically optimized detuning, the remaining coherent error is almost entirely leakage, i.e. $\varphi \approx 0$. By design, our optimal detunings are small, since $\varphi$ is already small at zero detuning (due to the destructive interference effect). For larger $\varphi$ values, e.g. with reduced inherited coupler nonlinearity (Appendix \ref{appendix_d_drag}), a larger detuning to set $\varphi \approx 0$ is needed, in which case population leakage into other nearby non-computational states becomes limiting.

\subsection{Leakage errors and DRAG-style pulse corrections}

For our parameters and gate durations $t_g \lesssim 50$ ns, the dominating source of population leakage is excited state population in the $\ket{1, 1, 1}$ and $\ket{1, 2, 1}$ states (Appendix \ref{appendix_d_drag}). We confirm that this residual population is due to evolution under the detuned driving of the $\ket{1, 1, 1}-\ket{1, 2, 1}$ transition, by implementing the Derivative Removal via Adiabatic Gate technique (DRAG) \cite{ribeiro_systematic_2017, chen_measuring_2016, gambetta_analytic_2011, theis_counteracting_2018, motzoi_improving_2013, motzoi_simple_2009} (Appendix \ref{appendix_d_drag}). With the addition of a second, in-quadrature tone proportional to the time derivative of the Gaussian envelope, we optimize all control parameters heuristically for minimal coherent error. 
 
In Appendix \ref{appendix_d_drag}, we observe reduction in the $\ket{1, 1, 1}$ and $\ket{1, 2, 1}$ excited state population after the gate, with total coherent error improvements within an order of magnitude compared to without the additional tone (Fig. \ref{fig:fig4}(b), black dots). 

However, we find empirically that the effectiveness of the DRAG technique is limited at short gate durations. In this regime, leakage into other non-computational states $\ket{i,1,j}$ can become comparable to the leakage into the $\ket{1, 1, 1}$ and $\ket{1, 2, 1}$ states. Since we do not expect the DRAG pulse to simultaneously suppress leakage into multiple states, the improvement is modest.
For simplicity, we continue analyzing our gate without DRAG, noting to the reader that further improvements could be made with additional pulse optimization.

\subsection{Scaling of coherent error}

Finally, we extract a scaling of coherent error with gate time, using only our simple Gaussian pulses (without any additional DRAG correction). We numerically optimize the drive amplitude and frequency for minimal coherent error $\epsilon_c$, for various gate durations $t_g$ 
(Fig.~\ref{fig:fig4}(b), orange dots), and fit the result to a power law 
(Fig.~\ref{fig:fig4}(b), blue line). Our finite amplitude and frequency sweeps for optimization results in some scatter about the fit, but the result is well described by a model $\epsilon_c \propto t_g^p$, where $p = -4.4 \pm 0.3$. This efficient polynomial scaling is approximately consistent with typical values found for leakage errors via a Magnus expansion \cite{ribeiro_systematic_2017}, and in particular, its improvement beyond $t_g^{-2}$ further shows that phase errors due to unwanted AC Stark shifts have largely been suppressed. By reducing the dominant sources of coherent error and improving its scaling with gate duration, we enable fast, high-fidelity gates with favorable scaling as the total gate duration increases.

\section{\label{sec:level2}Scalability and resonator impedance}

The flexibility in the resonator impedance to values well beyond that of a transmon should facilitate the integration of high-fidelity, two qubit gates into a larger circuit capable of running surface-code quantum error correction (QEC). For this, coupling to four nearest neighbors is required. However, in general, coupling to an increasing number of neighboring qubits comes with a cost; each coupling capacitance also contributes to the total effective capacitance of the qubit, as can be found when solving for the circuit Lagrangian \cite{devoret_quantum_1997, ciani_lecture_2024, rasmussen_superconducting_2021}. In general, as the coupling capacitance between circuit elements increases, the coupling strength between them increases, but the contribution of the coupling capacitance to their total effective capacitances also becomes larger. 

This presents a challenge in using fluxonium qubits in a QEC context, as the fluxonium qubit requires a capacitance several factors smaller than that of the transmon. Since strong couplings between the fluxonium qubits are required for fast, high-fidelity gates, there is generally a trade-off between gate speed/fidelity, and connectivity. Here, we propose to avoid this trade-off by using a high-impedance resonator as a coupling element between fluxonium qubits. Generally, a high impedance element can efficiently couple circuit elements, i.e., facilitate strong coupling even with a small coupling capacitance. 

This statement can be shown with a simple scaling argument. We consider the coupling between a resonator $c$ and a single fluxonium $i$ in an isolated circuit. As discussed above, strong hybridization with the fluxonium $\ket{1}\rightarrow \ket{2}$ transition is desired, in order to achieve a large inherited anharmonicity on the resonator and drive single-photon excitations. Therefore, we assume that the resonator frequency is fixed. Recall that the key matrix element of the Hamiltonian which generates the dispersive coupling $\chi$ and inherited anharmonicity $\alpha$ on the resonator, is $\langle 1, 1 | J_{A(B), c} \hat{n}_c \hat{n}_f | 0, 2\rangle $ (with the first index for the resonator state). This matrix element scales as
\begin{equation}\label{impedance_scaling}
J n_{c, 01} \propto \frac{C_c n_{c, 01}}{C_r C_f} \propto \omega_c \sqrt{Z_c} \frac{C_{c}}{C_f},
\end{equation}
where here the resonator angular frequency is $\omega_c$, the resonator impedance is $Z_c \equiv \sqrt{L_r/C_r}$, the resonator capacitance is $C_r$, the fluxonium capacitance is $C_f$, and the coupling capacitance between the fluxonium and resonator is $C_{c}$ with coupling strength $J$. In the last proportionality, we have used $n_{c, 01} \propto Z_c^{-1/2}$ for the resonator, and $C_r = 1/(\omega_c Z_c)$. Namely, for a given coupling capacitance, the coupling increases with impedance. In principle, the coupling $J n_{c, 01}$ can be increased to $\mathcal{O}(1)$ GHz by increasing the impedance to $\sim 4 \text{k}\Omega$ at a substantial loading $C_c/ C_{tot} \sim 1/2$. This maximum coupling is fundamentally limited by the capacitive loading on the resonator due to the couplings $C_c$, i.e., the intrinsic resonator capacitance $C_r$ approaches $C_c$. 

\begin{figure}[h]
\includegraphics[width=0.45\textwidth]{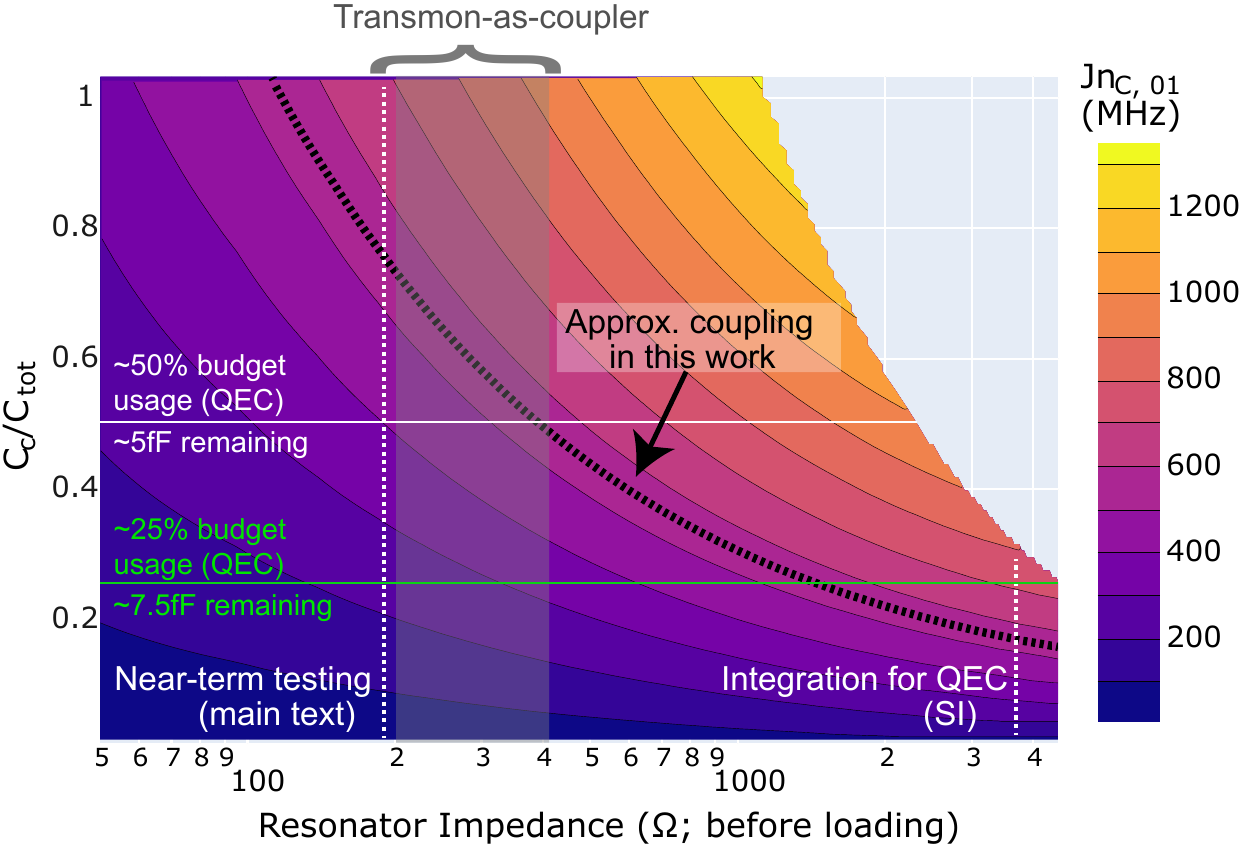}
\caption{{Advantages of high coupler impedances in high-connectivity architectures.  We plot the fluxonium-coupler couping strength $J n_{c,01}$ as a function of both the capacitive loading ratio $C_c / C_{tot}$ and the coupler impedance $Z_c$.  The resonator impedance is varied while keeping the (unloaded) resonator frequency fixed at 7.08 GHz.  We also approximately fix the total fluxonium capacitance $C_{tot} \simeq 10 $fF so that its charging energy is $E_C^{A(B)} \simeq 2$ GHz. These choices match parameter values in Table \ref{table:1}. We assume a differential qubit geometry for the fluxonium and a grounded resonator. The parasitic coupling to ground from the differential fluxoniums is 7 fF, and the joint capacitiance between the fluxoniums is $C_{AB} = 0.1$ fF. Horizontal lines: maximum capacitive loading that could be tolerated in a surface code architecture with degree-four connectivity (see text). Dashed black line: approximate coupling $J n_{c, 01}$ used for gate simulations, and which is large enough to enable fast, high-fidelity gates. We avoid the region in which the effective capacitance of the resonator (the diagonal element of the capacitance matrix for its node) is more than 50\% due to loading.  We see that to achieve large enough couplings for fast gates while having low enough capactive loading for a degree-four connectivity, high impedances $Z_c \gtrsim 1000 \Omega$ are needed.  This is can be achieved with a resonator coupler (as studied here), but is infeasible when using a typical transmon coupler.}}\label{fig:newfig}
\end{figure}

The above scaling argument is suggestive, but not complete, as it does not include other relevant details of the full circuit, such as the grounding configuration and parasitic capacitance to ground. For a more detailed understanding of the advantage of large-impedance resonator couplers, we numerically compute the coupling between fluxonium $A$ and the resonator as a function of the coupling capacitance and resonator impedance, by calculating and inverting the capacitance matrix (Appendix \ref{appendix_a_circuit}). Our results are shown in Fig. \ref{fig:newfig}. 

We consider a circuit configuration in which the resonator is directly connected to ground, and both fluxoniums are grounded through a parasitic capacitance $C_p$ (Appendix \ref{appendix_a_circuit}). For simplicity, we treat the circuit symmetrically, such that $J_{Ac} = J_{Bc} = J$, with coupling capacitance $C_c$. We vary both the resonator capacitance and inductance so that its (unloaded) frequency stays fixed at $\omega_c / (2\pi) = 7.08 $ GHz (matching the value in Table \ref{table:1}), while its impedance varies over a large range.  Note that after including the effects of coupling capacitance, the loaded frequency of the resonator decreases to a minimum of about 5.5 GHz when varying parameters over the range plotted in Fig.~\ref{fig:newfig}.      

For the fluxonium qubits, in order to maintain a fixed $E_C$ value for both fluxoniums within $\sim 10\%$ of 2 GHz (see Table \ref{table:1}) across the sweep range, we fix
\begin{equation}
    C_{tot} \equiv C_f + C_{AB}/4 + C_p/2 + C_c/4,
    \label{eq:Ctot}
\end{equation}    
which is the diagonal element of the capacitance matrix for the fluxoniums’ nodes (Appendix \ref{appendix_a_circuit}). We extract the coupling strength $J$ between either fluxonium and the resonator and multiply by the zero point fluctuation $n_{c, 01}$ of the resonator, calculated using the impedance after loading.

The last term in $C_{tot}$ represents the capacitive loading of the fluxoniums from the coupling capacitance. For integration in surface-code QEC with four nearest neighbors, the contribution from coupling capacitors to the effective qubit capacitance will approximately quadruple.  Using Eq.~(\ref{eq:Ctot}) this gives us a maximum allowable $C_c$:  it must be smaller than $C_{tot}$.  For $C_c$ larger than this value, it will be impossible to achieve a connectivity-four setup and keep $C_{tot}$ at our desired target value (i.e.~even if all other capacitances in 
Eq.~(\ref{eq:Ctot}) could be made vanishingly small).  Given this,  to roughly understand feasible coupling values when scaling, we consider loading in which $C_{c} \lesssim C_{tot}/2$, corresponding to a ``capacitance budget" of 50\% (white horizontal line, Fig. \ref{fig:newfig}). For this value,
when we move to a degree-4 architecture, we could maintain our desired target value of $C_{tot}$ for each fluxonium, with only half the contribution coming from coupling capacitances (thus leaving room for contributions from other elements).  

We find that for impedances of $\sim 500 \Omega$ and above, a coupling of $J n_{c, 01} \sim 600$ MHz (i.e., the same coupling used throughout this work to show fast, high-fidelity gates) is achievable 
{\it while still maintaining} an acceptable capacitive loading in the surface code context, i.e. with $C_c$, below $50\%$ of the total effective capacitance of the fluxonium (horizontal white line, Fig. \ref{fig:newfig}). This threshold impedance value $\sim 500\Omega$ is just barely at the range of impedances that can be achieved using a typical transmon coupler (grey shaded region in Fig.~\ref{fig:newfig}; we consider a transmon charging energy of $0.2$ GHz and transmon $E_J/E_C$ from 50 to 200). This indicates that the fluxonium's finite capacitance budget could limit the use of transmon couplers in a surface code QEC context, in contrast to high-impedance resonator couplers.

Considering that fluxoniums' charging energy is typically in the $\sim 1-2$ GHz regime, which corresponds to $10-20$ fF, additional margin in the capacitive budget is likely needed. For example, the self-capacitance of the small junctions in the fluxonium circuits and the parasitic capacitance to ground are typically a few fF or more. We expect that an impedance in the range of $\sim 1\text{k}\Omega$ to $\sim 5 \text{k}\Omega$ would be sufficient for practical implementation in a circuit when accounting for this additional margin. For example, as shown in Appendix \ref{alt_params}, we identify a set of circuit parameters in which the unloaded impedance of the resonator is $\sim$4 k$\Omega$, the coupling term between the fluxonium and resonator are $J n_{c, 01} \sim$0.54 MHz (the same value as here in the main text), the fluxoniums' charging energy is $E_C^{A(B)} = 1$ GHz, while the capacitive loading $C_{c}$ is only about 24\% of the effective capacitance of the fluxoniums. We expect that this would leave a sufficient capacitance budget for the other contributions discussed above. 

With caveats, this exercise shows regions of parameter space which may be feasible for integration in a surface code. While the quantitative values depend on many parameters, most of which we fix here to match Table \ref{table:1} as much as possible, we choose typical parameters we expect would be realized in an experimental setting. For example, lowering the fluxoniums' charging energy, below the 2 GHz used here, would result in weaker coupling but a larger capacitance of the fluxoniums and therefore additional margin for, e.g., parasitic coupling to ground (see Appendix \ref{alt_params} for an analysis using $E_C^{A(B)} = 1$ GHz). Also, increasing the resonator frequency increases the coupling, which may also be further optimized. Finally, the underlying circuit parameters in each pixel of Fig \ref{fig:newfig} are by no means optimized optimized for gate fidelity (for example, for some regions, the loaded coupler frequency crosses through the plasmon transition). Instead, we simply convey a relationship between the coupling, impedance, and capacitive loading on the fluxoniums. A full optimization across the many-dimensional parameter space is beyond the scope of this work.

In the main text, we analyze the gate with a relatively low impedance of 190 $\Omega$, corresponding to transmon values, in order to directly study the effects of replacing a transmon coupling element with a resonator, as well as facilitate immediate experimental testing, as low-impedance resonators are easily integrated into a design. However, we also repeat all of our simulations for the high-impedance case and show the results in the Supplementary Information.  We note to the reader that an increase in resonator impedance does not necessarily increase resonator loss, as shown in TiN thin films \cite{shearrow_atomic_2018, amin_loss_2022}. For example, in Ref.~\cite{shearrow_atomic_2018}, the authors observe quality factors of $\sim 10^{6}$ in 7 GHz resonators at an impedances of over $10\text{k}\Omega$. Conversely, as the resonator impedance increases, so does its capacitive loading from the fluxoniums, which may contribute to increased loss. This is a question for future experimental study. Nevertheless, in our analysis, we find gate fidelities are likely limited by the fluxonium 0-1 transition loss for both the high and low impedance cases (Appendix \ref{appendix_e_error_budget}). Assuming the same resonator loss rates, in both the low and high impedance cases, we predict gate fidelities of about $2 \times 10^{-4}$ in $\sim70$ns. We also observe robustness to junction mis-targeting and flux variation, as discussed below and in Appendix \ref{appendix_c_hamiltonian}.

\section{Robustness to parameter imperfections}

While a coherent error $\epsilon_c \lesssim 2 \times 10^{-6}$ at $t_g \sim 100$ns (as shown in Fig.~\ref{fig:fig4}) is encouraging for quantum computation, this result only has practical relevance if it has some degree of robustness to circuit parameter variations. 
Junction variations are effectively mitigated by control, while flux variations are suppressed by low $E_L$. 
Here we analyze both sources of error (neglecting the variation in $E_L$, assumed to be suppressed by $1/\sqrt{N}$ where $N$ is the number of junctions in a junction-based inductive shunt). We find that our approach are robust to both these mechanisms, considering typical experimental parameters and uncertainties. 
\begin{figure}[h] 
\includegraphics[width=0.5\textwidth]{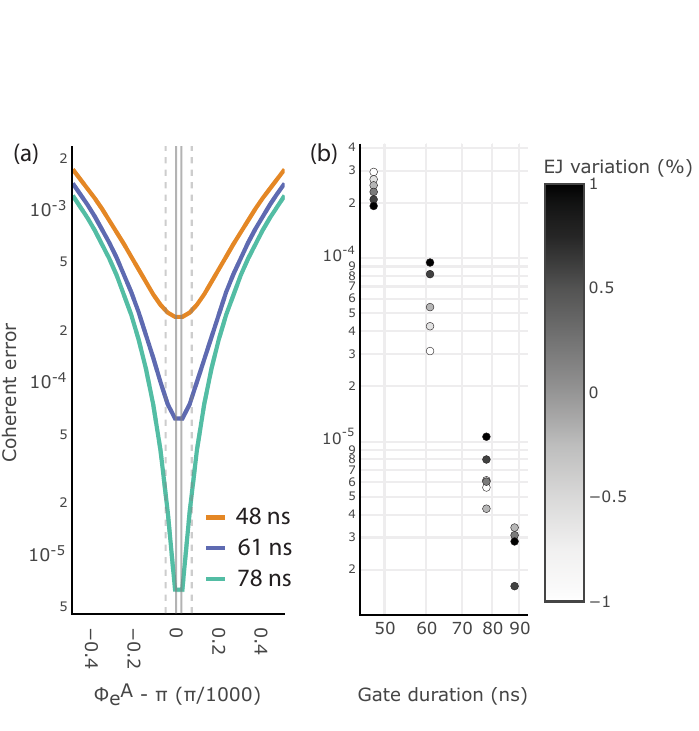}
\caption{Robustness of coherent error to parameter imperfections. (a) Coherent error as a function of static flux deviations of $\phi_e^A$, solid colored lines. Vertical grey lines represent the standard deviation of flux for an example laboratory setting \cite{zhang_universal_2021} (dashed lines), and rough state-of-the-art setting (solid line) \cite{sun_characterization_2023}, demonstrating that our gate should be robust to small changes in flux, for typical loop sizes and mutual inductance between the flux control line and qubit. (b) Coherent error robustness to $E_J$ mistargeting. We simulate our gate over a variety of $E_J$ values for fluxonium A, keeping the other circuit parameters fixed. We find less than a factor of $\sim 10$ change in the coherent error as we change the junction energy by a typical 2\% range. Because of the efficient scaling of coherent error with gate duration, such error increases can be efficiently mitigated by using a slightly longer gate, showing robustness of our approach to $E_J$ variation.}\label{fig:fig5}
\end{figure}
\subsection{Flux noise}
To approximate the effect of $1/f$ flux noise \cite{zhang_universal_2021, sun_characterization_2023} in the circuit, we simulate our gate time dynamics as a function of static flux with fixed, small offset from $\pi$, without allowing for drive re-calibration. This simulates low-frequency flux noise in the environment, since the flux would not change significantly on the timescale of the gate, but it would exhibit uncontrolled offsets from $\pi$ before recalibration is performed. Therefore, we repeat the simulation for many values of $\phi_e^A$ offsets around $\pi$, keeping $\phi_e^B$ fixed at $\pi$. (These results should be qualitatively representative for varying $\phi_e^B$, since both fluxoniums have similar circuit parameters and coupling strengths). For flux noise of both $S(\omega) = (5 \mu \Phi_0)^2 / \omega$, and $(1 \mu \Phi_0)^2/\omega$, corresponding to typical and state-of-the-art flux noise amplitudes found in the laboratory \cite{zhang_universal_2021, sun_characterization_2023}, respectively, we find that the variation in coherent error is negligible for the gate durations analyzed (see Fig. \ref{fig:fig5}(a)). To calculate the expected variation in flux for the typical and state-of-the-art values (assuming similar mutual inductance between the flux control line and qubit, as well as qubit loop area), we assume 1 hour time delays before re-calibration of the fluxoniums' external flux source back to its desired value of $\phi_e^A = \pi$, integrating the noise down to a low frequency cutoff of $1/ (1\text{ hour}) \sim 0.3$ mHz. The high frequency cutoff is the inverse of the gate time $\sim 1/100$ns. 
As discussed below, for typical dissipation rates, the short gate duration $t_g < 70$ ns is of most interest, corresponding to sufficient robustness in flux offset for the typical experimental setting.
\subsection{Junction mistargeting}
Fabrication imperfections and inhomogeneities, as well as chip aging, will result in variations in all junction energies, typically at the level of $\sim 2$\%. To study the impact on coherent gate error, we assume that the junction values are static and sweep $E_J$ of fluxonium A in a 2\% range about its desired 7 GHz value, recalibrating the drive for each $E_J^A$ and $t_g$ value. The results are shown in Fig. \ref{fig:fig5}(b), for various total gate durations $t_g$. We find that the coherent error stays within a factor of two or three within the value for the desired $E_J^A = 7$ GHz, which can easily be compensated by slightly increasing the gate duration, exploiting the efficient scaling. This robustness is expected and by design. The key parameters that enable our gate speed limits $\sim 1/\delta\chi_{ij}$ and $\sim 1/\alpha$ are polynomial in small $1/\Delta_{i}$, which shifts by order $\delta \omega_{1, 2}^i / (2\pi) \Delta_i \ll 1$. Therefore, small $E_J$ variation should result in small variations in gate speed limits and errors.

\section{Gate performance in the presence of dissipation}

Having analyzed both the impact of coherent errors and parameter variations, we now turn to the last crucial error channel: environment-induced dissipation. We simulate the effect of Markovian dissipation on our gate, using time-dependent master equation simulations.
We consider the impact of $T_1$ dissipation on both the fluxoniums and the cavity coupler, but neglect pure dephasing processes. This is a reasonable approximation, as the fluxoniums' coherence times are protected to first order in flux noise at the external flux values $\phi_{e} = \pi$, and generally pure dephasing rates in electromagnetic resonators are small \cite{gao_noise_2007}. 

Our simulations follow a master equation
\begin{equation}\label{master_eq}
\begin{aligned}
\dot{\rho} = -\frac{i}{h}[\mathcal{\hat{H}}, \rho] + \kappa_c \mathcal{D}(\hat{c})\rho + \\
\\ \sum_{i \in \{A, B\}} \kappa_{i}^{01} \Big((n_{i}^{01} + 1) \mathcal{D}(\ket{0}_i \bra{1}_i) + n_{i}^{01} \mathcal{D}(\ket{1}_i \bra{0}_i) \Big) \rho \\ + \kappa_{i}^{12} \mathcal{D}(\ket{1}_i \bra{2}_i) \rho,
\end{aligned}
\end{equation}
where $\kappa_c$ is the dissipation rate of the coupler with annihilation operator $c$, $\kappa_{i}^{kl}$ is the dissipation rate of the bare $\ket{k}_i-\ket{l}_i$ transition on fluxonium $i$, and $\ket{k}_i \bra{l}_i$ is the corresponding jump operator.
All jump operators act on individual, bare components, such that the Hamiltonian eigenstates inherit loss rates of linear combinations of the various $\kappa$ values. 
We include dissipation only for the fluxonium $\ket{0}-\ket{1}$ and $\ket{1}-\ket{2}$ transitions, because all other transitions are largely irrelevant to the gate dynamics.  Note that the loss process between $\ket{1}-\ket{2}$ enhances the effective loss rate of the coupler, as this transition is hybridized with the coupler at the 10-15\% level. For heating, we include raising jump operators on the qubit transitions of the fluxoniums, as the thermal occupation of the fluxoniums' first excited states are order unity for temperatures of $10-50$ mK. We neglect heating of the plasmon and coupler transitions, which are several GHz.

\begin{table}[t!]
\centering
\begin{tabular}{||c c c c c||}
 \hline
 Set & $\omega_c/\kappa_c$ & $1/\kappa_{01}^{i}$ & $1/\kappa_{12}^{i}$ & T \\ [0.5ex] 
 \hline\hline
 A & $5 \times 10^{6}$ & 1ms & 30$\mu$s & 30 mK \\ 
 \hline
 B & $5 \times 10^{6}$ & 1ms & 30$\mu$s & 50 mK \\ 
 \hline
 C & $5 \times 10^{6}$ & 1ms & 30$\mu$s & 100 mK \\ 
 \hline
 D & $10^6$ & 0.2ms & 10$\mu$s & 30 mK \\
 \hline
 E & $10^6$ & 0.2ms & 10$\mu$s & 50 mK \\
 \hline
 F & $10^6$ & 0.2ms & 10$\mu$s & 100 mK \\
 \hline
\end{tabular}
\caption{Parameter sets of dissipation rates and temperature for our simulations including dissipation. The sets A-C are most optimistic $\kappa$ values, for increasing temperatures, while the sets D-F are the most conservative $\kappa$ values. }
\label{table:2}
\end{table}
\begin{figure}[t!]
\includegraphics[width=0.45\textwidth]{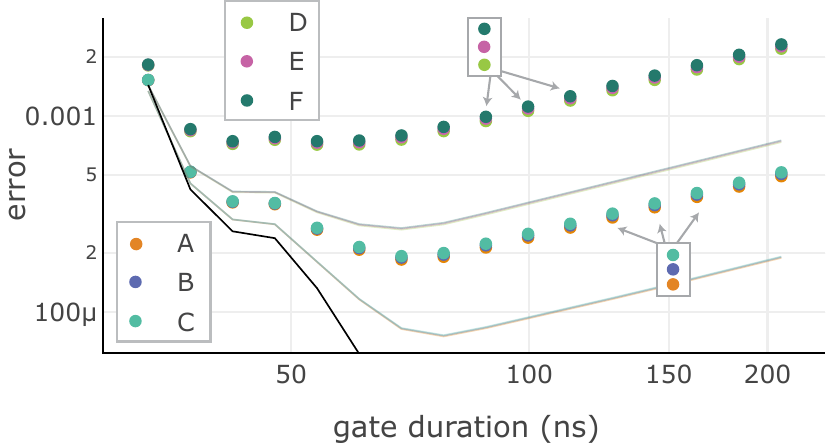}
\caption{Gate error including dissipation. We consider six parameter sets for dissipation rates and temperature, as described in Table \ref{table:2}. We observe an optimal gate duration $t_g$ for each parameter set, corresponding to roughly equal contributions of coherent and incoherent error (Appendix \ref{appendix_e_error_budget}). Grey lines are leakage errors, for parameter sets A-C (bottom grey line) and parameter sets D-F (top grey line). The coherent error is plotted as a black solid line.  For state-of-the-art dissipation rates, errors of $1.86 \times 10^{-4}$ in 70 ns are feasible.}\label{fig:fig6}
\end{figure}

 For computational feasibility, our dissipative simulations are performed using a smaller Hilbert space dimension than used for the coherent error analysis in the earlier sections; in particular, we go from a dimension of $d=45$ to $d=28$.  While the additional truncation does not omit any levels that have signficant population, it does lead to very small changes in non-resonant Stark shifts, and thus to the optimal drive detuning (see Sec. \ref{sec:drive_optimization}) that was used to mitigate the corresponding phases.  For this reason, for the simulations presented here, we re-optimize the drive detuning to account for the modified value of $d$.  With this re-optimization, we find that the purely coherent error of our gate is almost the same for $d=28$ and $d=45$ (i.e.~for all gate durations simulated, the coherent error predictions for the two values of $d$ agree to within a relative error of 7\%, see Appendix \ref{appendix_f_methods}).  
Here, we report the gate errors of the $d=28$ simulations, using the associated optimized drive parameters, as an estimate to the extracted error in a randomized benchmarking experiment. 

We present simulation results for six different sets of temperature and dissipation rate parameters, as listed in Table \ref{table:2}. We consider three values of temperature (30 mK, 50 mK and 100 mK), and both optimistic and pessimistic $\kappa$ values, sweeping from state-of-the-art lifetime for fluxonium $1/\kappa_{01} = 1$ ms \cite{ding_high-fidelity_2023, somoroff_millisecond_2021}, to a conservative $1/\kappa_{01} = 0.2$ ms. The resonator coupler should be fabricated from a high quality material, such as tantalum, to ensure that its quality factor is in the range of $1 \times 10^6$ to $5 \times 10^{6}$ \cite{place_new_2021, zmuidzinas_superconducting_2012}.

In Fig. \ref{fig:fig6} we plot the average gate infidelity $1-F$ following the standard definition in Ref. \cite{pedersen_fidelity_2007}, where the fidelity $F$ is
\begin{equation}
F = \frac{1}{d(d+1)}\left( \text{Tr}\left( \sum_k \hat{M}_k^\dagger \hat{M}_k \right) + \sum_k |\text{Tr}(\hat{M}_k)|^2 \right).
\end{equation}
Here, the computational dimension $d=4$, $M_k = P U_{\text{CZ}} G_k P $, $U_{\text{CZ}}$ is the ideal CZ unitary, $G_k$ are the Kraus operators, and $P$ is the projection to the computational subspace.
We show the error for all six parameter sets as a function of gate duration $t_g$, including all the incoherent error sources described above (colored dots). The leakage error (grey lines) are overlapping within each parameter set group, and remain small, while the coherent error in a unitary simulation is again plotted for comparison (black line). We observe an optimal $t_g$ for each parameter set, corresponding to approximately equal contributions of incoherent and coherent error. For the most pessimistic parameters (sets D-F), the optimized error of $6 \times 10^{-4}$ is predicted at $t_g = 55$ ns, while for the most optimistic (sets A-C), errors of $1.86 \times 10^{-4}$ for $t_g = 70$ ns are feasible. The incoherent error sources are well-approximated by simple ratios of gate duration and lifetime (Appendix \ref{appendix_e_error_budget}).

We also remark that the gate performance is robust to thermal heating effects (at the level of bath thermal factors $n^{01}_i$ in Eq.~\eqref{master_eq} that are order unity).  
This can be seen by the fact that the error is nearly identical for parameter sets A, B and C, and for D, E and F. We are only treating thermal heating effects during the gate, and are assuming that the initial state preparation of the system is not corrupted by thermal effects. In particular, at higher temperatures the resonator coupler must be reset, as any thermal population will quickly dephase the qubits, leading to a gate error. We also neglect additional quasiparticles, thermal photon shot noise in coupled readout resonators, or additional radiation from control and readout lines. In principle, if these additional mechanisms can be mitigated, our results suggest that high-fidelity gate operator at temperatures
 $\sim 100$mK should be possible.   
 

\section{Outlook}
Coherent errors govern effective gate speeds, limiting total fidelities when combined with incoherent dissipation rates. By understanding the dominating sources of coherent errors, i.e., dispersive phase shifts which result in errors scaling as $t_g^{-2}$, we design a fluxonium-based circuit to minimize them, enabling faster gates and higher fidelities for typical dissipation rates. Specifically, the implementation of a high-quality, linear resonator as a coupling element results in efficient scaling of coherent error with gate time through destructive interference. Our architecture enables reduced coupler loss, such that the fluxoniums' dissipation dominates the incoherent error budget. A perturbative analysis of the Hamiltonian guides our circuit parameter design to ensure that a given mixing between elements results in maximally contributions to gate fidelities. Strongly coupling our resonator to a single, highly-nonlinear transition of the fluxonium qubit gives rise to large, state-selective frequency shifts and nonlinearity within the resonator. However, the total hybridization can still be tuned to ensure inherited loss rates of the coupler from lossy non-computational states of the fluxonium are inconsequential.

Our resonator-as-coupler design also introduces an approach for scaling a two-qubit gate architecture with fluxonium qubits, proposing a solution for improved logical performance through fluxoniums' favorable lifetime and anharmonicities. One challenge in developing fluxonium-based quantum error correction is the decreased margins for capacitive loading from its larger $E_C$ values compared to current transmon architectures. While further analysis of our gate in a larger code is warranted, we identify a building-block for facilitating integration of fluxonium qubits in a highly connected circuit. This high-impedance resonator could result in more efficient coupling strength (per unit coupling capacitance) than a transmon coupler, and debut increased coordination number in a fluxonium-based architecture. Further work will include treatment of our gate in a realistic quantum error correction context: spectator errors, capacitance budgeting including readout resonators, and estimations of dephasing from their associated shot noise. In particular, an effective treatment of the correlated errors across a circuit, e.g. due to induced ZZZ interactions, presents a challenge in the context of such strong hybridization.

Our first-principles approach may also be applied to other quantum computing architectures. Identifying sources of the dominating coherent errors and engineering the circuit to natively cancel them may bear fruit in a transmon-coupler-transmon case, potentially identifying a path to microwave-activated gates with improved coherent error scaling and total fidelity. Our gate may also be executed with external flux values $\phi_{e} \neq \pi$, where the fluxoniums exhibit $T_1 > 100\text{ } T_2^*$, with $T_1$ approaching 10ms \cite{lin_protecting_2018, earnest_realization_2018, mizel_right-sizing_2020, zhu_asymptotic_2013, herrera-marti_tradeoff_2013, hassani_superconducting_2022, gyenis_moving_2021, lin_protecting_2018}. In this regime, our gate should preserve the $T_1 \gg T_2$ hierarchy; the additional noise channel, i.e. the coupler decay events, introduce additional dephasing of the qubits, but not bit-flips. Overall, our approach shows a path to record fidelities for two-qubit gates in a leading choice for superconducting circuits.
\begin{acknowledgments}
We thank Arne Grimsmo, Kyungjoo Noh, Colm Ryan and Mollie Schwartz for fruitful discussions. We thank Hanho Lee and Danyang Chen for facilitating adjustments after peer-review feedback, as well as technical insight and providing corrections. We thank Simone Severini, Bill Vass, Oskar Painter, Fernando Brand\~ao, Eric Chisholm and AWS for supporting the quantum computing program. 
\end{acknowledgments}

\appendix

\section{Circuit and Hamiltonian}\label{appendix_a_circuit}

\begin{thebibliography}{66}%
\makeatletter
\providecommand \@ifxundefined [1]{%
 \@ifx{#1\undefined}
}%
\providecommand \@ifnum [1]{%
 \ifnum #1\expandafter \@firstoftwo
 \else \expandafter \@secondoftwo
 \fi
}%
\providecommand \@ifx [1]{%
 \ifx #1\expandafter \@firstoftwo
 \else \expandafter \@secondoftwo
 \fi
}%
\providecommand \natexlab [1]{#1}%
\providecommand \enquote  [1]{``#1''}%
\providecommand \bibnamefont  [1]{#1}%
\providecommand \bibfnamefont [1]{#1}%
\providecommand \citenamefont [1]{#1}%
\providecommand \href@noop [0]{\@secondoftwo}%
\providecommand \href [0]{\begingroup \@sanitize@url \@href}%
\providecommand \@href[1]{\@@startlink{#1}\@@href}%
\providecommand \@@href[1]{\endgroup#1\@@endlink}%
\providecommand \@sanitize@url [0]{\catcode `\\12\catcode `\$12\catcode `\&12\catcode `\#12\catcode `\^12\catcode `\_12\catcode `\%12\relax}%
\providecommand \@@startlink[1]{}%
\providecommand \@@endlink[0]{}%
\providecommand \url  [0]{\begingroup\@sanitize@url \@url }%
\providecommand \@url [1]{\endgroup\@href {#1}{\urlprefix }}%
\providecommand \urlprefix  [0]{URL }%
\providecommand \Eprint [0]{\href }%
\providecommand \doibase [0]{https://doi.org/}%
\providecommand \selectlanguage [0]{\@gobble}%
\providecommand \bibinfo  [0]{\@secondoftwo}%
\providecommand \bibfield  [0]{\@secondoftwo}%
\providecommand \translation [1]{[#1]}%
\providecommand \BibitemOpen [0]{}%
\providecommand \bibitemStop [0]{}%
\providecommand \bibitemNoStop [0]{.\EOS\space}%
\providecommand \EOS [0]{\spacefactor3000\relax}%
\providecommand \BibitemShut  [1]{\csname bibitem#1\endcsname}%
\let\auto@bib@innerbib\@empty
\bibitem [{\citenamefont {Koch}\ \emph {et~al.}(2009)\citenamefont {Koch}, \citenamefont {Manucharyan}, \citenamefont {Devoret},\ and\ \citenamefont {Glazman}}]{koch_charging_2009}%
  \BibitemOpen
  \bibfield  {author} {\bibinfo {author} {\bibfnamefont {J.}~\bibnamefont {Koch}}, \bibinfo {author} {\bibfnamefont {V.}~\bibnamefont {Manucharyan}}, \bibinfo {author} {\bibfnamefont {M.~H.}\ \bibnamefont {Devoret}},\ and\ \bibinfo {author} {\bibfnamefont {L.~I.}\ \bibnamefont {Glazman}},\ }\bibfield  {title} {\bibinfo {title} {Charging effects in the inductively shunted {Josephson} junction},\ }\href {https://doi.org/10.1103/PhysRevLett.103.217004} {\bibfield  {journal} {\bibinfo  {journal} {Physical Review Letters}\ }\textbf {\bibinfo {volume} {103}},\ \bibinfo {pages} {217004} (\bibinfo {year} {2009})}\BibitemShut {NoStop}%
\bibitem [{\citenamefont {Manucharyan}\ \emph {et~al.}(2009)\citenamefont {Manucharyan}, \citenamefont {Koch}, \citenamefont {Glazman},\ and\ \citenamefont {Devoret}}]{manucharyan_fluxonium_2009}%
  \BibitemOpen
  \bibfield  {author} {\bibinfo {author} {\bibfnamefont {V.~E.}\ \bibnamefont {Manucharyan}}, \bibinfo {author} {\bibfnamefont {J.}~\bibnamefont {Koch}}, \bibinfo {author} {\bibfnamefont {L.}~\bibnamefont {Glazman}},\ and\ \bibinfo {author} {\bibfnamefont {M.}~\bibnamefont {Devoret}},\ }\bibfield  {title} {\bibinfo {title} {Fluxonium: single {Cooper} pair circuit free of charge offsets},\ }\href {https://doi.org/10.1126/science.1175552} {\bibfield  {journal} {\bibinfo  {journal} {Science}\ }\textbf {\bibinfo {volume} {326}},\ \bibinfo {pages} {113} (\bibinfo {year} {2009})}\BibitemShut {NoStop}%
\bibitem [{\citenamefont {Masluk}\ \emph {et~al.}(2012)\citenamefont {Masluk}, \citenamefont {Pop}, \citenamefont {Kamal}, \citenamefont {Minev},\ and\ \citenamefont {Devoret}}]{masluk_microwave_2012}%
  \BibitemOpen
  \bibfield  {author} {\bibinfo {author} {\bibfnamefont {N.~A.}\ \bibnamefont {Masluk}}, \bibinfo {author} {\bibfnamefont {I.~M.}\ \bibnamefont {Pop}}, \bibinfo {author} {\bibfnamefont {A.}~\bibnamefont {Kamal}}, \bibinfo {author} {\bibfnamefont {Z.~K.}\ \bibnamefont {Minev}},\ and\ \bibinfo {author} {\bibfnamefont {M.~H.}\ \bibnamefont {Devoret}},\ }\bibfield  {title} {\bibinfo {title} {Microwave {Characterization} of {Josephson} {Junction} {Arrays}: {Implementing} a {Low} {Loss} {Superinductance}},\ }\href {https://doi.org/10.1103/PhysRevLett.109.137002} {\bibfield  {journal} {\bibinfo  {journal} {Physical Review Letters}\ }\textbf {\bibinfo {volume} {109}},\ \bibinfo {pages} {137002} (\bibinfo {year} {2012})}\BibitemShut {NoStop}%
\bibitem [{\citenamefont {Manucharyan}\ \emph {et~al.}(2012)\citenamefont {Manucharyan}, \citenamefont {Masluk}, \citenamefont {Kamal}, \citenamefont {Koch}, \citenamefont {Glazman},\ and\ \citenamefont {Devoret}}]{manucharyan_evidence_2012}%
  \BibitemOpen
  \bibfield  {author} {\bibinfo {author} {\bibfnamefont {V.~E.}\ \bibnamefont {Manucharyan}}, \bibinfo {author} {\bibfnamefont {N.~A.}\ \bibnamefont {Masluk}}, \bibinfo {author} {\bibfnamefont {A.}~\bibnamefont {Kamal}}, \bibinfo {author} {\bibfnamefont {J.}~\bibnamefont {Koch}}, \bibinfo {author} {\bibfnamefont {L.~I.}\ \bibnamefont {Glazman}},\ and\ \bibinfo {author} {\bibfnamefont {M.~H.}\ \bibnamefont {Devoret}},\ }\bibfield  {title} {\bibinfo {title} {Evidence for coherent quantum phase-slips across a {Josephson} junction array},\ }\href {https://doi.org/10.1103/PhysRevB.85.024521} {\bibfield  {journal} {\bibinfo  {journal} {Physical Review B}\ }\textbf {\bibinfo {volume} {85}},\ \bibinfo {pages} {024521} (\bibinfo {year} {2012})}\BibitemShut {NoStop}%
\bibitem [{\citenamefont {Nguyen}\ \emph {et~al.}(2019)\citenamefont {Nguyen}, \citenamefont {Lin}, \citenamefont {Somoroff}, \citenamefont {Mencia}, \citenamefont {Grabon},\ and\ \citenamefont {Manucharyan}}]{nguyen_high-coherence_2019}%
  \BibitemOpen
  \bibfield  {author} {\bibinfo {author} {\bibfnamefont {L.~B.}\ \bibnamefont {Nguyen}}, \bibinfo {author} {\bibfnamefont {Y.-H.}\ \bibnamefont {Lin}}, \bibinfo {author} {\bibfnamefont {A.}~\bibnamefont {Somoroff}}, \bibinfo {author} {\bibfnamefont {R.}~\bibnamefont {Mencia}}, \bibinfo {author} {\bibfnamefont {N.}~\bibnamefont {Grabon}},\ and\ \bibinfo {author} {\bibfnamefont {V.~E.}\ \bibnamefont {Manucharyan}},\ }\bibfield  {title} {\bibinfo {title} {The high-coherence fluxonium qubit},\ }\href {https://doi.org/10.1103/PhysRevX.9.041041} {\bibfield  {journal} {\bibinfo  {journal} {Physical Review X}\ }\textbf {\bibinfo {volume} {9}},\ \bibinfo {pages} {041041} (\bibinfo {year} {2019})}\BibitemShut {NoStop}%
\bibitem [{\citenamefont {Somoroff}\ \emph {et~al.}(2021)\citenamefont {Somoroff}, \citenamefont {Ficheux}, \citenamefont {Mencia}, \citenamefont {Xiong}, \citenamefont {Kuzmin},\ and\ \citenamefont {Manucharyan}}]{somoroff_millisecond_2021}%
  \BibitemOpen
  \bibfield  {author} {\bibinfo {author} {\bibfnamefont {A.}~\bibnamefont {Somoroff}}, \bibinfo {author} {\bibfnamefont {Q.}~\bibnamefont {Ficheux}}, \bibinfo {author} {\bibfnamefont {R.~A.}\ \bibnamefont {Mencia}}, \bibinfo {author} {\bibfnamefont {H.}~\bibnamefont {Xiong}}, \bibinfo {author} {\bibfnamefont {R.~V.}\ \bibnamefont {Kuzmin}},\ and\ \bibinfo {author} {\bibfnamefont {V.~E.}\ \bibnamefont {Manucharyan}},\ }\href {http://arxiv.org/abs/2103.08578} {\bibinfo {title} {Millisecond coherence in a superconducting qubit}} (\bibinfo {year} {2021})\BibitemShut {NoStop}%
\bibitem [{\citenamefont {Kapit}\ and\ \citenamefont {Oganesyan}(2022)}]{kapit_small_2022}%
  \BibitemOpen
  \bibfield  {author} {\bibinfo {author} {\bibfnamefont {E.}~\bibnamefont {Kapit}}\ and\ \bibinfo {author} {\bibfnamefont {V.}~\bibnamefont {Oganesyan}},\ }\href {http://arxiv.org/abs/2212.04588} {\bibinfo {title} {Small logical qubit architecture based on strong interactions and many-body dynamical decoupling}} (\bibinfo {year} {2022})\BibitemShut {NoStop}%
\bibitem [{\citenamefont {Bao}\ \emph {et~al.}(2022)\citenamefont {Bao}, \citenamefont {Deng}, \citenamefont {Ding}, \citenamefont {Gao}, \citenamefont {Gao}, \citenamefont {Huang}, \citenamefont {Jiang}, \citenamefont {Ku}, \citenamefont {Li}, \citenamefont {Ma}, \citenamefont {Ni}, \citenamefont {Qin}, \citenamefont {Song}, \citenamefont {Sun}, \citenamefont {Tang}, \citenamefont {Wang}, \citenamefont {Wu}, \citenamefont {Xia}, \citenamefont {Yu}, \citenamefont {Zhang}, \citenamefont {Zhang}, \citenamefont {Zhang}, \citenamefont {Zhou}, \citenamefont {Zhu}, \citenamefont {Shi}, \citenamefont {Chen}, \citenamefont {Zhao},\ and\ \citenamefont {Deng}}]{bao_fluxonium_2022}%
  \BibitemOpen
  \bibfield  {author} {\bibinfo {author} {\bibfnamefont {F.}~\bibnamefont {Bao}}, \bibinfo {author} {\bibfnamefont {H.}~\bibnamefont {Deng}}, \bibinfo {author} {\bibfnamefont {D.}~\bibnamefont {Ding}}, \bibinfo {author} {\bibfnamefont {R.}~\bibnamefont {Gao}}, \bibinfo {author} {\bibfnamefont {X.}~\bibnamefont {Gao}}, \bibinfo {author} {\bibfnamefont {C.}~\bibnamefont {Huang}}, \bibinfo {author} {\bibfnamefont {X.}~\bibnamefont {Jiang}}, \bibinfo {author} {\bibfnamefont {H.-S.}\ \bibnamefont {Ku}}, \bibinfo {author} {\bibfnamefont {Z.}~\bibnamefont {Li}}, \bibinfo {author} {\bibfnamefont {X.}~\bibnamefont {Ma}}, \bibinfo {author} {\bibfnamefont {X.}~\bibnamefont {Ni}}, \bibinfo {author} {\bibfnamefont {J.}~\bibnamefont {Qin}}, \bibinfo {author} {\bibfnamefont {Z.}~\bibnamefont {Song}}, \bibinfo {author} {\bibfnamefont {H.}~\bibnamefont {Sun}}, \bibinfo {author} {\bibfnamefont {C.}~\bibnamefont {Tang}}, \bibinfo {author} {\bibfnamefont {T.}~\bibnamefont {Wang}}, \bibinfo {author} {\bibfnamefont
  {F.}~\bibnamefont {Wu}}, \bibinfo {author} {\bibfnamefont {T.}~\bibnamefont {Xia}}, \bibinfo {author} {\bibfnamefont {W.}~\bibnamefont {Yu}}, \bibinfo {author} {\bibfnamefont {F.}~\bibnamefont {Zhang}}, \bibinfo {author} {\bibfnamefont {G.}~\bibnamefont {Zhang}}, \bibinfo {author} {\bibfnamefont {X.}~\bibnamefont {Zhang}}, \bibinfo {author} {\bibfnamefont {J.}~\bibnamefont {Zhou}}, \bibinfo {author} {\bibfnamefont {X.}~\bibnamefont {Zhu}}, \bibinfo {author} {\bibfnamefont {Y.}~\bibnamefont {Shi}}, \bibinfo {author} {\bibfnamefont {J.}~\bibnamefont {Chen}}, \bibinfo {author} {\bibfnamefont {H.-H.}\ \bibnamefont {Zhao}},\ and\ \bibinfo {author} {\bibfnamefont {C.}~\bibnamefont {Deng}},\ }\bibfield  {title} {\bibinfo {title} {Fluxonium: an alternative qubit platform for high-fidelity operations},\ }\href {https://doi.org/10.1103/PhysRevLett.129.010502} {\bibfield  {journal} {\bibinfo  {journal} {Physical Review Letters}\ }\textbf {\bibinfo {volume} {129}},\ \bibinfo {pages} {010502} (\bibinfo {year}
  {2022})}\BibitemShut {NoStop}%
\bibitem [{\citenamefont {Ficheux}\ \emph {et~al.}(2021)\citenamefont {Ficheux}, \citenamefont {Nguyen}, \citenamefont {Somoroff}, \citenamefont {Xiong}, \citenamefont {Nesterov}, \citenamefont {Vavilov},\ and\ \citenamefont {Manucharyan}}]{ficheux_fast_2021}%
  \BibitemOpen
  \bibfield  {author} {\bibinfo {author} {\bibfnamefont {Q.}~\bibnamefont {Ficheux}}, \bibinfo {author} {\bibfnamefont {L.~B.}\ \bibnamefont {Nguyen}}, \bibinfo {author} {\bibfnamefont {A.}~\bibnamefont {Somoroff}}, \bibinfo {author} {\bibfnamefont {H.}~\bibnamefont {Xiong}}, \bibinfo {author} {\bibfnamefont {K.~N.}\ \bibnamefont {Nesterov}}, \bibinfo {author} {\bibfnamefont {M.~G.}\ \bibnamefont {Vavilov}},\ and\ \bibinfo {author} {\bibfnamefont {V.~E.}\ \bibnamefont {Manucharyan}},\ }\bibfield  {title} {\bibinfo {title} {Fast logic with slow qubits: microwave-activated controlled-{Z} gate on low-frequency fluxoniums},\ }\href {https://doi.org/10.1103/PhysRevX.11.021026} {\bibfield  {journal} {\bibinfo  {journal} {Physical Review X}\ }\textbf {\bibinfo {volume} {11}},\ \bibinfo {pages} {021026} (\bibinfo {year} {2021})}\BibitemShut {NoStop}%
\bibitem [{\citenamefont {Xiong}\ \emph {et~al.}(2021)\citenamefont {Xiong}, \citenamefont {Ficheux}, \citenamefont {Somoroff}, \citenamefont {Nguyen}, \citenamefont {Dogan}, \citenamefont {Rosenstock}, \citenamefont {Wang}, \citenamefont {Nesterov}, \citenamefont {Vavilov},\ and\ \citenamefont {Manucharyan}}]{xiong_arbitrary_2021}%
  \BibitemOpen
  \bibfield  {author} {\bibinfo {author} {\bibfnamefont {H.}~\bibnamefont {Xiong}}, \bibinfo {author} {\bibfnamefont {Q.}~\bibnamefont {Ficheux}}, \bibinfo {author} {\bibfnamefont {A.}~\bibnamefont {Somoroff}}, \bibinfo {author} {\bibfnamefont {L.~B.}\ \bibnamefont {Nguyen}}, \bibinfo {author} {\bibfnamefont {E.}~\bibnamefont {Dogan}}, \bibinfo {author} {\bibfnamefont {D.}~\bibnamefont {Rosenstock}}, \bibinfo {author} {\bibfnamefont {C.}~\bibnamefont {Wang}}, \bibinfo {author} {\bibfnamefont {K.~N.}\ \bibnamefont {Nesterov}}, \bibinfo {author} {\bibfnamefont {M.~G.}\ \bibnamefont {Vavilov}},\ and\ \bibinfo {author} {\bibfnamefont {V.~E.}\ \bibnamefont {Manucharyan}},\ }\href {http://arxiv.org/abs/2103.04491} {\bibinfo {title} {Arbitrary controlled-phase gate on fluxonium qubits using differential ac-{Stark} shifts}} (\bibinfo {year} {2021})\BibitemShut {NoStop}%
\bibitem [{\citenamefont {Nesterov}\ \emph {et~al.}(2022)\citenamefont {Nesterov}, \citenamefont {Wang}, \citenamefont {Manucharyan},\ and\ \citenamefont {Vavilov}}]{nesterov_cnot_2022}%
  \BibitemOpen
  \bibfield  {author} {\bibinfo {author} {\bibfnamefont {K.~N.}\ \bibnamefont {Nesterov}}, \bibinfo {author} {\bibfnamefont {C.}~\bibnamefont {Wang}}, \bibinfo {author} {\bibfnamefont {V.~E.}\ \bibnamefont {Manucharyan}},\ and\ \bibinfo {author} {\bibfnamefont {M.~G.}\ \bibnamefont {Vavilov}},\ }\bibfield  {title} {\bibinfo {title} {{CNOT} gates for fluxonium qubits via selective darkening of transitions},\ }\href {https://doi.org/10.1103/PhysRevApplied.18.034063} {\bibfield  {journal} {\bibinfo  {journal} {Physical Review Applied}\ }\textbf {\bibinfo {volume} {18}},\ \bibinfo {pages} {034063} (\bibinfo {year} {2022})}\BibitemShut {NoStop}%
\bibitem [{\citenamefont {Dogan}\ \emph {et~al.}(2022)\citenamefont {Dogan}, \citenamefont {Rosenstock}, \citenamefont {Guevel}, \citenamefont {Xiong}, \citenamefont {Mencia}, \citenamefont {Somoroff}, \citenamefont {Nesterov}, \citenamefont {Vavilov}, \citenamefont {Manucharyan},\ and\ \citenamefont {Wang}}]{dogan_demonstration_2022}%
  \BibitemOpen
  \bibfield  {author} {\bibinfo {author} {\bibfnamefont {E.}~\bibnamefont {Dogan}}, \bibinfo {author} {\bibfnamefont {D.}~\bibnamefont {Rosenstock}}, \bibinfo {author} {\bibfnamefont {L.~L.}\ \bibnamefont {Guevel}}, \bibinfo {author} {\bibfnamefont {H.}~\bibnamefont {Xiong}}, \bibinfo {author} {\bibfnamefont {R.~A.}\ \bibnamefont {Mencia}}, \bibinfo {author} {\bibfnamefont {A.}~\bibnamefont {Somoroff}}, \bibinfo {author} {\bibfnamefont {K.~N.}\ \bibnamefont {Nesterov}}, \bibinfo {author} {\bibfnamefont {M.~G.}\ \bibnamefont {Vavilov}}, \bibinfo {author} {\bibfnamefont {V.~E.}\ \bibnamefont {Manucharyan}},\ and\ \bibinfo {author} {\bibfnamefont {C.}~\bibnamefont {Wang}},\ }\href {http://arxiv.org/abs/2204.11829} {\bibinfo {title} {Demonstration of the {Two}-{Fluxonium} {Cross}-{Resonance} {Gate}}} (\bibinfo {year} {2022})\BibitemShut {NoStop}%
\bibitem [{\citenamefont {Nesterov}\ \emph {et~al.}(2021)\citenamefont {Nesterov}, \citenamefont {Ficheux}, \citenamefont {Manucharyan},\ and\ \citenamefont {Vavilov}}]{nesterov_proposal_2021}%
  \BibitemOpen
  \bibfield  {author} {\bibinfo {author} {\bibfnamefont {K.~N.}\ \bibnamefont {Nesterov}}, \bibinfo {author} {\bibfnamefont {Q.}~\bibnamefont {Ficheux}}, \bibinfo {author} {\bibfnamefont {V.~E.}\ \bibnamefont {Manucharyan}},\ and\ \bibinfo {author} {\bibfnamefont {M.~G.}\ \bibnamefont {Vavilov}},\ }\bibfield  {title} {\bibinfo {title} {Proposal for entangling gates on fluxonium qubits via a two-photon transition},\ }\href {https://doi.org/10.1103/PRXQuantum.2.020345} {\bibfield  {journal} {\bibinfo  {journal} {PRX Quantum}\ }\textbf {\bibinfo {volume} {2}},\ \bibinfo {pages} {020345} (\bibinfo {year} {2021})}\BibitemShut {NoStop}%
\bibitem [{\citenamefont {Chen}\ \emph {et~al.}(2022)\citenamefont {Chen}, \citenamefont {Nesterov}, \citenamefont {Manucharyan},\ and\ \citenamefont {Vavilov}}]{chen_fast_2022}%
  \BibitemOpen
  \bibfield  {author} {\bibinfo {author} {\bibfnamefont {Y.}~\bibnamefont {Chen}}, \bibinfo {author} {\bibfnamefont {K.~N.}\ \bibnamefont {Nesterov}}, \bibinfo {author} {\bibfnamefont {V.~E.}\ \bibnamefont {Manucharyan}},\ and\ \bibinfo {author} {\bibfnamefont {M.~G.}\ \bibnamefont {Vavilov}},\ }\bibfield  {title} {\bibinfo {title} {Fast {Flux} {Entangling} {Gate} for {Fluxonium} {Circuits}},\ }\href {https://doi.org/10.1103/PhysRevApplied.18.034027} {\bibfield  {journal} {\bibinfo  {journal} {Physical Review Applied}\ }\textbf {\bibinfo {volume} {18}},\ \bibinfo {pages} {034027} (\bibinfo {year} {2022})}\BibitemShut {NoStop}%
\bibitem [{\citenamefont {Moskalenko}\ \emph {et~al.}(2022)\citenamefont {Moskalenko}, \citenamefont {Simakov}, \citenamefont {Abramov}, \citenamefont {Grigorev}, \citenamefont {Moskalev}, \citenamefont {Pishchimova}, \citenamefont {Smirnov}, \citenamefont {Zikiy}, \citenamefont {Rodionov},\ and\ \citenamefont {Besedin}}]{moskalenko_high_2022}%
  \BibitemOpen
  \bibfield  {author} {\bibinfo {author} {\bibfnamefont {I.~N.}\ \bibnamefont {Moskalenko}}, \bibinfo {author} {\bibfnamefont {I.~A.}\ \bibnamefont {Simakov}}, \bibinfo {author} {\bibfnamefont {N.~N.}\ \bibnamefont {Abramov}}, \bibinfo {author} {\bibfnamefont {A.~A.}\ \bibnamefont {Grigorev}}, \bibinfo {author} {\bibfnamefont {D.~O.}\ \bibnamefont {Moskalev}}, \bibinfo {author} {\bibfnamefont {A.~A.}\ \bibnamefont {Pishchimova}}, \bibinfo {author} {\bibfnamefont {N.~S.}\ \bibnamefont {Smirnov}}, \bibinfo {author} {\bibfnamefont {E.~V.}\ \bibnamefont {Zikiy}}, \bibinfo {author} {\bibfnamefont {I.~A.}\ \bibnamefont {Rodionov}},\ and\ \bibinfo {author} {\bibfnamefont {I.~S.}\ \bibnamefont {Besedin}},\ }\bibfield  {title} {\bibinfo {title} {High fidelity two-qubit gates on fluxoniums using a tunable coupler},\ }\href {https://doi.org/10.1038/s41534-022-00644-x} {\bibfield  {journal} {\bibinfo  {journal} {npj Quantum Information}\ }\textbf {\bibinfo {volume} {8}},\ \bibinfo {pages} {130} (\bibinfo {year}
  {2022})}\BibitemShut {NoStop}%
\bibitem [{\citenamefont {Simakov}\ \emph {et~al.}(2023{\natexlab{a}})\citenamefont {Simakov}, \citenamefont {Mazhorin}, \citenamefont {Moskalenko}, \citenamefont {Abramov}, \citenamefont {Grigorev}, \citenamefont {Moskalev}, \citenamefont {Pishchimova}, \citenamefont {Smirnov}, \citenamefont {Zikiy}, \citenamefont {Rodionov},\ and\ \citenamefont {Besedin}}]{simakov_coupler_2023}%
  \BibitemOpen
  \bibfield  {author} {\bibinfo {author} {\bibfnamefont {I.~A.}\ \bibnamefont {Simakov}}, \bibinfo {author} {\bibfnamefont {G.~S.}\ \bibnamefont {Mazhorin}}, \bibinfo {author} {\bibfnamefont {I.~N.}\ \bibnamefont {Moskalenko}}, \bibinfo {author} {\bibfnamefont {N.~N.}\ \bibnamefont {Abramov}}, \bibinfo {author} {\bibfnamefont {A.~A.}\ \bibnamefont {Grigorev}}, \bibinfo {author} {\bibfnamefont {D.~O.}\ \bibnamefont {Moskalev}}, \bibinfo {author} {\bibfnamefont {A.~A.}\ \bibnamefont {Pishchimova}}, \bibinfo {author} {\bibfnamefont {N.~S.}\ \bibnamefont {Smirnov}}, \bibinfo {author} {\bibfnamefont {E.~V.}\ \bibnamefont {Zikiy}}, \bibinfo {author} {\bibfnamefont {I.~A.}\ \bibnamefont {Rodionov}},\ and\ \bibinfo {author} {\bibfnamefont {I.~S.}\ \bibnamefont {Besedin}},\ }\href {http://arxiv.org/abs/2302.09819} {\bibinfo {title} {Coupler microwave-activated controlled phase gate on fluxonium qubits}} (\bibinfo {year} {2023}{\natexlab{a}})\BibitemShut {NoStop}%
\bibitem [{\citenamefont {Setiawan}\ \emph {et~al.}(2023)\citenamefont {Setiawan}, \citenamefont {Groszkowski},\ and\ \citenamefont {Clerk}}]{setiawan_fast_2023}%
  \BibitemOpen
  \bibfield  {author} {\bibinfo {author} {\bibfnamefont {F.}~\bibnamefont {Setiawan}}, \bibinfo {author} {\bibfnamefont {P.}~\bibnamefont {Groszkowski}},\ and\ \bibinfo {author} {\bibfnamefont {A.~A.}\ \bibnamefont {Clerk}},\ }\bibfield  {title} {\bibinfo {title} {Fast and {Robust} {Geometric} {Two}-{Qubit} {Gates} for {Superconducting} {Qubits} and beyond},\ }\href {https://doi.org/10.1103/PhysRevApplied.19.034071} {\bibfield  {journal} {\bibinfo  {journal} {Physical Review Applied}\ }\textbf {\bibinfo {volume} {19}},\ \bibinfo {pages} {034071} (\bibinfo {year} {2023})}\BibitemShut {NoStop}%
\bibitem [{\citenamefont {Weiss}\ \emph {et~al.}(2022)\citenamefont {Weiss}, \citenamefont {Zhang}, \citenamefont {Ding}, \citenamefont {Ma}, \citenamefont {Schuster},\ and\ \citenamefont {Koch}}]{weiss_fast_2022}%
  \BibitemOpen
  \bibfield  {author} {\bibinfo {author} {\bibfnamefont {D.~K.}\ \bibnamefont {Weiss}}, \bibinfo {author} {\bibfnamefont {H.}~\bibnamefont {Zhang}}, \bibinfo {author} {\bibfnamefont {C.}~\bibnamefont {Ding}}, \bibinfo {author} {\bibfnamefont {Y.}~\bibnamefont {Ma}}, \bibinfo {author} {\bibfnamefont {D.~I.}\ \bibnamefont {Schuster}},\ and\ \bibinfo {author} {\bibfnamefont {J.}~\bibnamefont {Koch}},\ }\bibfield  {title} {\bibinfo {title} {Fast high-fidelity gates for galvanically-coupled fluxonium qubits using strong flux modulation},\ }\href {https://doi.org/10.1103/PRXQuantum.3.040336} {\bibfield  {journal} {\bibinfo  {journal} {PRX Quantum}\ }\textbf {\bibinfo {volume} {3}},\ \bibinfo {pages} {040336} (\bibinfo {year} {2022})}\BibitemShut {NoStop}%
\bibitem [{\citenamefont {Ciani}\ \emph {et~al.}(2022)\citenamefont {Ciani}, \citenamefont {Varbanov}, \citenamefont {Jolly}, \citenamefont {Andersen},\ and\ \citenamefont {Terhal}}]{ciani_microwave-activated_2022}%
  \BibitemOpen
  \bibfield  {author} {\bibinfo {author} {\bibfnamefont {A.}~\bibnamefont {Ciani}}, \bibinfo {author} {\bibfnamefont {B.~M.}\ \bibnamefont {Varbanov}}, \bibinfo {author} {\bibfnamefont {N.}~\bibnamefont {Jolly}}, \bibinfo {author} {\bibfnamefont {C.~K.}\ \bibnamefont {Andersen}},\ and\ \bibinfo {author} {\bibfnamefont {B.~M.}\ \bibnamefont {Terhal}},\ }\href {http://arxiv.org/abs/2206.06203} {\bibinfo {title} {Microwave-activated gates between a fluxonium and a transmon qubit}} (\bibinfo {year} {2022})\BibitemShut {NoStop}%
\bibitem [{\citenamefont {Moskalenko}\ \emph {et~al.}(2021)\citenamefont {Moskalenko}, \citenamefont {Besedin}, \citenamefont {Simakov},\ and\ \citenamefont {Ustinov}}]{moskalenko_tunable_2021}%
  \BibitemOpen
  \bibfield  {author} {\bibinfo {author} {\bibfnamefont {I.~N.}\ \bibnamefont {Moskalenko}}, \bibinfo {author} {\bibfnamefont {I.~S.}\ \bibnamefont {Besedin}}, \bibinfo {author} {\bibfnamefont {I.~A.}\ \bibnamefont {Simakov}},\ and\ \bibinfo {author} {\bibfnamefont {A.~V.}\ \bibnamefont {Ustinov}},\ }\bibfield  {title} {\bibinfo {title} {Tunable coupling scheme for implementing two-qubit gates on fluxonium qubits},\ }\href {https://doi.org/10.1063/5.0064800} {\bibfield  {journal} {\bibinfo  {journal} {Applied Physics Letters}\ }\textbf {\bibinfo {volume} {119}},\ \bibinfo {pages} {194001} (\bibinfo {year} {2021})}\BibitemShut {NoStop}%
\bibitem [{\citenamefont {Nesterov}\ \emph {et~al.}(2018)\citenamefont {Nesterov}, \citenamefont {Pechenezhskiy}, \citenamefont {Wang}, \citenamefont {Manucharyan},\ and\ \citenamefont {Vavilov}}]{nesterov_microwave-activated_2018}%
  \BibitemOpen
  \bibfield  {author} {\bibinfo {author} {\bibfnamefont {K.~N.}\ \bibnamefont {Nesterov}}, \bibinfo {author} {\bibfnamefont {I.~V.}\ \bibnamefont {Pechenezhskiy}}, \bibinfo {author} {\bibfnamefont {C.}~\bibnamefont {Wang}}, \bibinfo {author} {\bibfnamefont {V.~E.}\ \bibnamefont {Manucharyan}},\ and\ \bibinfo {author} {\bibfnamefont {M.~G.}\ \bibnamefont {Vavilov}},\ }\bibfield  {title} {\bibinfo {title} {Microwave-{Activated} {Controlled}-{Z} {Gate} for {Fixed}-{Frequency} {Fluxonium} {Qubits}},\ }\href {https://doi.org/10.1103/PhysRevA.98.030301} {\bibfield  {journal} {\bibinfo  {journal} {Physical Review A}\ }\textbf {\bibinfo {volume} {98}},\ \bibinfo {pages} {030301} (\bibinfo {year} {2018})}\BibitemShut {NoStop}%
\bibitem [{\citenamefont {Simakov}\ \emph {et~al.}(2023{\natexlab{b}})\citenamefont {Simakov}, \citenamefont {Mazhorin}, \citenamefont {Moskalenko}, \citenamefont {Seidov},\ and\ \citenamefont {Besedin}}]{simakov_high-fidelity_2023}%
  \BibitemOpen
  \bibfield  {author} {\bibinfo {author} {\bibfnamefont {I.~A.}\ \bibnamefont {Simakov}}, \bibinfo {author} {\bibfnamefont {G.~S.}\ \bibnamefont {Mazhorin}}, \bibinfo {author} {\bibfnamefont {I.~N.}\ \bibnamefont {Moskalenko}}, \bibinfo {author} {\bibfnamefont {S.~S.}\ \bibnamefont {Seidov}},\ and\ \bibinfo {author} {\bibfnamefont {I.~S.}\ \bibnamefont {Besedin}},\ }\href {http://arxiv.org/abs/2308.15229} {\bibinfo {title} {High-fidelity transmon coupler activated {CCZ} gate on fluxonium qubits}} (\bibinfo {year} {2023}{\natexlab{b}})\BibitemShut {NoStop}%
\bibitem [{\citenamefont {Ma}\ \emph {et~al.}(2023)\citenamefont {Ma}, \citenamefont {Zhang}, \citenamefont {Wu}, \citenamefont {Bao}, \citenamefont {Chang}, \citenamefont {Chen}, \citenamefont {Deng}, \citenamefont {Gao}, \citenamefont {Gao}, \citenamefont {Hu}, \citenamefont {Ji}, \citenamefont {Ku}, \citenamefont {Lu}, \citenamefont {Ma}, \citenamefont {Mao}, \citenamefont {Song}, \citenamefont {Sun}, \citenamefont {Tang}, \citenamefont {Wang}, \citenamefont {Wang}, \citenamefont {Wang}, \citenamefont {Xia}, \citenamefont {Ying}, \citenamefont {Zhan}, \citenamefont {Zhou}, \citenamefont {Zhu}, \citenamefont {Zhu}, \citenamefont {Shi}, \citenamefont {Zhao},\ and\ \citenamefont {Deng}}]{ma_native_2023}%
  \BibitemOpen
  \bibfield  {author} {\bibinfo {author} {\bibfnamefont {X.}~\bibnamefont {Ma}}, \bibinfo {author} {\bibfnamefont {G.}~\bibnamefont {Zhang}}, \bibinfo {author} {\bibfnamefont {F.}~\bibnamefont {Wu}}, \bibinfo {author} {\bibfnamefont {F.}~\bibnamefont {Bao}}, \bibinfo {author} {\bibfnamefont {X.}~\bibnamefont {Chang}}, \bibinfo {author} {\bibfnamefont {J.}~\bibnamefont {Chen}}, \bibinfo {author} {\bibfnamefont {H.}~\bibnamefont {Deng}}, \bibinfo {author} {\bibfnamefont {R.}~\bibnamefont {Gao}}, \bibinfo {author} {\bibfnamefont {X.}~\bibnamefont {Gao}}, \bibinfo {author} {\bibfnamefont {L.}~\bibnamefont {Hu}}, \bibinfo {author} {\bibfnamefont {H.}~\bibnamefont {Ji}}, \bibinfo {author} {\bibfnamefont {H.-S.}\ \bibnamefont {Ku}}, \bibinfo {author} {\bibfnamefont {K.}~\bibnamefont {Lu}}, \bibinfo {author} {\bibfnamefont {L.}~\bibnamefont {Ma}}, \bibinfo {author} {\bibfnamefont {L.}~\bibnamefont {Mao}}, \bibinfo {author} {\bibfnamefont {Z.}~\bibnamefont {Song}}, \bibinfo {author} {\bibfnamefont {H.}~\bibnamefont
  {Sun}}, \bibinfo {author} {\bibfnamefont {C.}~\bibnamefont {Tang}}, \bibinfo {author} {\bibfnamefont {F.}~\bibnamefont {Wang}}, \bibinfo {author} {\bibfnamefont {H.}~\bibnamefont {Wang}}, \bibinfo {author} {\bibfnamefont {T.}~\bibnamefont {Wang}}, \bibinfo {author} {\bibfnamefont {T.}~\bibnamefont {Xia}}, \bibinfo {author} {\bibfnamefont {M.}~\bibnamefont {Ying}}, \bibinfo {author} {\bibfnamefont {H.}~\bibnamefont {Zhan}}, \bibinfo {author} {\bibfnamefont {T.}~\bibnamefont {Zhou}}, \bibinfo {author} {\bibfnamefont {M.}~\bibnamefont {Zhu}}, \bibinfo {author} {\bibfnamefont {Q.}~\bibnamefont {Zhu}}, \bibinfo {author} {\bibfnamefont {Y.}~\bibnamefont {Shi}}, \bibinfo {author} {\bibfnamefont {H.-H.}\ \bibnamefont {Zhao}},\ and\ \bibinfo {author} {\bibfnamefont {C.}~\bibnamefont {Deng}},\ }\href {http://arxiv.org/abs/2308.16040} {\bibinfo {title} {Native approach to controlled-{Z} gates in inductively coupled fluxonium qubits}} (\bibinfo {year} {2023})\BibitemShut {NoStop}%
\bibitem [{\citenamefont {Nguyen}\ \emph {et~al.}(2022)\citenamefont {Nguyen}, \citenamefont {Koolstra}, \citenamefont {Kim}, \citenamefont {Morvan}, \citenamefont {Chistolini}, \citenamefont {Singh}, \citenamefont {Nesterov}, \citenamefont {J{\"u}nger}, \citenamefont {Chen}, \citenamefont {Pedramrazi}, \citenamefont {Mitchell}, \citenamefont {Kreikebaum}, \citenamefont {Puri}, \citenamefont {Santiago},\ and\ \citenamefont {Siddiqi}}]{nguyen_scalable_2022}%
  \BibitemOpen
  \bibfield  {author} {\bibinfo {author} {\bibfnamefont {L.~B.}\ \bibnamefont {Nguyen}}, \bibinfo {author} {\bibfnamefont {G.}~\bibnamefont {Koolstra}}, \bibinfo {author} {\bibfnamefont {Y.}~\bibnamefont {Kim}}, \bibinfo {author} {\bibfnamefont {A.}~\bibnamefont {Morvan}}, \bibinfo {author} {\bibfnamefont {T.}~\bibnamefont {Chistolini}}, \bibinfo {author} {\bibfnamefont {S.}~\bibnamefont {Singh}}, \bibinfo {author} {\bibfnamefont {K.~N.}\ \bibnamefont {Nesterov}}, \bibinfo {author} {\bibfnamefont {C.}~\bibnamefont {J{\"u}nger}}, \bibinfo {author} {\bibfnamefont {L.}~\bibnamefont {Chen}}, \bibinfo {author} {\bibfnamefont {Z.}~\bibnamefont {Pedramrazi}}, \bibinfo {author} {\bibfnamefont {B.~K.}\ \bibnamefont {Mitchell}}, \bibinfo {author} {\bibfnamefont {J.~M.}\ \bibnamefont {Kreikebaum}}, \bibinfo {author} {\bibfnamefont {S.}~\bibnamefont {Puri}}, \bibinfo {author} {\bibfnamefont {D.~I.}\ \bibnamefont {Santiago}},\ and\ \bibinfo {author} {\bibfnamefont {I.}~\bibnamefont {Siddiqi}},\ }\bibfield  {title}
  {\bibinfo {title} {Scalable {High}-{Performance} {Fluxonium} {Quantum} {Processor}},\ }\href {https://doi.org/10.1103/PRXQuantum.3.037001} {\bibfield  {journal} {\bibinfo  {journal} {PRX Quantum}\ }\textbf {\bibinfo {volume} {3}},\ \bibinfo {pages} {037001} (\bibinfo {year} {2022})}\BibitemShut {NoStop}%
\bibitem [{\citenamefont {Zhang}\ \emph {et~al.}(2023)\citenamefont {Zhang}, \citenamefont {Ding}, \citenamefont {Weiss}, \citenamefont {Huang}, \citenamefont {Ma}, \citenamefont {Guinn}, \citenamefont {Sussman}, \citenamefont {Chitta}, \citenamefont {Chen}, \citenamefont {Houck}, \citenamefont {Koch},\ and\ \citenamefont {Schuster}}]{zhang_tunable_2023}%
  \BibitemOpen
  \bibfield  {author} {\bibinfo {author} {\bibfnamefont {H.}~\bibnamefont {Zhang}}, \bibinfo {author} {\bibfnamefont {C.}~\bibnamefont {Ding}}, \bibinfo {author} {\bibfnamefont {D.~K.}\ \bibnamefont {Weiss}}, \bibinfo {author} {\bibfnamefont {Z.}~\bibnamefont {Huang}}, \bibinfo {author} {\bibfnamefont {Y.}~\bibnamefont {Ma}}, \bibinfo {author} {\bibfnamefont {C.}~\bibnamefont {Guinn}}, \bibinfo {author} {\bibfnamefont {S.}~\bibnamefont {Sussman}}, \bibinfo {author} {\bibfnamefont {S.~P.}\ \bibnamefont {Chitta}}, \bibinfo {author} {\bibfnamefont {D.}~\bibnamefont {Chen}}, \bibinfo {author} {\bibfnamefont {A.~A.}\ \bibnamefont {Houck}}, \bibinfo {author} {\bibfnamefont {J.}~\bibnamefont {Koch}},\ and\ \bibinfo {author} {\bibfnamefont {D.~I.}\ \bibnamefont {Schuster}},\ }\href {http://arxiv.org/abs/2309.05720} {\bibinfo {title} {Tunable inductive coupler for high fidelity gates between fluxonium qubits}} (\bibinfo {year} {2023})\BibitemShut {NoStop}%
\bibitem [{\citenamefont {Ding}\ \emph {et~al.}(2023)\citenamefont {Ding}, \citenamefont {Hays}, \citenamefont {Sung}, \citenamefont {Kannan}, \citenamefont {An}, \citenamefont {Di~Paolo}, \citenamefont {Karamlou}, \citenamefont {Hazard}, \citenamefont {Azar}, \citenamefont {Kim}, \citenamefont {Niedzielski}, \citenamefont {Melville}, \citenamefont {Schwartz}, \citenamefont {Yoder}, \citenamefont {Orlando}, \citenamefont {Gustavsson}, \citenamefont {Grover}, \citenamefont {Serniak},\ and\ \citenamefont {Oliver}}]{ding_high-fidelity_2023}%
  \BibitemOpen
  \bibfield  {author} {\bibinfo {author} {\bibfnamefont {L.}~\bibnamefont {Ding}}, \bibinfo {author} {\bibfnamefont {M.}~\bibnamefont {Hays}}, \bibinfo {author} {\bibfnamefont {Y.}~\bibnamefont {Sung}}, \bibinfo {author} {\bibfnamefont {B.}~\bibnamefont {Kannan}}, \bibinfo {author} {\bibfnamefont {J.}~\bibnamefont {An}}, \bibinfo {author} {\bibfnamefont {A.}~\bibnamefont {Di~Paolo}}, \bibinfo {author} {\bibfnamefont {A.~H.}\ \bibnamefont {Karamlou}}, \bibinfo {author} {\bibfnamefont {T.~M.}\ \bibnamefont {Hazard}}, \bibinfo {author} {\bibfnamefont {K.}~\bibnamefont {Azar}}, \bibinfo {author} {\bibfnamefont {D.~K.}\ \bibnamefont {Kim}}, \bibinfo {author} {\bibfnamefont {B.~M.}\ \bibnamefont {Niedzielski}}, \bibinfo {author} {\bibfnamefont {A.}~\bibnamefont {Melville}}, \bibinfo {author} {\bibfnamefont {M.~E.}\ \bibnamefont {Schwartz}}, \bibinfo {author} {\bibfnamefont {J.~L.}\ \bibnamefont {Yoder}}, \bibinfo {author} {\bibfnamefont {T.~P.}\ \bibnamefont {Orlando}}, \bibinfo {author} {\bibfnamefont
  {S.}~\bibnamefont {Gustavsson}}, \bibinfo {author} {\bibfnamefont {J.~A.}\ \bibnamefont {Grover}}, \bibinfo {author} {\bibfnamefont {K.}~\bibnamefont {Serniak}},\ and\ \bibinfo {author} {\bibfnamefont {W.~D.}\ \bibnamefont {Oliver}},\ }\href {http://arxiv.org/abs/2304.06087} {\bibinfo {title} {High-{Fidelity}, {Frequency}-{Flexible} {Two}-{Qubit} {Fluxonium} {Gates} with a {Transmon} {Coupler}}} (\bibinfo {year} {2023})\BibitemShut {NoStop}%
\bibitem [{\citenamefont {Mizel}\ and\ \citenamefont {Yanay}(2020)}]{mizel_right-sizing_2020}%
  \BibitemOpen
  \bibfield  {author} {\bibinfo {author} {\bibfnamefont {A.}~\bibnamefont {Mizel}}\ and\ \bibinfo {author} {\bibfnamefont {Y.}~\bibnamefont {Yanay}},\ }\bibfield  {title} {\bibinfo {title} {Right-sizing fluxonium against charge noise},\ }\href {https://doi.org/10.1103/PhysRevB.102.014512} {\bibfield  {journal} {\bibinfo  {journal} {Physical Review B}\ }\textbf {\bibinfo {volume} {102}},\ \bibinfo {pages} {014512} (\bibinfo {year} {2020})}\BibitemShut {NoStop}%
\bibitem [{\citenamefont {Zhu}\ and\ \citenamefont {Koch}(2013)}]{zhu_asymptotic_2013}%
  \BibitemOpen
  \bibfield  {author} {\bibinfo {author} {\bibfnamefont {G.}~\bibnamefont {Zhu}}\ and\ \bibinfo {author} {\bibfnamefont {J.}~\bibnamefont {Koch}},\ }\bibfield  {title} {\bibinfo {title} {Asymptotic {Expressions} for {Charge} {Matrix} {Elements} of the {Fluxonium} {Circuit}},\ }\href {https://doi.org/10.1103/PhysRevB.87.144518} {\bibfield  {journal} {\bibinfo  {journal} {Physical Review B}\ }\textbf {\bibinfo {volume} {87}},\ \bibinfo {pages} {144518} (\bibinfo {year} {2013})}\BibitemShut {NoStop}%
\bibitem [{\citenamefont {Herrera-Mart{\'i}}\ \emph {et~al.}(2013)\citenamefont {Herrera-Mart{\'i}}, \citenamefont {Nazir},\ and\ \citenamefont {Barrett}}]{herrera-marti_tradeoff_2013}%
  \BibitemOpen
  \bibfield  {author} {\bibinfo {author} {\bibfnamefont {D.~A.}\ \bibnamefont {Herrera-Mart{\'i}}}, \bibinfo {author} {\bibfnamefont {A.}~\bibnamefont {Nazir}},\ and\ \bibinfo {author} {\bibfnamefont {S.~D.}\ \bibnamefont {Barrett}},\ }\bibfield  {title} {\bibinfo {title} {Tradeoff between {Leakage} and {Dephasing} {Errors} in the {Fluxonium} {Qubit}},\ }\href {https://doi.org/10.1103/PhysRevB.88.094512} {\bibfield  {journal} {\bibinfo  {journal} {Physical Review B}\ }\textbf {\bibinfo {volume} {88}},\ \bibinfo {pages} {094512} (\bibinfo {year} {2013})}\BibitemShut {NoStop}%
\bibitem [{\citenamefont {Hassani}\ \emph {et~al.}(2022)\citenamefont {Hassani}, \citenamefont {Peruzzo}, \citenamefont {Kapoor}, \citenamefont {Trioni}, \citenamefont {Zemlicka},\ and\ \citenamefont {Fink}}]{hassani_superconducting_2022}%
  \BibitemOpen
  \bibfield  {author} {\bibinfo {author} {\bibfnamefont {F.}~\bibnamefont {Hassani}}, \bibinfo {author} {\bibfnamefont {M.}~\bibnamefont {Peruzzo}}, \bibinfo {author} {\bibfnamefont {L.~N.}\ \bibnamefont {Kapoor}}, \bibinfo {author} {\bibfnamefont {A.}~\bibnamefont {Trioni}}, \bibinfo {author} {\bibfnamefont {M.}~\bibnamefont {Zemlicka}},\ and\ \bibinfo {author} {\bibfnamefont {J.~M.}\ \bibnamefont {Fink}},\ }\href {http://arxiv.org/abs/2202.13917} {\bibinfo {title} {A superconducting qubit with noise-insensitive plasmon levels and decay-protected fluxon states}} (\bibinfo {year} {2022})\BibitemShut {NoStop}%
\bibitem [{\citenamefont {Gyenis}\ \emph {et~al.}(2021)\citenamefont {Gyenis}, \citenamefont {Di~Paolo}, \citenamefont {Koch}, \citenamefont {Blais}, \citenamefont {Houck},\ and\ \citenamefont {Schuster}}]{gyenis_moving_2021}%
  \BibitemOpen
  \bibfield  {author} {\bibinfo {author} {\bibfnamefont {A.}~\bibnamefont {Gyenis}}, \bibinfo {author} {\bibfnamefont {A.}~\bibnamefont {Di~Paolo}}, \bibinfo {author} {\bibfnamefont {J.}~\bibnamefont {Koch}}, \bibinfo {author} {\bibfnamefont {A.}~\bibnamefont {Blais}}, \bibinfo {author} {\bibfnamefont {A.~A.}\ \bibnamefont {Houck}},\ and\ \bibinfo {author} {\bibfnamefont {D.~I.}\ \bibnamefont {Schuster}},\ }\bibfield  {title} {\bibinfo {title} {Moving beyond the {Transmon}: {Noise}-{Protected} {Superconducting} {Quantum} {Circuits}},\ }\href {https://doi.org/10.1103/PRXQuantum.2.030101} {\bibfield  {journal} {\bibinfo  {journal} {PRX Quantum}\ }\textbf {\bibinfo {volume} {2}},\ \bibinfo {pages} {030101} (\bibinfo {year} {2021})}\BibitemShut {NoStop}%
\bibitem [{\citenamefont {Lin}\ \emph {et~al.}(2018)\citenamefont {Lin}, \citenamefont {Nguyen}, \citenamefont {Grabon}, \citenamefont {Miguel}, \citenamefont {Pankratova},\ and\ \citenamefont {Manucharyan}}]{lin_protecting_2018}%
  \BibitemOpen
  \bibfield  {author} {\bibinfo {author} {\bibfnamefont {Y.-H.}\ \bibnamefont {Lin}}, \bibinfo {author} {\bibfnamefont {L.~B.}\ \bibnamefont {Nguyen}}, \bibinfo {author} {\bibfnamefont {N.}~\bibnamefont {Grabon}}, \bibinfo {author} {\bibfnamefont {J.~S.}\ \bibnamefont {Miguel}}, \bibinfo {author} {\bibfnamefont {N.}~\bibnamefont {Pankratova}},\ and\ \bibinfo {author} {\bibfnamefont {V.~E.}\ \bibnamefont {Manucharyan}},\ }\bibfield  {title} {\bibinfo {title} {Protecting a superconducting qubit from energy decay by selection rule engineering},\ }\href {https://doi.org/10.1103/PhysRevLett.120.150503} {\bibfield  {journal} {\bibinfo  {journal} {Physical Review Letters}\ }\textbf {\bibinfo {volume} {120}},\ \bibinfo {pages} {150503} (\bibinfo {year} {2018})}\BibitemShut {NoStop}%
\bibitem [{\citenamefont {Sung}\ \emph {et~al.}(2021)\citenamefont {Sung}, \citenamefont {Ding}, \citenamefont {Braum{\"u}ller}, \citenamefont {Veps{\"a}l{\"a}inen}, \citenamefont {Kannan}, \citenamefont {Kjaergaard}, \citenamefont {Greene}, \citenamefont {Samach}, \citenamefont {McNally}, \citenamefont {Kim}, \citenamefont {Melville}, \citenamefont {Niedzielski}, \citenamefont {Schwartz}, \citenamefont {Yoder}, \citenamefont {Orlando}, \citenamefont {Gustavsson},\ and\ \citenamefont {Oliver}}]{sung_realization_2021}%
  \BibitemOpen
  \bibfield  {author} {\bibinfo {author} {\bibfnamefont {Y.}~\bibnamefont {Sung}}, \bibinfo {author} {\bibfnamefont {L.}~\bibnamefont {Ding}}, \bibinfo {author} {\bibfnamefont {J.}~\bibnamefont {Braum{\"u}ller}}, \bibinfo {author} {\bibfnamefont {A.}~\bibnamefont {Veps{\"a}l{\"a}inen}}, \bibinfo {author} {\bibfnamefont {B.}~\bibnamefont {Kannan}}, \bibinfo {author} {\bibfnamefont {M.}~\bibnamefont {Kjaergaard}}, \bibinfo {author} {\bibfnamefont {A.}~\bibnamefont {Greene}}, \bibinfo {author} {\bibfnamefont {G.~O.}\ \bibnamefont {Samach}}, \bibinfo {author} {\bibfnamefont {C.}~\bibnamefont {McNally}}, \bibinfo {author} {\bibfnamefont {D.}~\bibnamefont {Kim}}, \bibinfo {author} {\bibfnamefont {A.}~\bibnamefont {Melville}}, \bibinfo {author} {\bibfnamefont {B.~M.}\ \bibnamefont {Niedzielski}}, \bibinfo {author} {\bibfnamefont {M.~E.}\ \bibnamefont {Schwartz}}, \bibinfo {author} {\bibfnamefont {J.~L.}\ \bibnamefont {Yoder}}, \bibinfo {author} {\bibfnamefont {T.~P.}\ \bibnamefont {Orlando}}, \bibinfo {author}
  {\bibfnamefont {S.}~\bibnamefont {Gustavsson}},\ and\ \bibinfo {author} {\bibfnamefont {W.~D.}\ \bibnamefont {Oliver}},\ }\bibfield  {title} {\bibinfo {title} {Realization of {High}-{Fidelity} {CZ} and \${ZZ}\$-{Free} {iSWAP} {Gates} with a {Tunable} {Coupler}},\ }\href {https://doi.org/10.1103/PhysRevX.11.021058} {\bibfield  {journal} {\bibinfo  {journal} {Physical Review X}\ }\textbf {\bibinfo {volume} {11}},\ \bibinfo {pages} {021058} (\bibinfo {year} {2021})},\ \bibinfo {note} {publisher: American Physical Society}\BibitemShut {NoStop}%
\bibitem [{\citenamefont {Yan}\ \emph {et~al.}(2018)\citenamefont {Yan}, \citenamefont {Krantz}, \citenamefont {Sung}, \citenamefont {Kjaergaard}, \citenamefont {Campbell}, \citenamefont {Wang}, \citenamefont {Orlando}, \citenamefont {Gustavsson},\ and\ \citenamefont {Oliver}}]{yan_tunable_2018}%
  \BibitemOpen
  \bibfield  {author} {\bibinfo {author} {\bibfnamefont {F.}~\bibnamefont {Yan}}, \bibinfo {author} {\bibfnamefont {P.}~\bibnamefont {Krantz}}, \bibinfo {author} {\bibfnamefont {Y.}~\bibnamefont {Sung}}, \bibinfo {author} {\bibfnamefont {M.}~\bibnamefont {Kjaergaard}}, \bibinfo {author} {\bibfnamefont {D.}~\bibnamefont {Campbell}}, \bibinfo {author} {\bibfnamefont {J.~I.~J.}\ \bibnamefont {Wang}}, \bibinfo {author} {\bibfnamefont {T.~P.}\ \bibnamefont {Orlando}}, \bibinfo {author} {\bibfnamefont {S.}~\bibnamefont {Gustavsson}},\ and\ \bibinfo {author} {\bibfnamefont {W.~D.}\ \bibnamefont {Oliver}},\ }\bibfield  {title} {\bibinfo {title} {A tunable coupling scheme for implementing high-fidelity two-qubit gates},\ }\href {https://doi.org/10.1103/PhysRevApplied.10.054062} {\bibfield  {journal} {\bibinfo  {journal} {Physical Review Applied}\ }\textbf {\bibinfo {volume} {10}},\ \bibinfo {pages} {054062} (\bibinfo {year} {2018})},\ \bibinfo {note} {arXiv:1803.09813 [quant-ph]}\BibitemShut {NoStop}%
\bibitem [{\citenamefont {Paik}\ \emph {et~al.}(2016)\citenamefont {Paik}, \citenamefont {Mezzacapo}, \citenamefont {Sandberg}, \citenamefont {McClure}, \citenamefont {Abdo}, \citenamefont {Corcoles}, \citenamefont {Dial}, \citenamefont {Bogorin}, \citenamefont {Plourde}, \citenamefont {Steffen}, \citenamefont {Cross}, \citenamefont {Gambetta},\ and\ \citenamefont {Chow}}]{paik_experimental_2016}%
  \BibitemOpen
  \bibfield  {author} {\bibinfo {author} {\bibfnamefont {H.}~\bibnamefont {Paik}}, \bibinfo {author} {\bibfnamefont {A.}~\bibnamefont {Mezzacapo}}, \bibinfo {author} {\bibfnamefont {M.}~\bibnamefont {Sandberg}}, \bibinfo {author} {\bibfnamefont {D.~T.}\ \bibnamefont {McClure}}, \bibinfo {author} {\bibfnamefont {B.}~\bibnamefont {Abdo}}, \bibinfo {author} {\bibfnamefont {A.~D.}\ \bibnamefont {Corcoles}}, \bibinfo {author} {\bibfnamefont {O.}~\bibnamefont {Dial}}, \bibinfo {author} {\bibfnamefont {D.~F.}\ \bibnamefont {Bogorin}}, \bibinfo {author} {\bibfnamefont {B.~L.~T.}\ \bibnamefont {Plourde}}, \bibinfo {author} {\bibfnamefont {M.}~\bibnamefont {Steffen}}, \bibinfo {author} {\bibfnamefont {A.~W.}\ \bibnamefont {Cross}}, \bibinfo {author} {\bibfnamefont {J.~M.}\ \bibnamefont {Gambetta}},\ and\ \bibinfo {author} {\bibfnamefont {J.~M.}\ \bibnamefont {Chow}},\ }\href {https://doi.org/10.1103/PhysRevLett.117.250502} {\bibinfo {title} {Experimental demonstration of a resonator-induced phase gate in a multi-qubit
  circuit {QED} system}} (\bibinfo {year} {2016}),\ \bibinfo {note} {publication Title: arXiv.org}\BibitemShut {NoStop}%
\bibitem [{\citenamefont {Puri}\ and\ \citenamefont {Blais}(2016)}]{puri_high-fidelity_2016}%
  \BibitemOpen
  \bibfield  {author} {\bibinfo {author} {\bibfnamefont {S.}~\bibnamefont {Puri}}\ and\ \bibinfo {author} {\bibfnamefont {A.}~\bibnamefont {Blais}},\ }\bibfield  {title} {\bibinfo {title} {High-{Fidelity} {Resonator}-{Induced} {Phase} {Gate} with {Single}-{Mode} {Squeezing}},\ }\href {https://doi.org/10.1103/PhysRevLett.116.180501} {\bibfield  {journal} {\bibinfo  {journal} {Physical Review Letters}\ }\textbf {\bibinfo {volume} {116}},\ \bibinfo {pages} {180501} (\bibinfo {year} {2016})},\ \bibinfo {note} {publisher: American Physical Society}\BibitemShut {NoStop}%
\bibitem [{\citenamefont {Shearrow}\ \emph {et~al.}(2018)\citenamefont {Shearrow}, \citenamefont {Koolstra}, \citenamefont {Whiteley}, \citenamefont {Earnest}, \citenamefont {Barry}, \citenamefont {Heremans}, \citenamefont {Awschalom}, \citenamefont {Shirokoff},\ and\ \citenamefont {Schuster}}]{shearrow_atomic_2018}%
  \BibitemOpen
  \bibfield  {author} {\bibinfo {author} {\bibfnamefont {A.}~\bibnamefont {Shearrow}}, \bibinfo {author} {\bibfnamefont {G.}~\bibnamefont {Koolstra}}, \bibinfo {author} {\bibfnamefont {S.~J.}\ \bibnamefont {Whiteley}}, \bibinfo {author} {\bibfnamefont {N.}~\bibnamefont {Earnest}}, \bibinfo {author} {\bibfnamefont {P.~S.}\ \bibnamefont {Barry}}, \bibinfo {author} {\bibfnamefont {F.~J.}\ \bibnamefont {Heremans}}, \bibinfo {author} {\bibfnamefont {D.~D.}\ \bibnamefont {Awschalom}}, \bibinfo {author} {\bibfnamefont {E.}~\bibnamefont {Shirokoff}},\ and\ \bibinfo {author} {\bibfnamefont {D.~I.}\ \bibnamefont {Schuster}},\ }\bibfield  {title} {\bibinfo {title} {Atomic layer deposition of titanium nitride for quantum circuits},\ }\href {https://doi.org/10.1063/1.5053461} {\bibfield  {journal} {\bibinfo  {journal} {Applied Physics Letters}\ }\textbf {\bibinfo {volume} {113}},\ \bibinfo {pages} {212601} (\bibinfo {year} {2018})}\BibitemShut {NoStop}%
\bibitem [{\citenamefont {Gao}(2008)}]{gao_physics_2008}%
  \BibitemOpen
  \bibfield  {author} {\bibinfo {author} {\bibfnamefont {J.}~\bibnamefont {Gao}},\ }\emph {\bibinfo {title} {The {Physics} of {Superconducting} {Microwave} {Resonators}}},\ \href {https://doi.org/10.7907/RAT0-VM75} {\bibinfo {type} {phd}},\ \bibinfo  {school} {California Institute of Technology} (\bibinfo {year} {2008})\BibitemShut {NoStop}%
\bibitem [{\citenamefont {Amin}\ \emph {et~al.}(2022)\citenamefont {Amin}, \citenamefont {Ladner}, \citenamefont {Jourdan}, \citenamefont {Hentz}, \citenamefont {Roch},\ and\ \citenamefont {Renard}}]{amin_loss_2022}%
  \BibitemOpen
  \bibfield  {author} {\bibinfo {author} {\bibfnamefont {K.~R.}\ \bibnamefont {Amin}}, \bibinfo {author} {\bibfnamefont {C.}~\bibnamefont {Ladner}}, \bibinfo {author} {\bibfnamefont {G.}~\bibnamefont {Jourdan}}, \bibinfo {author} {\bibfnamefont {S.}~\bibnamefont {Hentz}}, \bibinfo {author} {\bibfnamefont {N.}~\bibnamefont {Roch}},\ and\ \bibinfo {author} {\bibfnamefont {J.}~\bibnamefont {Renard}},\ }\bibfield  {title} {\bibinfo {title} {Loss mechanisms in {TiN} high impedance superconducting microwave circuits},\ }\href {https://doi.org/10.1063/5.0086019} {\bibfield  {journal} {\bibinfo  {journal} {Applied Physics Letters}\ }\textbf {\bibinfo {volume} {120}},\ \bibinfo {pages} {164001} (\bibinfo {year} {2022})}\BibitemShut {NoStop}%
\bibitem [{\citenamefont {Shi}\ \emph {et~al.}(2022)\citenamefont {Shi}, \citenamefont {Guo}, \citenamefont {Su}, \citenamefont {Chi}, \citenamefont {Sheng}, \citenamefont {Jiang}, \citenamefont {Cao}, \citenamefont {Wu}, \citenamefont {Tu}, \citenamefont {Sun}, \citenamefont {Chen},\ and\ \citenamefont {Wu}}]{shi_tantalum_2022}%
  \BibitemOpen
  \bibfield  {author} {\bibinfo {author} {\bibfnamefont {L.}~\bibnamefont {Shi}}, \bibinfo {author} {\bibfnamefont {T.}~\bibnamefont {Guo}}, \bibinfo {author} {\bibfnamefont {R.}~\bibnamefont {Su}}, \bibinfo {author} {\bibfnamefont {T.}~\bibnamefont {Chi}}, \bibinfo {author} {\bibfnamefont {Y.}~\bibnamefont {Sheng}}, \bibinfo {author} {\bibfnamefont {J.}~\bibnamefont {Jiang}}, \bibinfo {author} {\bibfnamefont {C.}~\bibnamefont {Cao}}, \bibinfo {author} {\bibfnamefont {J.}~\bibnamefont {Wu}}, \bibinfo {author} {\bibfnamefont {X.}~\bibnamefont {Tu}}, \bibinfo {author} {\bibfnamefont {G.}~\bibnamefont {Sun}}, \bibinfo {author} {\bibfnamefont {J.}~\bibnamefont {Chen}},\ and\ \bibinfo {author} {\bibfnamefont {P.}~\bibnamefont {Wu}},\ }\bibfield  {title} {\bibinfo {title} {Tantalum microwave resonators with ultra-high intrinsic quality factors},\ }\href {https://doi.org/10.1063/5.0124821} {\bibfield  {journal} {\bibinfo  {journal} {Applied Physics Letters}\ }\textbf {\bibinfo {volume} {121}},\ \bibinfo {pages}
  {242601} (\bibinfo {year} {2022})}\BibitemShut {NoStop}%
\bibitem [{\citenamefont {Koll{\'a}r}\ \emph {et~al.}(2019)\citenamefont {Koll{\'a}r}, \citenamefont {Fitzpatrick},\ and\ \citenamefont {Houck}}]{kollar_hyperbolic_2019}%
  \BibitemOpen
  \bibfield  {author} {\bibinfo {author} {\bibfnamefont {A.~J.}\ \bibnamefont {Koll{\'a}r}}, \bibinfo {author} {\bibfnamefont {M.}~\bibnamefont {Fitzpatrick}},\ and\ \bibinfo {author} {\bibfnamefont {A.~A.}\ \bibnamefont {Houck}},\ }\bibfield  {title} {\bibinfo {title} {Hyperbolic lattices in circuit quantum electrodynamics},\ }\href {https://doi.org/10.1038/s41586-019-1348-3} {\bibfield  {journal} {\bibinfo  {journal} {Nature}\ }\textbf {\bibinfo {volume} {571}},\ \bibinfo {pages} {45} (\bibinfo {year} {2019})},\ \bibinfo {note} {publisher: Nature Publishing Group}\BibitemShut {NoStop}%
\bibitem [{\citenamefont {Schafer}\ and\ \citenamefont {Adkins}(1991)}]{schafer_annealing_1991}%
  \BibitemOpen
  \bibfield  {author} {\bibinfo {author} {\bibfnamefont {J.}~\bibnamefont {Schafer}}\ and\ \bibinfo {author} {\bibfnamefont {C.~J.}\ \bibnamefont {Adkins}},\ }\bibfield  {title} {\bibinfo {title} {Annealing effects and oxide structure in alumina tunnelling barriers},\ }\href {https://doi.org/10.1088/0953-8984/3/17/008} {\bibfield  {journal} {\bibinfo  {journal} {Journal of Physics: Condensed Matter}\ }\textbf {\bibinfo {volume} {3}},\ \bibinfo {pages} {2907} (\bibinfo {year} {1991})}\BibitemShut {NoStop}%
\bibitem [{\citenamefont {Svensson}\ \emph {et~al.}(2018)\citenamefont {Svensson}, \citenamefont {Pierre}, \citenamefont {Simoen}, \citenamefont {Wustmann}, \citenamefont {Krantz}, \citenamefont {Bengtsson}, \citenamefont {Johansson}, \citenamefont {Bylander}, \citenamefont {Shumeiko},\ and\ \citenamefont {Delsing}}]{svensson_microwave_2018}%
  \BibitemOpen
  \bibfield  {author} {\bibinfo {author} {\bibfnamefont {I.-M.}\ \bibnamefont {Svensson}}, \bibinfo {author} {\bibfnamefont {M.}~\bibnamefont {Pierre}}, \bibinfo {author} {\bibfnamefont {M.}~\bibnamefont {Simoen}}, \bibinfo {author} {\bibfnamefont {W.}~\bibnamefont {Wustmann}}, \bibinfo {author} {\bibfnamefont {P.}~\bibnamefont {Krantz}}, \bibinfo {author} {\bibfnamefont {A.}~\bibnamefont {Bengtsson}}, \bibinfo {author} {\bibfnamefont {G.}~\bibnamefont {Johansson}}, \bibinfo {author} {\bibfnamefont {J.}~\bibnamefont {Bylander}}, \bibinfo {author} {\bibfnamefont {V.}~\bibnamefont {Shumeiko}},\ and\ \bibinfo {author} {\bibfnamefont {P.}~\bibnamefont {Delsing}},\ }\bibfield  {title} {\bibinfo {title} {Microwave photon generation in a doubly tunable superconducting resonator},\ }\href {https://doi.org/10.1088/1742-6596/969/1/012146} {\bibfield  {journal} {\bibinfo  {journal} {Journal of Physics: Conference Series}\ }\textbf {\bibinfo {volume} {969}},\ \bibinfo {pages} {012146} (\bibinfo {year}
  {2018})}\BibitemShut {NoStop}%
\bibitem [{\citenamefont {Vissers}\ \emph {et~al.}(2015)\citenamefont {Vissers}, \citenamefont {Hubmayr}, \citenamefont {Sandberg}, \citenamefont {Chaudhuri}, \citenamefont {Bockstiegel},\ and\ \citenamefont {Gao}}]{vissers_frequency-tunable_2015}%
  \BibitemOpen
  \bibfield  {author} {\bibinfo {author} {\bibfnamefont {M.~R.}\ \bibnamefont {Vissers}}, \bibinfo {author} {\bibfnamefont {J.}~\bibnamefont {Hubmayr}}, \bibinfo {author} {\bibfnamefont {M.}~\bibnamefont {Sandberg}}, \bibinfo {author} {\bibfnamefont {S.}~\bibnamefont {Chaudhuri}}, \bibinfo {author} {\bibfnamefont {C.}~\bibnamefont {Bockstiegel}},\ and\ \bibinfo {author} {\bibfnamefont {J.}~\bibnamefont {Gao}},\ }\bibfield  {title} {\bibinfo {title} {Frequency-tunable superconducting resonators via nonlinear kinetic inductance},\ }\href {https://doi.org/10.1063/1.4927444} {\bibfield  {journal} {\bibinfo  {journal} {Applied Physics Letters}\ }\textbf {\bibinfo {volume} {107}},\ \bibinfo {pages} {062601} (\bibinfo {year} {2015})}\BibitemShut {NoStop}%
\bibitem [{\citenamefont {Vall{\'e}s-Sanclemente}\ \emph {et~al.}(2023)\citenamefont {Vall{\'e}s-Sanclemente}, \citenamefont {van~der Meer}, \citenamefont {Finkel}, \citenamefont {Muthusubramanian}, \citenamefont {Beekman}, \citenamefont {Ali}, \citenamefont {Marques}, \citenamefont {Zachariadis}, \citenamefont {Veen}, \citenamefont {Stavenga}, \citenamefont {Haider},\ and\ \citenamefont {DiCarlo}}]{valles-sanclemente_post-fabrication_2023}%
  \BibitemOpen
  \bibfield  {author} {\bibinfo {author} {\bibfnamefont {S.}~\bibnamefont {Vall{\'e}s-Sanclemente}}, \bibinfo {author} {\bibfnamefont {S.~L.~M.}\ \bibnamefont {van~der Meer}}, \bibinfo {author} {\bibfnamefont {M.}~\bibnamefont {Finkel}}, \bibinfo {author} {\bibfnamefont {N.}~\bibnamefont {Muthusubramanian}}, \bibinfo {author} {\bibfnamefont {M.}~\bibnamefont {Beekman}}, \bibinfo {author} {\bibfnamefont {H.}~\bibnamefont {Ali}}, \bibinfo {author} {\bibfnamefont {J.~F.}\ \bibnamefont {Marques}}, \bibinfo {author} {\bibfnamefont {C.}~\bibnamefont {Zachariadis}}, \bibinfo {author} {\bibfnamefont {H.~M.}\ \bibnamefont {Veen}}, \bibinfo {author} {\bibfnamefont {T.}~\bibnamefont {Stavenga}}, \bibinfo {author} {\bibfnamefont {N.}~\bibnamefont {Haider}},\ and\ \bibinfo {author} {\bibfnamefont {L.}~\bibnamefont {DiCarlo}},\ }\bibfield  {title} {\bibinfo {title} {Post-fabrication frequency trimming of coplanar-waveguide resonators in circuit {QED} quantum processors},\ }\href {https://doi.org/10.1063/5.0148222}
  {\bibfield  {journal} {\bibinfo  {journal} {Applied Physics Letters}\ }\textbf {\bibinfo {volume} {123}},\ \bibinfo {pages} {034004} (\bibinfo {year} {2023})}\BibitemShut {NoStop}%
\bibitem [{\citenamefont {Zhang}\ \emph {et~al.}(2021)\citenamefont {Zhang}, \citenamefont {Chakram}, \citenamefont {Roy}, \citenamefont {Earnest}, \citenamefont {Lu}, \citenamefont {Huang}, \citenamefont {Weiss}, \citenamefont {Koch},\ and\ \citenamefont {Schuster}}]{zhang_universal_2021}%
  \BibitemOpen
  \bibfield  {author} {\bibinfo {author} {\bibfnamefont {H.}~\bibnamefont {Zhang}}, \bibinfo {author} {\bibfnamefont {S.}~\bibnamefont {Chakram}}, \bibinfo {author} {\bibfnamefont {T.}~\bibnamefont {Roy}}, \bibinfo {author} {\bibfnamefont {N.}~\bibnamefont {Earnest}}, \bibinfo {author} {\bibfnamefont {Y.}~\bibnamefont {Lu}}, \bibinfo {author} {\bibfnamefont {Z.}~\bibnamefont {Huang}}, \bibinfo {author} {\bibfnamefont {D.}~\bibnamefont {Weiss}}, \bibinfo {author} {\bibfnamefont {J.}~\bibnamefont {Koch}},\ and\ \bibinfo {author} {\bibfnamefont {D.~I.}\ \bibnamefont {Schuster}},\ }\bibfield  {title} {\bibinfo {title} {Universal fast flux control of a coherent, low-frequency qubit},\ }\href {https://doi.org/10.1103/PhysRevX.11.011010} {\bibfield  {journal} {\bibinfo  {journal} {Physical Review X}\ }\textbf {\bibinfo {volume} {11}},\ \bibinfo {pages} {011010} (\bibinfo {year} {2021})}\BibitemShut {NoStop}%
\bibitem [{\citenamefont {Sun}\ \emph {et~al.}(2023)\citenamefont {Sun}, \citenamefont {Wu}, \citenamefont {Ku}, \citenamefont {Ma}, \citenamefont {Qin}, \citenamefont {Song}, \citenamefont {Wang}, \citenamefont {Zhang}, \citenamefont {Zhou}, \citenamefont {Shi}, \citenamefont {Zhao},\ and\ \citenamefont {Deng}}]{sun_characterization_2023}%
  \BibitemOpen
  \bibfield  {author} {\bibinfo {author} {\bibfnamefont {H.}~\bibnamefont {Sun}}, \bibinfo {author} {\bibfnamefont {F.}~\bibnamefont {Wu}}, \bibinfo {author} {\bibfnamefont {H.-S.}\ \bibnamefont {Ku}}, \bibinfo {author} {\bibfnamefont {X.}~\bibnamefont {Ma}}, \bibinfo {author} {\bibfnamefont {J.}~\bibnamefont {Qin}}, \bibinfo {author} {\bibfnamefont {Z.}~\bibnamefont {Song}}, \bibinfo {author} {\bibfnamefont {T.}~\bibnamefont {Wang}}, \bibinfo {author} {\bibfnamefont {G.}~\bibnamefont {Zhang}}, \bibinfo {author} {\bibfnamefont {J.}~\bibnamefont {Zhou}}, \bibinfo {author} {\bibfnamefont {Y.}~\bibnamefont {Shi}}, \bibinfo {author} {\bibfnamefont {H.-H.}\ \bibnamefont {Zhao}},\ and\ \bibinfo {author} {\bibfnamefont {C.}~\bibnamefont {Deng}},\ }\href {http://arxiv.org/abs/2302.08110} {\bibinfo {title} {Characterization of loss mechanisms in a fluxonium qubit}} (\bibinfo {year} {2023})\BibitemShut {NoStop}%
\bibitem [{\citenamefont {Chang}\ \emph {et~al.}(2013)\citenamefont {Chang}, \citenamefont {Vissers}, \citenamefont {Corcoles}, \citenamefont {Sandberg}, \citenamefont {Gao}, \citenamefont {Abraham}, \citenamefont {Chow}, \citenamefont {Gambetta}, \citenamefont {Rothwell}, \citenamefont {Keefe}, \citenamefont {Steffen},\ and\ \citenamefont {Pappas}}]{chang_improved_2013}%
  \BibitemOpen
  \bibfield  {author} {\bibinfo {author} {\bibfnamefont {J.}~\bibnamefont {Chang}}, \bibinfo {author} {\bibfnamefont {M.~R.}\ \bibnamefont {Vissers}}, \bibinfo {author} {\bibfnamefont {A.~D.}\ \bibnamefont {Corcoles}}, \bibinfo {author} {\bibfnamefont {M.}~\bibnamefont {Sandberg}}, \bibinfo {author} {\bibfnamefont {J.}~\bibnamefont {Gao}}, \bibinfo {author} {\bibfnamefont {D.~W.}\ \bibnamefont {Abraham}}, \bibinfo {author} {\bibfnamefont {J.~M.}\ \bibnamefont {Chow}}, \bibinfo {author} {\bibfnamefont {J.~M.}\ \bibnamefont {Gambetta}}, \bibinfo {author} {\bibfnamefont {M.~B.}\ \bibnamefont {Rothwell}}, \bibinfo {author} {\bibfnamefont {G.~A.}\ \bibnamefont {Keefe}}, \bibinfo {author} {\bibfnamefont {M.}~\bibnamefont {Steffen}},\ and\ \bibinfo {author} {\bibfnamefont {D.~P.}\ \bibnamefont {Pappas}},\ }\bibfield  {title} {\bibinfo {title} {Improved superconducting qubit coherence using titanium nitride},\ }\href {https://doi.org/10.1063/1.4813269} {\bibfield  {journal} {\bibinfo  {journal} {Applied Physics
  Letters}\ }\textbf {\bibinfo {volume} {103}},\ \bibinfo {pages} {012602} (\bibinfo {year} {2013})}\BibitemShut {NoStop}%
\bibitem [{\citenamefont {Vissers}\ \emph {et~al.}(2010)\citenamefont {Vissers}, \citenamefont {Gao}, \citenamefont {Wisbey}, \citenamefont {Hite}, \citenamefont {Tsuei}, \citenamefont {Corcoles}, \citenamefont {Steffen},\ and\ \citenamefont {Pappas}}]{vissers_low_2010}%
  \BibitemOpen
  \bibfield  {author} {\bibinfo {author} {\bibfnamefont {M.~R.}\ \bibnamefont {Vissers}}, \bibinfo {author} {\bibfnamefont {J.}~\bibnamefont {Gao}}, \bibinfo {author} {\bibfnamefont {D.~S.}\ \bibnamefont {Wisbey}}, \bibinfo {author} {\bibfnamefont {D.~A.}\ \bibnamefont {Hite}}, \bibinfo {author} {\bibfnamefont {C.~C.}\ \bibnamefont {Tsuei}}, \bibinfo {author} {\bibfnamefont {A.~D.}\ \bibnamefont {Corcoles}}, \bibinfo {author} {\bibfnamefont {M.}~\bibnamefont {Steffen}},\ and\ \bibinfo {author} {\bibfnamefont {D.~P.}\ \bibnamefont {Pappas}},\ }\bibfield  {title} {\bibinfo {title} {Low {Loss} {Superconducting} {Titanium} {Nitride} {Coplanar} {Waveguide} {Resonators}},\ }\href {https://doi.org/10.1063/1.3517252} {\bibfield  {journal} {\bibinfo  {journal} {Applied Physics Letters}\ }\textbf {\bibinfo {volume} {97}},\ \bibinfo {pages} {232509} (\bibinfo {year} {2010})}\BibitemShut {NoStop}%
\bibitem [{\citenamefont {Place}\ \emph {et~al.}(2021)\citenamefont {Place}, \citenamefont {Rodgers}, \citenamefont {Mundada}, \citenamefont {Smitham}, \citenamefont {Fitzpatrick}, \citenamefont {Leng}, \citenamefont {Premkumar}, \citenamefont {Bryon}, \citenamefont {Vrajitoarea}, \citenamefont {Sussman}, \citenamefont {Cheng}, \citenamefont {Madhavan}, \citenamefont {Babla}, \citenamefont {Le}, \citenamefont {Gang}, \citenamefont {J{\"a}ck}, \citenamefont {Gyenis}, \citenamefont {Yao}, \citenamefont {Cava}, \citenamefont {de~Leon},\ and\ \citenamefont {Houck}}]{place_new_2021}%
  \BibitemOpen
  \bibfield  {author} {\bibinfo {author} {\bibfnamefont {A.~P.~M.}\ \bibnamefont {Place}}, \bibinfo {author} {\bibfnamefont {L.~V.~H.}\ \bibnamefont {Rodgers}}, \bibinfo {author} {\bibfnamefont {P.}~\bibnamefont {Mundada}}, \bibinfo {author} {\bibfnamefont {B.~M.}\ \bibnamefont {Smitham}}, \bibinfo {author} {\bibfnamefont {M.}~\bibnamefont {Fitzpatrick}}, \bibinfo {author} {\bibfnamefont {Z.}~\bibnamefont {Leng}}, \bibinfo {author} {\bibfnamefont {A.}~\bibnamefont {Premkumar}}, \bibinfo {author} {\bibfnamefont {J.}~\bibnamefont {Bryon}}, \bibinfo {author} {\bibfnamefont {A.}~\bibnamefont {Vrajitoarea}}, \bibinfo {author} {\bibfnamefont {S.}~\bibnamefont {Sussman}}, \bibinfo {author} {\bibfnamefont {G.}~\bibnamefont {Cheng}}, \bibinfo {author} {\bibfnamefont {T.}~\bibnamefont {Madhavan}}, \bibinfo {author} {\bibfnamefont {H.~K.}\ \bibnamefont {Babla}}, \bibinfo {author} {\bibfnamefont {X.~H.}\ \bibnamefont {Le}}, \bibinfo {author} {\bibfnamefont {Y.}~\bibnamefont {Gang}}, \bibinfo {author} {\bibfnamefont
  {B.}~\bibnamefont {J{\"a}ck}}, \bibinfo {author} {\bibfnamefont {A.}~\bibnamefont {Gyenis}}, \bibinfo {author} {\bibfnamefont {N.}~\bibnamefont {Yao}}, \bibinfo {author} {\bibfnamefont {R.~J.}\ \bibnamefont {Cava}}, \bibinfo {author} {\bibfnamefont {N.~P.}\ \bibnamefont {de~Leon}},\ and\ \bibinfo {author} {\bibfnamefont {A.~A.}\ \bibnamefont {Houck}},\ }\bibfield  {title} {\bibinfo {title} {New material platform for superconducting transmon qubits with coherence times exceeding 0.3 milliseconds},\ }\href {https://doi.org/10.1038/s41467-021-22030-5} {\bibfield  {journal} {\bibinfo  {journal} {Nature Communications}\ }\textbf {\bibinfo {volume} {12}},\ \bibinfo {pages} {1779} (\bibinfo {year} {2021})}\BibitemShut {NoStop}%
\bibitem [{\citenamefont {Chitta}\ \emph {et~al.}(2022)\citenamefont {Chitta}, \citenamefont {Zhao}, \citenamefont {Huang}, \citenamefont {Mondragon-Shem},\ and\ \citenamefont {Koch}}]{chitta_computer-aided_2022}%
  \BibitemOpen
  \bibfield  {author} {\bibinfo {author} {\bibfnamefont {S.~P.}\ \bibnamefont {Chitta}}, \bibinfo {author} {\bibfnamefont {T.}~\bibnamefont {Zhao}}, \bibinfo {author} {\bibfnamefont {Z.}~\bibnamefont {Huang}}, \bibinfo {author} {\bibfnamefont {I.}~\bibnamefont {Mondragon-Shem}},\ and\ \bibinfo {author} {\bibfnamefont {J.}~\bibnamefont {Koch}},\ }\href {http://arxiv.org/abs/2206.08320} {\bibinfo {title} {Computer-aided quantization and numerical analysis of superconducting circuits}} (\bibinfo {year} {2022})\BibitemShut {NoStop}%
\bibitem [{\citenamefont {Groszkowski}\ and\ \citenamefont {Koch}(2021)}]{groszkowski_scqubits_2021}%
  \BibitemOpen
  \bibfield  {author} {\bibinfo {author} {\bibfnamefont {P.}~\bibnamefont {Groszkowski}}\ and\ \bibinfo {author} {\bibfnamefont {J.}~\bibnamefont {Koch}},\ }\bibfield  {title} {\bibinfo {title} {Scqubits: a {Python} package for superconducting qubits},\ }\href {https://doi.org/10.22331/q-2021-11-17-583} {\bibfield  {journal} {\bibinfo  {journal} {Quantum}\ }\textbf {\bibinfo {volume} {5}},\ \bibinfo {pages} {583} (\bibinfo {year} {2021})}\BibitemShut {NoStop}%
\bibitem [{\citenamefont {Pedersen}\ \emph {et~al.}(2007)\citenamefont {Pedersen}, \citenamefont {Molmer},\ and\ \citenamefont {Moller}}]{pedersen_fidelity_2007}%
  \BibitemOpen
  \bibfield  {author} {\bibinfo {author} {\bibfnamefont {L.~H.}\ \bibnamefont {Pedersen}}, \bibinfo {author} {\bibfnamefont {K.}~\bibnamefont {Molmer}},\ and\ \bibinfo {author} {\bibfnamefont {N.~M.}\ \bibnamefont {Moller}},\ }\bibfield  {title} {\bibinfo {title} {Fidelity of quantum operations},\ }\href {https://doi.org/10.1016/j.physleta.2007.02.069} {\bibfield  {journal} {\bibinfo  {journal} {Physics Letters A}\ }\textbf {\bibinfo {volume} {367}},\ \bibinfo {pages} {47} (\bibinfo {year} {2007})}\BibitemShut {NoStop}%
\bibitem [{\citenamefont {Ribeiro}\ \emph {et~al.}(2017)\citenamefont {Ribeiro}, \citenamefont {Baksic},\ and\ \citenamefont {Clerk}}]{ribeiro_systematic_2017}%
  \BibitemOpen
  \bibfield  {author} {\bibinfo {author} {\bibfnamefont {H.}~\bibnamefont {Ribeiro}}, \bibinfo {author} {\bibfnamefont {A.}~\bibnamefont {Baksic}},\ and\ \bibinfo {author} {\bibfnamefont {A.~A.}\ \bibnamefont {Clerk}},\ }\bibfield  {title} {\bibinfo {title} {Systematic {Magnus}-based approach for suppressing leakage and non-adiabatic errors in quantum dynamics},\ }\href {https://doi.org/10.1103/PhysRevX.7.011021} {\bibfield  {journal} {\bibinfo  {journal} {Physical Review X}\ }\textbf {\bibinfo {volume} {7}},\ \bibinfo {pages} {011021} (\bibinfo {year} {2017})}\BibitemShut {NoStop}%
\bibitem [{\citenamefont {Setiawan}\ \emph {et~al.}(2021)\citenamefont {Setiawan}, \citenamefont {Groszkowski}, \citenamefont {Ribeiro},\ and\ \citenamefont {Clerk}}]{setiawan_analytic_2021}%
  \BibitemOpen
  \bibfield  {author} {\bibinfo {author} {\bibfnamefont {F.}~\bibnamefont {Setiawan}}, \bibinfo {author} {\bibfnamefont {P.}~\bibnamefont {Groszkowski}}, \bibinfo {author} {\bibfnamefont {H.}~\bibnamefont {Ribeiro}},\ and\ \bibinfo {author} {\bibfnamefont {A.~A.}\ \bibnamefont {Clerk}},\ }\bibfield  {title} {\bibinfo {title} {Analytic {Design} of {Accelerated} {Adiabatic} {Gates} in {Realistic} {Qubits}: {General} {Theory} and {Applications} to {Superconducting} {Circuits}},\ }\href {https://doi.org/10.1103/PRXQuantum.2.030306} {\bibfield  {journal} {\bibinfo  {journal} {PRX Quantum}\ }\textbf {\bibinfo {volume} {2}},\ \bibinfo {pages} {030306} (\bibinfo {year} {2021})}\BibitemShut {NoStop}%
\bibitem [{\citenamefont {Chen}\ \emph {et~al.}(2016)\citenamefont {Chen}, \citenamefont {Kelly}, \citenamefont {Quintana}, \citenamefont {Barends}, \citenamefont {Campbell}, \citenamefont {Chen}, \citenamefont {Chiaro}, \citenamefont {Dunsworth}, \citenamefont {Fowler}, \citenamefont {Lucero}, \citenamefont {Jeffrey}, \citenamefont {Megrant}, \citenamefont {Mutus}, \citenamefont {Neeley}, \citenamefont {Neill}, \citenamefont {O{\textquoteright}Malley}, \citenamefont {Roushan}, \citenamefont {Sank}, \citenamefont {Vainsencher}, \citenamefont {Wenner}, \citenamefont {White}, \citenamefont {Korotkov},\ and\ \citenamefont {Martinis}}]{chen_measuring_2016}%
  \BibitemOpen
  \bibfield  {author} {\bibinfo {author} {\bibfnamefont {Z.}~\bibnamefont {Chen}}, \bibinfo {author} {\bibfnamefont {J.}~\bibnamefont {Kelly}}, \bibinfo {author} {\bibfnamefont {C.}~\bibnamefont {Quintana}}, \bibinfo {author} {\bibfnamefont {R.}~\bibnamefont {Barends}}, \bibinfo {author} {\bibfnamefont {B.}~\bibnamefont {Campbell}}, \bibinfo {author} {\bibfnamefont {Y.}~\bibnamefont {Chen}}, \bibinfo {author} {\bibfnamefont {B.}~\bibnamefont {Chiaro}}, \bibinfo {author} {\bibfnamefont {A.}~\bibnamefont {Dunsworth}}, \bibinfo {author} {\bibfnamefont {A.~G.}\ \bibnamefont {Fowler}}, \bibinfo {author} {\bibfnamefont {E.}~\bibnamefont {Lucero}}, \bibinfo {author} {\bibfnamefont {E.}~\bibnamefont {Jeffrey}}, \bibinfo {author} {\bibfnamefont {A.}~\bibnamefont {Megrant}}, \bibinfo {author} {\bibfnamefont {J.}~\bibnamefont {Mutus}}, \bibinfo {author} {\bibfnamefont {M.}~\bibnamefont {Neeley}}, \bibinfo {author} {\bibfnamefont {C.}~\bibnamefont {Neill}}, \bibinfo {author} {\bibfnamefont {P.~J.~J.}\ \bibnamefont
  {O{\textquoteright}Malley}}, \bibinfo {author} {\bibfnamefont {P.}~\bibnamefont {Roushan}}, \bibinfo {author} {\bibfnamefont {D.}~\bibnamefont {Sank}}, \bibinfo {author} {\bibfnamefont {A.}~\bibnamefont {Vainsencher}}, \bibinfo {author} {\bibfnamefont {J.}~\bibnamefont {Wenner}}, \bibinfo {author} {\bibfnamefont {T.~C.}\ \bibnamefont {White}}, \bibinfo {author} {\bibfnamefont {A.~N.}\ \bibnamefont {Korotkov}},\ and\ \bibinfo {author} {\bibfnamefont {J.~M.}\ \bibnamefont {Martinis}},\ }\bibfield  {title} {\bibinfo {title} {Measuring and {Suppressing} {Quantum} {State} {Leakage} in a {Superconducting} {Qubit}},\ }\href {https://doi.org/10.1103/PhysRevLett.116.020501} {\bibfield  {journal} {\bibinfo  {journal} {Physical Review Letters}\ }\textbf {\bibinfo {volume} {116}},\ \bibinfo {pages} {020501} (\bibinfo {year} {2016})}\BibitemShut {NoStop}%
\bibitem [{\citenamefont {Gambetta}\ \emph {et~al.}(2011)\citenamefont {Gambetta}, \citenamefont {Motzoi}, \citenamefont {Merkel},\ and\ \citenamefont {Wilhelm}}]{gambetta_analytic_2011}%
  \BibitemOpen
  \bibfield  {author} {\bibinfo {author} {\bibfnamefont {J.~M.}\ \bibnamefont {Gambetta}}, \bibinfo {author} {\bibfnamefont {F.}~\bibnamefont {Motzoi}}, \bibinfo {author} {\bibfnamefont {S.~T.}\ \bibnamefont {Merkel}},\ and\ \bibinfo {author} {\bibfnamefont {F.~K.}\ \bibnamefont {Wilhelm}},\ }\bibfield  {title} {\bibinfo {title} {Analytic control methods for high fidelity unitary operations in a weakly nonlinear oscillator},\ }\href {https://doi.org/10.1103/PhysRevA.83.012308} {\bibfield  {journal} {\bibinfo  {journal} {Physical Review A}\ }\textbf {\bibinfo {volume} {83}},\ \bibinfo {pages} {012308} (\bibinfo {year} {2011})}\BibitemShut {NoStop}%
\bibitem [{\citenamefont {Theis}\ \emph {et~al.}(2018)\citenamefont {Theis}, \citenamefont {Motzoi}, \citenamefont {Machnes},\ and\ \citenamefont {Wilhelm}}]{theis_counteracting_2018}%
  \BibitemOpen
  \bibfield  {author} {\bibinfo {author} {\bibfnamefont {L.~S.}\ \bibnamefont {Theis}}, \bibinfo {author} {\bibfnamefont {F.}~\bibnamefont {Motzoi}}, \bibinfo {author} {\bibfnamefont {S.}~\bibnamefont {Machnes}},\ and\ \bibinfo {author} {\bibfnamefont {F.~K.}\ \bibnamefont {Wilhelm}},\ }\bibfield  {title} {\bibinfo {title} {Counteracting systems of diabaticities using {DRAG} controls: {The} status after 10 years},\ }\href {https://doi.org/10.1209/0295-5075/123/60001} {\bibfield  {journal} {\bibinfo  {journal} {EPL (Europhysics Letters)}\ }\textbf {\bibinfo {volume} {123}},\ \bibinfo {pages} {60001} (\bibinfo {year} {2018})}\BibitemShut {NoStop}%
\bibitem [{\citenamefont {Motzoi}\ and\ \citenamefont {Wilhelm}(2013)}]{motzoi_improving_2013}%
  \BibitemOpen
  \bibfield  {author} {\bibinfo {author} {\bibfnamefont {F.}~\bibnamefont {Motzoi}}\ and\ \bibinfo {author} {\bibfnamefont {F.~K.}\ \bibnamefont {Wilhelm}},\ }\bibfield  {title} {\bibinfo {title} {Improving frequency selection of driven pulses using derivative-based transition suppression},\ }\href {https://doi.org/10.1103/PhysRevA.88.062318} {\bibfield  {journal} {\bibinfo  {journal} {Physical Review A}\ }\textbf {\bibinfo {volume} {88}},\ \bibinfo {pages} {062318} (\bibinfo {year} {2013})}\BibitemShut {NoStop}%
\bibitem [{\citenamefont {Motzoi}\ \emph {et~al.}(2009)\citenamefont {Motzoi}, \citenamefont {Gambetta}, \citenamefont {Rebentrost},\ and\ \citenamefont {Wilhelm}}]{motzoi_simple_2009}%
  \BibitemOpen
  \bibfield  {author} {\bibinfo {author} {\bibfnamefont {F.}~\bibnamefont {Motzoi}}, \bibinfo {author} {\bibfnamefont {J.~M.}\ \bibnamefont {Gambetta}}, \bibinfo {author} {\bibfnamefont {P.}~\bibnamefont {Rebentrost}},\ and\ \bibinfo {author} {\bibfnamefont {F.~K.}\ \bibnamefont {Wilhelm}},\ }\bibfield  {title} {\bibinfo {title} {Simple pulses for elimination of leakage in weakly nonlinear qubits},\ }\href {https://doi.org/10.1103/PhysRevLett.103.110501} {\bibfield  {journal} {\bibinfo  {journal} {Physical Review Letters}\ }\textbf {\bibinfo {volume} {103}},\ \bibinfo {pages} {110501} (\bibinfo {year} {2009})}\BibitemShut {NoStop}%
\bibitem [{\citenamefont {Devoret}(1997)}]{devoret_quantum_1997}%
  \BibitemOpen
  \bibfield  {author} {\bibinfo {author} {\bibfnamefont {M.}~\bibnamefont {Devoret}},\ }\href@noop {} {\emph {\bibinfo {title} {Quantum fluctuations in electrical circuits}}}\ (\bibinfo  {publisher} {Edition de Physique},\ \bibinfo {address} {France},\ \bibinfo {year} {1997})\ \bibinfo {note} {iNIS Reference Number: 29063476}\BibitemShut {NoStop}%
\bibitem [{\citenamefont {Ciani}\ \emph {et~al.}(2024)\citenamefont {Ciani}, \citenamefont {DiVincenzo},\ and\ \citenamefont {Terhal}}]{ciani_lecture_2024}%
  \BibitemOpen
  \bibfield  {author} {\bibinfo {author} {\bibfnamefont {A.}~\bibnamefont {Ciani}}, \bibinfo {author} {\bibfnamefont {D.~P.}\ \bibnamefont {DiVincenzo}},\ and\ \bibinfo {author} {\bibfnamefont {B.~M.}\ \bibnamefont {Terhal}},\ }\href {https://doi.org/10.59490/tb.85} {\emph {\bibinfo {title} {Lecture {Notes} on {Quantum} {Electrical} {Circuits}}}}\ (\bibinfo  {publisher} {TU Delft OPEN Textbooks},\ \bibinfo {year} {2024})\ \bibinfo {note} {publication Title: TU Delft OPEN Textbooks}\BibitemShut {NoStop}%
\bibitem [{\citenamefont {Rasmussen}\ \emph {et~al.}(2021)\citenamefont {Rasmussen}, \citenamefont {Christensen}, \citenamefont {Pedersen}, \citenamefont {Kristensen}, \citenamefont {B{\ae}kkegaard}, \citenamefont {Loft},\ and\ \citenamefont {Zinner}}]{rasmussen_superconducting_2021}%
  \BibitemOpen
  \bibfield  {author} {\bibinfo {author} {\bibfnamefont {S.}~\bibnamefont {Rasmussen}}, \bibinfo {author} {\bibfnamefont {K.}~\bibnamefont {Christensen}}, \bibinfo {author} {\bibfnamefont {S.}~\bibnamefont {Pedersen}}, \bibinfo {author} {\bibfnamefont {L.}~\bibnamefont {Kristensen}}, \bibinfo {author} {\bibfnamefont {T.}~\bibnamefont {B{\ae}kkegaard}}, \bibinfo {author} {\bibfnamefont {N.}~\bibnamefont {Loft}},\ and\ \bibinfo {author} {\bibfnamefont {N.}~\bibnamefont {Zinner}},\ }\bibfield  {title} {\bibinfo {title} {Superconducting {Circuit} {Companion}---an {Introduction} with {Worked} {Examples}},\ }\href {https://doi.org/10.1103/PRXQuantum.2.040204} {\bibfield  {journal} {\bibinfo  {journal} {PRX Quantum}\ }\textbf {\bibinfo {volume} {2}},\ \bibinfo {pages} {040204} (\bibinfo {year} {2021})},\ \bibinfo {note} {publisher: American Physical Society}\BibitemShut {NoStop}%
\bibitem [{\citenamefont {Gao}\ \emph {et~al.}(2007)\citenamefont {Gao}, \citenamefont {Mazin}, \citenamefont {Day}, \citenamefont {Zmuidzinas},\ and\ \citenamefont {LeDuc}}]{gao_noise_2007}%
  \BibitemOpen
  \bibfield  {author} {\bibinfo {author} {\bibfnamefont {J.}~\bibnamefont {Gao}}, \bibinfo {author} {\bibfnamefont {B.~A.}\ \bibnamefont {Mazin}}, \bibinfo {author} {\bibfnamefont {P.~K.}\ \bibnamefont {Day}}, \bibinfo {author} {\bibfnamefont {J.}~\bibnamefont {Zmuidzinas}},\ and\ \bibinfo {author} {\bibfnamefont {H.~G.}\ \bibnamefont {LeDuc}},\ }\bibfield  {title} {\bibinfo {title} {Noise {Properties} of {Superconducting} {Coplanar} {Waveguide} {Microwave} {Resonators}},\ }\href {https://doi.org/10.1063/1.2711770} {\bibfield  {journal} {\bibinfo  {journal} {Applied Physics Letters}\ }\textbf {\bibinfo {volume} {90}},\ \bibinfo {pages} {102507} (\bibinfo {year} {2007})}\BibitemShut {NoStop}%
\bibitem [{\citenamefont {Zmuidzinas}(2012)}]{zmuidzinas_superconducting_2012}%
  \BibitemOpen
  \bibfield  {author} {\bibinfo {author} {\bibfnamefont {J.}~\bibnamefont {Zmuidzinas}},\ }\bibfield  {title} {\bibinfo {title} {Superconducting {Microresonators}: {Physics} and {Applications}},\ }\href {https://doi.org/10.1146/annurev-conmatphys-020911-125022} {\bibfield  {journal} {\bibinfo  {journal} {Annual Review of Condensed Matter Physics}\ }\textbf {\bibinfo {volume} {3}},\ \bibinfo {pages} {169} (\bibinfo {year} {2012})}\BibitemShut {NoStop}%
\bibitem [{\citenamefont {Earnest}\ \emph {et~al.}(2018)\citenamefont {Earnest}, \citenamefont {Chakram}, \citenamefont {Lu}, \citenamefont {Irons}, \citenamefont {Naik}, \citenamefont {Leung}, \citenamefont {Ocola}, \citenamefont {Czaplewski}, \citenamefont {Baker}, \citenamefont {Lawrence}, \citenamefont {Koch},\ and\ \citenamefont {Schuster}}]{earnest_realization_2018}%
  \BibitemOpen
  \bibfield  {author} {\bibinfo {author} {\bibfnamefont {N.}~\bibnamefont {Earnest}}, \bibinfo {author} {\bibfnamefont {S.}~\bibnamefont {Chakram}}, \bibinfo {author} {\bibfnamefont {Y.}~\bibnamefont {Lu}}, \bibinfo {author} {\bibfnamefont {N.}~\bibnamefont {Irons}}, \bibinfo {author} {\bibfnamefont {R.~K.}\ \bibnamefont {Naik}}, \bibinfo {author} {\bibfnamefont {N.}~\bibnamefont {Leung}}, \bibinfo {author} {\bibfnamefont {L.}~\bibnamefont {Ocola}}, \bibinfo {author} {\bibfnamefont {D.~A.}\ \bibnamefont {Czaplewski}}, \bibinfo {author} {\bibfnamefont {B.}~\bibnamefont {Baker}}, \bibinfo {author} {\bibfnamefont {J.}~\bibnamefont {Lawrence}}, \bibinfo {author} {\bibfnamefont {J.}~\bibnamefont {Koch}},\ and\ \bibinfo {author} {\bibfnamefont {D.~I.}\ \bibnamefont {Schuster}},\ }\bibfield  {title} {\bibinfo {title} {Realization of a \{\vphantom{\}}{\textbackslash}{textbackslashLambda}{\textbackslash} system with metastable states of a capacitively-shunted fluxonium},\ }\href
  {https://doi.org/10.1103/PhysRevLett.120.150504} {\bibfield  {journal} {\bibinfo  {journal} {Physical Review Letters}\ }\textbf {\bibinfo {volume} {120}},\ \bibinfo {pages} {150504} (\bibinfo {year} {2018})}\BibitemShut {NoStop}%
\end{thebibliography}
Here we introduce the fluxonium-resonator-fluxonium circuit, derive the capacitance matrices for different grounding configurations, and analyze the resulting Hamiltonian.
\subsection{Capacitance matrix: grounded fluxoniums, grounded resonator}
\begin{figure}[ht]
\includegraphics[width=0.4\textwidth]{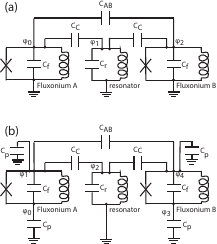}
\caption{Grounding configurations and mode labeling. (a) Grounded fluxoniums and resonator, (b) differential fluxoniums and resonator. The differential configuration is preferable for reduced loading of the coupling capacitors, as discussed in \cite{ding_high-fidelity_2023}. Mode indices $\varphi_i$ refer to the mode ordering of the indices in the capacitance matrices in this work.}\label{fig:grounding_configs}
\end{figure}

The simplest grounding configuration includes grounded fluxoniums and a grounded resonator, with a dedicated capacitance for each pairwise coupling (which may be from desired \textit{or} parasitic cross-capacitances). For a circuit diagram and mode labeling, which follows \cite{ding_high-fidelity_2023}, see Fig. \ref{fig:grounding_configs}(a). In this case, the capacitance matrix is 

\begin{equation}\label{cap_matrix_simple}
C = \begin{pmatrix}
C_{f} + C_{c} + C_{AB} & - C_c & -C_{AB} \\
-C_c & C_r + 2C_c & -C_c \\
-C_{AB} & -C_c & C_f + C_c + C_{AB}
\end{pmatrix}
\end{equation}
where $C_c$ is the coupling capacitance between each fluxonium and the resonator, assumed to be equal for both fluxoniums here for simplicity. The capacitance $C_f$ is the capacitance across the fluxonium to ground. The capacitance $C_{AB}$ is the coupling capacitance between the two fluxoniums, and the resonator capacitance is $C_r$. 

The grounding capacitance for the fluxoniums $C_f$ includes the contribution from any capacitor pads as well as the small phase slip junction itself. (The contribution from a junction array inductor is generally negligible since it scales as 1/$N$, the number of junctions). Typically, the junction capacitance for 7 GHz is order $\sim 3$ fF, setting a lower bound on $C_f$ given by the critical current density of the junction fabrication process. We assume the differential capacitance across the pads $C_f$ can be tuned to be small, on the order of 2 fF. This sets a maximum of roughly 4 fF for any additional parasitic capacitances to ground, for a total effective capacitance $C_{\text{eff}} = 6.6 \text{ fF} + C_{c} + C_{AB}$. For integration with a higher impedance resonator with circuit parameters designed for integration in a larger circuit, see parameters in Tables \ref{cap_matrix_gnd_inputs} and \ref{cap_matrix_gnd_outputs} below. We find in both impedance cases, very small capacitances between the two fluxonium qubits are required to maintain small $\eta$, which may be difficult to target accurately. This motivates a change in grounding geometry to differential qubits as discussed below.
\begin{table*}[t]
\begin{tabular}{||c c c c c c||} 
 \hline
 Descriptor & $C_f$ (fF) & $Z_{c, in}$ ($\Omega$) & $\omega_c / (2\pi)$ (GHz) & $C_c$ (fF) & $C_{AB}$ (fF) \\ 
 \hline\hline
 main text & 7.27 & 193 & 7.08 & 2.45 & 0.01($< 0.04$) \\
 high impedance & 9.0 & 2132 & 7.08 & 0.76 & 0.01($< 0.04$) \\
 \hline\hline
 \end{tabular}
\caption{\label{cap_matrix_gnd_inputs}Capacitances in 
 Eq.~\eqref{cap_matrix_simple} for capacitive energies and coupling strengths shown in Table \ref{cap_matrix_gnd_outputs}. We assume grounded fluxonium qubits and a grounded resonator. The frequency $\omega_c$ corresponds to the loaded resonator frequency after integration into the circuit. Recall that the capacitance matrix sets the spurious ZZ interaction $\eta$ between the qubits; for more details and expressions see the perturbative analysis below. Here, the bound on $C_{AB}$ refers to a 4 kHz ZZ interaction $\eta$ between the fluxonium qubits, below which $\eta$ is smaller than 4 kHz. The extremely small $C_{AB}$ value required for ZZ cancellation motivates a differential fluxonium qubit as discussed in the next section.}
\vspace{1cm}
\begin{tabular}{||c c c c c c ||} 
 \hline
 Descriptor & $E_C^{A(B)}$ (GHz) & $Z_c$ ($\Omega$) & $J_c$ (GHz) 
 & $J_c n_{c, 01}$ (GHz) & $J_{AB}$ (GHz) \\
 \hline\hline
 main text & 2.0 & 191 & 0.33 & 0.54 & 0.1 \\
  high impedance & 2.0 & 2000 & 1.07 & 0.54 & 0.1 \\
 \hline\hline
 \end{tabular}
 \caption{\label{cap_matrix_gnd_outputs}Capacitive energies and coupling strengths from input capacitances in Table \ref{cap_matrix_gnd_inputs}, i.e., the grounded qubit geometry. For both parameter sets, we recover the same Hamiltonian since they have identical $J_{c}n_{c, 01}$. Recall here $J_{Ac} = J_{Bc} = J_c$.}
\end{table*}

\subsection{Capacitance matrix: differential fluxoniums, grounded resonator}
Here, we assume differential fluxoniums and a grounded resonator, as in Fig. \ref{fig:grounding_configs}(b). There are several motivating reasons for using a differential qubit geometry: (i) assuming an inductive shunt fabricated from an array of Josephson Junctions (JJA) for the fluxoniums, a differential geometry exhibits $\sim$2x higher collective mode frequencies in the JJA, which should push them to the $10+$ GHz regime (outside the scale of interest for the gate). (ii) As referenced in \cite{ding_high-fidelity_2023}, Appendix B, differential fluxoniums enable smaller capacitive loading per unit coupling capacitance, and (iii) using a grounded coupler enables reduced sensitivity of the ZZ cancellation condition to imperfections in the coupling capacitance targeting. We start by assuming the fluxonium-resonator-fluxonium circuit shown in Fig. \ref{fig:grounding_configs}(b), with the fluxonium modes differential and the resonator grounded. The capacitance matrix is
\begin{equation}\label{cap_matrix_differential}
C = \begin{pmatrix}
C_f + C_{p} & -C_{f} & 0 & 0 & 0 \\
-C_{f} & C_{tot} & -C_c & 0 & -C_{AB} \\
0 & -C_c & C_r + 2 C_c & 0 & -C_c \\
0 & 0 & 0 & C_{p} + C_f & -C_{f} \\
0 & -C_{AB} & -C_c & -C_{f} & C_{tot} \\
\end{pmatrix},
\end{equation}
where $C_f$ is the differential capacitance across each fluxonium (taken here to be equal for both fluxoniums, for simplicity), $C_{p}$ is each fluxoniums' parasitic coupling to ground, one on each side of the qubit, $C_r$ is the capacitance of the resonator (to ground, by definition), and $C_{tot} = C_f + C_{p} + C_c + C_{AB}$. Here, $C_c$ is the coupling capacitance between each fluxonium and the resonator, again taken to be equal for both fluxoniums here for simplicity. 

In the index ordering for the capacitance matrix above, the fluxonium A corresponds to the first and second index, fluxonium B to the last two indices, and the resonator to the third one. Since we use a differential mode for the qubits, we follow \cite{ding_high-fidelity_2023} by defining the qubit modes as sums and differences of the original node phases. The authors define the new capacitance matrix as $\tilde{C}\equiv (M^T)^{-1} C M^{-1}$ where $M$ defines the sums and differences of each fluxonium mode: 
\begin{equation}\label{M_matrix}
M = \begin{pmatrix}
1 & 1 & 0 & 0 & 0 \\
1 & -1 & 0 & 0 & 0 \\
0 & 0 & 1 & 0 & 0 \\
0 & 0 & 0 & 1 & 1 \\
0 & 0 & 0 & 1 & -1 \\
\end{pmatrix}.
\end{equation}
This results in the new capacitance matrix of 
\begin{widetext}
\begin{multline}
\tilde{C} = \left(
\begin{matrix}
C_{AB} / 4 + C_c /4 + C_p / 2 & - C_{AB} / 4 - C_c /4 & -C_c /2 &  - C_{AB} / 4  & C_{AB} / 4 \\
- C_{AB} / 4 - C_c /4  & C_{tot} & C_c / 2 & C_{AB} / 4  & - C_{AB} / 4\\
-C_c /2 & C_c /2 & 2 C_c + C_r &  -C_c /2 & C_c /2\\
- C_{AB} / 4  & C_{AB} / 4 &  -C_c / 2 & C_{AB} / 4 + C_c /4 + C_p / 2  & - C_{AB} / 4 - C_c /4  \\
 C_{AB} / 4  & -C_{AB} / 4 &  C_c / 2 &- C_{AB} / 4 - C_c /4 & C_{tot}  \\
\end{matrix}\right).
\end{multline}
\end{widetext}
where $C_{tot} =  C_{AB} / 4 + C_c /4 + C_p / 2 + C_f$.
\begin{table*}[t]
\begin{tabular}{||c c c c c c c ||} 
 \hline
 Descriptor & $C_f$ (fF) & $C_{p}$ (fF) & $Z_{c, in}$ ($\Omega$) & $\omega_c / (2\pi)$ (GHz) & $C_c$ (fF) & $C_{AB}$ (fF) \\ 
 \hline\hline
 main text & 4.7 & 7.4 & 193 & 7.08 & 7.9 & 0.06($<$0.3) \\
 high impedance & 8.23 & 2.42 & 4462 & 7.08 & 1.46 & 0.09($ < 0.25$) \\
 \hline\hline
 \end{tabular}
\caption{\label{cap_matrix_diff_inputs}Capacitances in 
Eq.~\eqref{cap_matrix_differential} for capacitive energies and coupling strengths shown in Table \ref{cap_matrix_diff_outputs}. We assume differential fluxonium qubits and a grounded resonator. The bound on $C_{AB}$ corresponds to 4 kHz ZZ coupling $\eta$ between the two fluxonium qubits. The frequency $\omega_c$ corresponds to the loaded resonator frequency after integration into the circuit. In the low impedance case, we find a large coupling capacitance $C_c > C_f$ is needed, such that in a surface code context, this parameter set can be used only in conjunction with high impedance resonators to maintain qubit $E_C \sim 2$ GHz. For the high impedance case, the coupling capacitance $C_c$ is well below the self-capacitance $C_f$ of the fluxoniums, implying an extension to degree-4 connected circuits using similar parameters. However, small parasitic capacitances to ground are required. These observations motivate parameter sets for scaling with lower $E_C$ values, as discussed below. The two rows correspond to identical Hamiltonians, within 2\%.}

\vspace{1cm}

\begin{tabular}{||c c c c c c ||} 
 \hline
 Descriptor & $E_C^{A(B)}$ (GHz) & $Z_c$ ($\Omega$) & $J_c$ (GHz) & $J_{c}n_{c, 01}$ (GHz)  & $J_{AB}$ (GHz) \\
 \hline\hline
 main text & 2.0 & 191 & 0.33 & 0.54 & 0.1 \\
  high impedance & 2.0 & 3600 & 1.43 & 0.54 & 0.1 \\
 \hline\hline
 \end{tabular}\caption{\label{cap_matrix_diff_outputs}Capacitive energies and coupling strengths from input capacitances in Table \ref{cap_matrix_diff_inputs}. For both parameter sets, we recover the same Hamiltonian since they have identical $J_{c}n_{c, 01}$. Recall here $J_{Ac} = J_{Bc} = J_c$.}
\end{table*}
The qubit modes now correspond to the second and last indices, while the first and fourth modes correspond to the sums of those differences. 

We estimate the capacitance values needed for implementation of the Hamiltonian in the main text. These are shown in Tables \ref{cap_matrix_diff_inputs} and \ref{cap_matrix_diff_outputs}. We find that while the tolerance on parasitic cross-capacitance $C_{AB}$ is larger than the grounded case, the self-capacitance $C_f$ is somewhat small, only a few times larger than the capacitance of the junction itself, leading to stringent requirements on the lead geometry. This motivates studying our gate with lower $E_C$ values of the fluxoniums as discussed below.

\section{Alternate circuit parameters for reduced capacitive loading}\label{alt_params}
For facilitating integration with larger parasitic capacitances to ground, as well as non-trivial capacitances from the junctions themselves, we study capacitance matrices here with lower $E_C = 1$ GHz for the fluxoniums. A larger impedance enables relaxed capacitive loading even with similar coupling compared the main text. For our exemplary circuit parameters, see Tables \ref{cap_matrix_diff_inputs_low_EC_2} and \ref{cap_matrix_diff_outputs_low_EC_2}. Descriptive remarks about the circuit include:
\begin{itemize}
\item For inductive elements, we use $E_L^{A(B)} = E_C^{A(B)} = 1$ GHz, and $E_J^{A(B)} = 7.0(6.9)$ GHz.
\item Here, the coupling capacitors only contribute 6\% to each of the fluxoniums' capacitance budgets, implying feasible capacitive budgets for connecting to a degree-4 coordination number, as required for a surface code.
\item We additionally focus on a differential fluxonium mode geometry to suppress stray ZZ interactions from parasitic cross-capacitance between the two fluxoniums. For example, for a capacitance of 0.15fF, the ZZ coupling strength is $\eta = -3.6$ kHz, with reduced ZZ coupling at smaller cross-capacitances. 
\item The hybridization between the fluxonium plasmon transitions and the coupler is about the same as the main text, at $h_A = 14\%$ and $h_B = 18\%$. 
\item In practice, the maximum impedance we expect to achieve is limited by the increased drive strength required for a fast gate (here a Rabi frequency of $\Omega \sim 70$ MHz for a 70ns gate), capacitive loading on the \textit{coupler}, and increased sensitivity of the ZZ interaction $\eta$ to parasitic cross-capacitances between the two fluxoniums.
\end{itemize}

\begin{figure}[ht]
\centering
\includegraphics[width=0.4\textwidth]{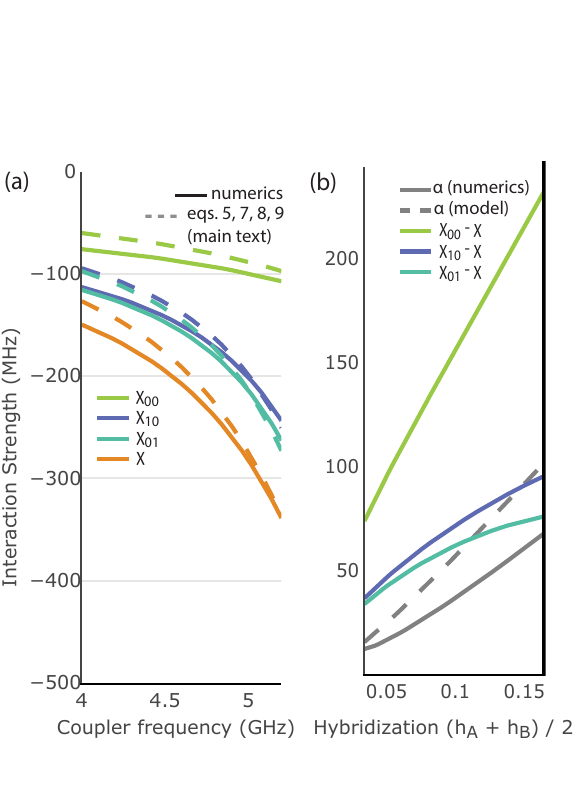}
\caption{Key terms in the Hamiltonian for the circuit parameter set described in Tables \ref{cap_matrix_diff_inputs_low_EC_2} and \ref{cap_matrix_diff_outputs_low_EC_2}. We observe similar $\chi_{ij}$ and $\alpha$ values as a function of plasmon-coupler hybridization, since we maintain similar effective coupling strength $J_{A(B), c} n_{A(B), 12} n_{c, 01}$ as in the main text.}\label{fig:SI_system_4_fig2}
\end{figure}

\begin{figure}[ht]
\centering
\includegraphics[width=0.22\textwidth]{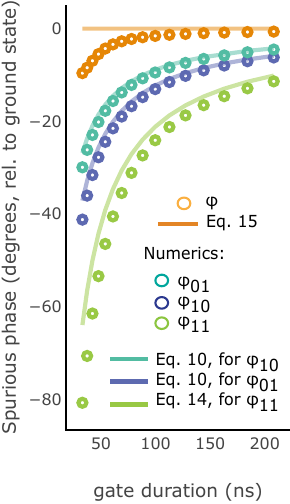}
\caption{Spurious phase and modeling prediction for the computational states, and for the circuit parameter set described in Tables \ref{cap_matrix_diff_inputs_low_EC_2} and \ref{cap_matrix_diff_outputs_low_EC_2}. We observe similar spurious phases as the main text parameters. The modeling of the $\varphi_{ij}$ values (green and blue lines) is accurate to about 2 degrees for all $t_g$ (green and blue circles), leading to substantial disagreement in the relative phase $\varphi$ numerics (orange crosses) and prediction (orange line). We attribute the deviation to detuned driving of the plasmon transitions, e.g. the $\ket{0, 0, 1}-\ket{0, 0, 2}$ transition shifts $\varphi_{01}$. This both renders our simple model for coherent error mechanisms reported in the main text inaccurate, and can be expected when going to higher coupler impedance; a stronger drive is required compared to the main text parameters.}\label{fig:SI_system_4_spurious_phase}
\end{figure}

We repeat the same analysis as in the main text and show results below. 
In Fig. \ref{fig:SI_system_4_fig2}, we show the key interaction terms in the Hamiltonian as done in the main text. While we observe similar hybridization, a larger $J_{A(B), c}$ value, even beyond what we report here, should be feasible, as the coupling capacitances are only 6\% of the fluxoniums' capacitance budget. Consequently, in Fig. \ref{fig:SI_system_4_spurious_phase}, we observe qualitative agreement between the numerically-extracted $\varphi_{ij}$ values and our analytic estimates.

\begin{figure}[ht]
\centering
\includegraphics[width=0.4\textwidth]{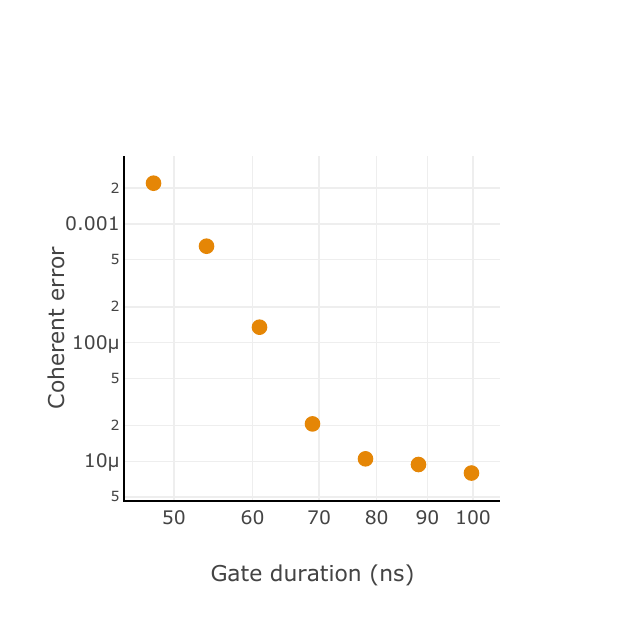}
\caption{Coherent error as a function of gate time for the parameter set described in Tables \ref{cap_matrix_diff_inputs_low_EC_2} and \ref{cap_matrix_diff_outputs_low_EC_2}. We observe increased scaling of coherent error with gate duration at short gate times, which we attribute to increased leakage from detuned driving of the plasmon transitions. Nevertheless, we recover coherent errors within $2 \times 10^{-5}$ at gate durations of $\sim 70$ns, similar performance for the main text parameters. }\label{fig:system_4_fig_3}
\end{figure}

Next, in Fig. \ref{fig:system_4_fig_3}, we illustrate the coherent error as a function of gate time, after drive optimization. We find that similarly low coherent error compared to the main text is possible at a similar gate time of $t_g \approx 70$ ns. However, the scaling is much more efficient at short gate times, which we attribute to increased leakage of the plasmon excitations (e.g. to the $\ket{0, 0, 2}, \ket{2, 0, 0}$ states). 
\begin{table*}[t]
\begin{tabular}{||c c c c c c c ||} 
 \hline
 Descriptor & $C_f$ (fF) & $C_{p}$ (fF) & $Z_{c, in}$ ($\Omega$) & $\omega_c / (2\pi)$ (GHz) & $C_c$ (fF) & $C_{AB}$ (fF) \\ 
 \hline\hline
 high impedance & 15.2 & 6.9 & 3812 & 5.18 & 4.67 & 0.66($<$0.7)  \\
 \hline\hline
 \end{tabular}
\caption{\label{cap_matrix_diff_inputs_low_EC_2}Capacitances in Eq.~\eqref{cap_matrix_differential} for capacitive energies and coupling strengths shown in Table \ref{cap_matrix_diff_outputs_low_EC_2}. We assume differential fluxonium qubits and a grounded resonator. The frequency $\omega_c$ corresponds to the loaded resonator frequency after integration into the circuit. This case should facilitate scaling to a surface code in the presence of large parasitic capacitances to ground. In particular, the contribution of the coupling capacitance, $C_c/4$, to the total effective capacitance $C_f + C_{p}/2 + C_c/4 + C_{AB}/4 \approx 20 \text{ fF}$ is about 6\%, such that scaling to four coupled resonators coupled to each fluxonium should be possible, even with substantial margin for coupling to readout resonators or other circuit elements. The the bound on $C_{AB}$ corresponds to 4 kHz ZZ coupling $\eta$ between the two fluxoniums.}
\vspace{1cm}
\begin{tabular}{||c c c c c c ||} 
 \hline
 Descriptor & $E_C^{A(B)}$ (GHz) & $Z_c$ ($\Omega$) & $J_c$ (GHz) & $J_{c}n_{c, 01}$ (GHz)  & $J_{AB}$ (GHz) \\
 \hline\hline
  high impedance & 1 & 2546
  & 1.15 & 0.58 & 0.138 \\

 \hline\hline
 \end{tabular}
 \caption{\label{cap_matrix_diff_outputs_low_EC_2}Capacitive energies and coupling strengths from input capacitances in Table \ref{cap_matrix_diff_inputs_low_EC_2}. For the inductive components, we assume $E_L^{A(B)} = 1.0(1.0)$ GHz, and $E_J^{A(B)} = 7.0(6.9)$ GHz.}
\end{table*}

\begin{figure}[ht]
\centering
\includegraphics[width=0.5\textwidth]{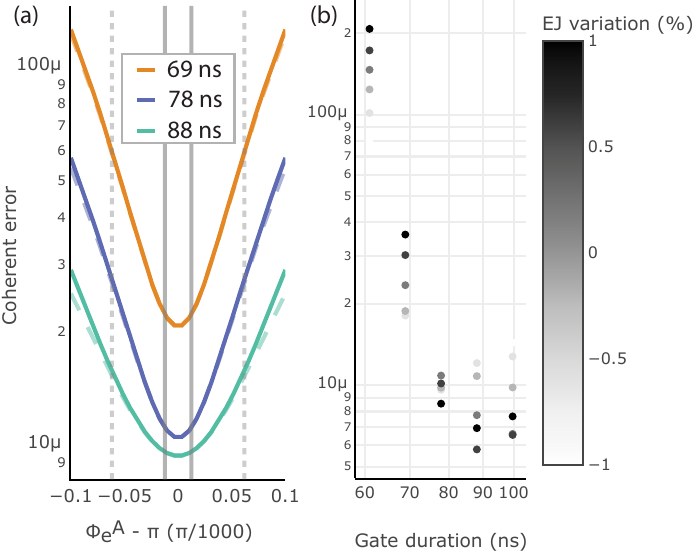}
\caption{Robustness of coherent error to circuit imperfections, for the parameter set described in Tables \ref{cap_matrix_diff_inputs_low_EC_2} and \ref{cap_matrix_diff_outputs_low_EC_2}. (a) The sensitivity to flux variation is increased compared to the main text. This is because of the increased $E_L$ values of the fluxoniums, which are in turn required to maintain large $\chi_{ij}$ values when paired with the lower $E_C = 1 $ GHz (for further details see section \ref{omega_23_section}). Vertical grey lines correspond to the same flux noise values as in the main text. Despite the increased sensitivity, the relative variation in coherent error for state-of-the-art flux noise  is still below $\sim 20$\%. (b) Variation of coherent error with $E_J$ offsets. As also found in the main text, the variation in coherent error with $E_J$ offsets is within a factor of $\sim 2$, which can be efficiently mitigated with a slight increase in gate duration $t_g$ if needed. }\label{fig:system_4_fig_5}
\end{figure}

In Fig. \ref{fig:system_4_fig_5}, we show robustness of the coherent error to variations in circuit parameters. While we observe an increased sensitivity to flux noise, as expected from the higher $E_L^{A(B)}$ values, within state-of-the-art flux noise the variation in coherent error is still of order 20\% for our gate durations of interest $t_g \approx 70$ns. Similarly, we observe similar robustness of coherent error to $E_J^{A(B)}$ variation, which can be efficiently mitigated with a slight increase in gate duration if required. 

\begin{figure}[ht]
\centering
\includegraphics[width=0.45\textwidth]{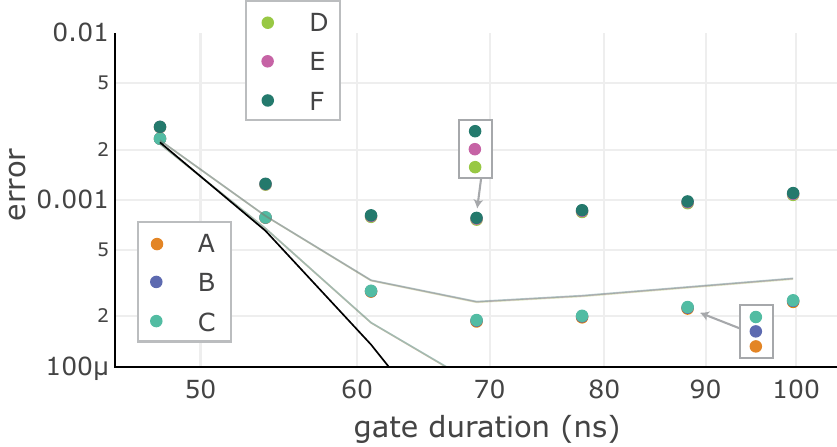}
\caption{Total gate error, for the parameter set described in Tables \ref{cap_matrix_diff_inputs_low_EC_2} and \ref{cap_matrix_diff_outputs_low_EC_2}, and parameter sets A-F for loss rates reported in the main text. We observe near-identical errors as found in the main text, suggesting that our gate scheme can be extended to the ultrahigh impedance regime.}\label{fig:system_4_fig_6}
\end{figure}

Finally, we report the total error including the same loss mechanisms and loss rates as reported in the main text. In Fig. \ref{fig:system_4_fig_6}, we show nearly identical errors as the values in the main text, suggesting that our gate approach may be extended to the ultrahigh impedance regime and therefore enable quantum error correction with high-fidelity gates.

\section{Hamiltonian}\label{appendix_c_hamiltonian}
The system Hamiltonian is $\mathcal{H} \equiv \mathcal{H}_0 + \mathcal{H}_{int}$ where the bare Hamiltonian is
\begin{equation}\label{Hamiltonian_0}
\mathcal{\hat{H}}_0/h = \sum_{i = A, B} 4 E_{C, i} \hat{n}_i^2 + \left( \frac{E_{L, i}}{2} \right) \hat{\phi}_i^2 + E_{J, i} \cos \left(\hat{\phi}_i \right)
    + \frac{\omega_c}{2\pi}\hat{c}^\dagger \hat{c},
\end{equation}
and the interaction Hamiltonian is 
\begin{equation}\label{H_int}
\mathcal{\hat{H}}_{int}/h =
    J_{A, c} \hat{n}_A \hat{n}_c + J_{B, c} \hat{n}_B \hat{n}_c + J_{A, B} \hat{n}_A \hat{n}_B.
\end{equation}
As stated in the main text, the external flux for both fluxoniums is set so that the fluxoniums are at their sweet spot, $\phi_e^{(A)} = \phi_e^{(B)} = \pi$. 
In this section, we study the perturbative terms that are responsible for $\chi_{ij}$ and $\alpha$. 

\subsection{Summary of Hamiltonian analysis}\label{summary}
We summarize the findings from our analysis of the Hamiltonian which is detailed below. Following our notation for the main text, $n_{O, ij} \equiv \bra{i} \hat{n}_O \ket{j}$ is the charge operator for circuit element $O \in {A, B, c}$ on transition $\ket{i} - \ket{j}$. We assume an approximate heirarchy of energy scales and matrix elements given in section \ref{heirarchy}, which also contains our approximations.
\begin{enumerate}
    \item The fluxonium ZZ interaction $\eta$ is given by virtual excitations to non-computational states, particularly the fluxoniums' second and third excited state. The perturbative contributions to $\eta$ at order $i$ are $\eta^{(i)}$, where: 
    
    $2\pi \eta^{(2)} \approx -2\pi J_{A, B}^2 \sum_{i, j \in \{A, B\}, j\neq i} \frac{|n_{i, 12} n_{j, 12}|^2}{\omega_{i, 12} + \omega_{j, 12}} - \frac{|n_{i, 12} n_{j, 03}|^2}{\omega_{i, 12} + \omega_{j, 03}} + \frac{|n_{i, 03} n_{j, 03}|^2}{\omega_{i, 03} + \omega_{j, 03}}\\$
    $2\pi \eta^{(3)} = 4\pi J_{A, c} J_{B, c} |n_{c, 01}|^2 \left( \frac{1}{\omega_{A, 12} + \omega_c} + \frac{1}{\omega_{B, 12} + \omega_c} \right)\\$
    
    \item To second order in $\mathcal{H}_{int}$, first order in $1/\Delta_{A(B)}$ where, $2\pi \Delta_{A(B)} \equiv \omega_{A(B)12} - \omega_c$,  and zeroth order in $n_{A(B), 01}$, the dispersive coupling strengths can be estimated as follows: 
    
    $$\chi^{(2)} = -\frac{J_{A, c}^2 |n_{A, 12}  n_{c, 01}|^2}{\Delta_A} - \frac{J_{B, c}^2| n_{B, 12} n_{c, 01}|^2}{\Delta_B} 
    \\$$
    $$\chi_{0, 1}^{(2)} =  -\frac{2\pi J_{A, c}^2 |n_{A, 03} n_{c,01}|^2}{\omega_{A, 03} -\omega_c} - \frac{J_{B, c}^2 |n_{B, 12} n_{c, 01}|^2 }{\Delta_B} \\$$
    $$\chi_{1, 0}^{(2)} = -\frac{2\pi J_{B, c}^2 |n_{B, 03} n_{c,01}|^2}{\omega_{B, 03}-\omega_c} -\frac{J_{A, c}^2| n_{A, 12} n_{c, 01}|^2 }{\Delta_A} \\$$
    $$\chi_{0, 0}^{(2)} =  -\frac{2\pi J_{A, c}^2 |n_{A, 03}^2 n_{c,   01}|^2}{\omega_{A, 03}-\omega_c} - \frac{2\pi J_{B, c}^2 |n_{B, 03} n_{c, 01}|^2}{\omega_{B, 03}-\omega_c} \\$$
    
    We estimate the third order terms and find they are $\sim 10$ MHz for our heirarchy described in section \ref{heirarchy}. When $2\pi \Delta_{A, B} \gg \omega_{23}$, the plasmon transition terms will dominate the dispersive interaction, yielding $\delta \chi_{ij}  = \chi - \chi_{ij}$ of $\sim 100$ MHz. Also, we choose $\Delta_{A(B)}$ to be small, with $J_{A(B)} n_{c, 01} n_{A(B), 12} / \Delta_{A(B)} \lesssim 1$, and the plasmon transition above the coupler transition ($\Delta_{A, B} > 0$). This allocation will maintain a large detuning between coupler frequency and the fluxonium 0-3 transitions, as required for maximal $|\delta\chi_{ij}|$. 
    \item The second order shifts to computational states are of order 1-10 MHz. The largest contributions are virtual excitations through a plasmon excitation and a coupler excitation ($\ket{2, 1, 1}$ and $\ket{1, 1, 2}$). The next highest term corresponds to three-photon excitation $\propto n_{03}^2 / \omega_{03}$.
    \item Our estimates of $\chi_{ij}$ do not rely on any intrinsic coupler nonlinearity. The large dispersive interaction emerges because of the substantial charge matrix elements of the coupler and the fluxonium 1-2 transition, and by tuning $J_{A(B)} n_{c, 01} n_{A(B), 12} / \Delta_{A(B)} \sim 1$. 
    \item The large $\chi_{ij}$ can be engineered while also choosing a capacitance matrix such that the ZZ interaction in the computational manifold is suppressed. The energy shifts in $\eta^{(n)}$ from $n$-order perturbation theory all have large, negative energy gaps. Therefore, the perturbative contributions to the ZZ coupling in the computational manifold $\propto \left(1/(\omega_{A(B), 12} + \omega_{c})\right)^{n-1}$, where $n \geq 2$. Conversely, $\delta \chi_{ij}$ is proportional to $1/\Delta_{A, B}$, which can be tuned independently such that $\Delta_{A(B)}$ is small even if $\omega_{12} + \omega_c$ is large.
    \item The anharmonicity of the coupler is significant at fourth as well as higher orders. 
    While our analytic estimations and the exact perturbative shifts for the coupler nonlinearity agree within $\sim 30\%$, including up to fourth order, neither agree with the exact numeric result within a factor of 2 to 3. 
\end{enumerate}

\subsection{ZZ interaction strength numerical analysis}\label{summary}

The ZZ interaction, $\eta$, is particularly important for our microwave-activated approach. When the gate is off, the qubits will become entangled at rate $\eta$, resulting in correlated errors. In our work as well as \cite{ding_high-fidelity_2023}, a low $\eta$ value is achieved through a careful tuning of the coupling strengths in the circuit.  For particular values of the couplings $J_{ij}$ and spectra, $\eta$ can be tuned to be small, on the order of $\mathcal{O}(1)$ kHz. We deem this scale to be sufficient for initial quantum error correction applications, because it would very roughly correspond to correlated errors in 50ns gates of $\sim 10^{-4}$.

In Sec. \ref{perturbative_anaysis}, we derive the condition for small $\eta$ using a perturbative expansion in the coupling terms between circuit elements. We find that our perturbative estimate agrees with numerically-identified values of $\eta$ within $\sim 5\%$, implying robustness to changes in the qubit spectra and coupling strengths. Here, we report explicit values of $\eta$ for various choices of $J_{A(B), c}$ calculated by numerically diagonalizing the Hamiltonian.

\begin{figure}[h!]
\centering
    \includegraphics[width=0.5\textwidth]{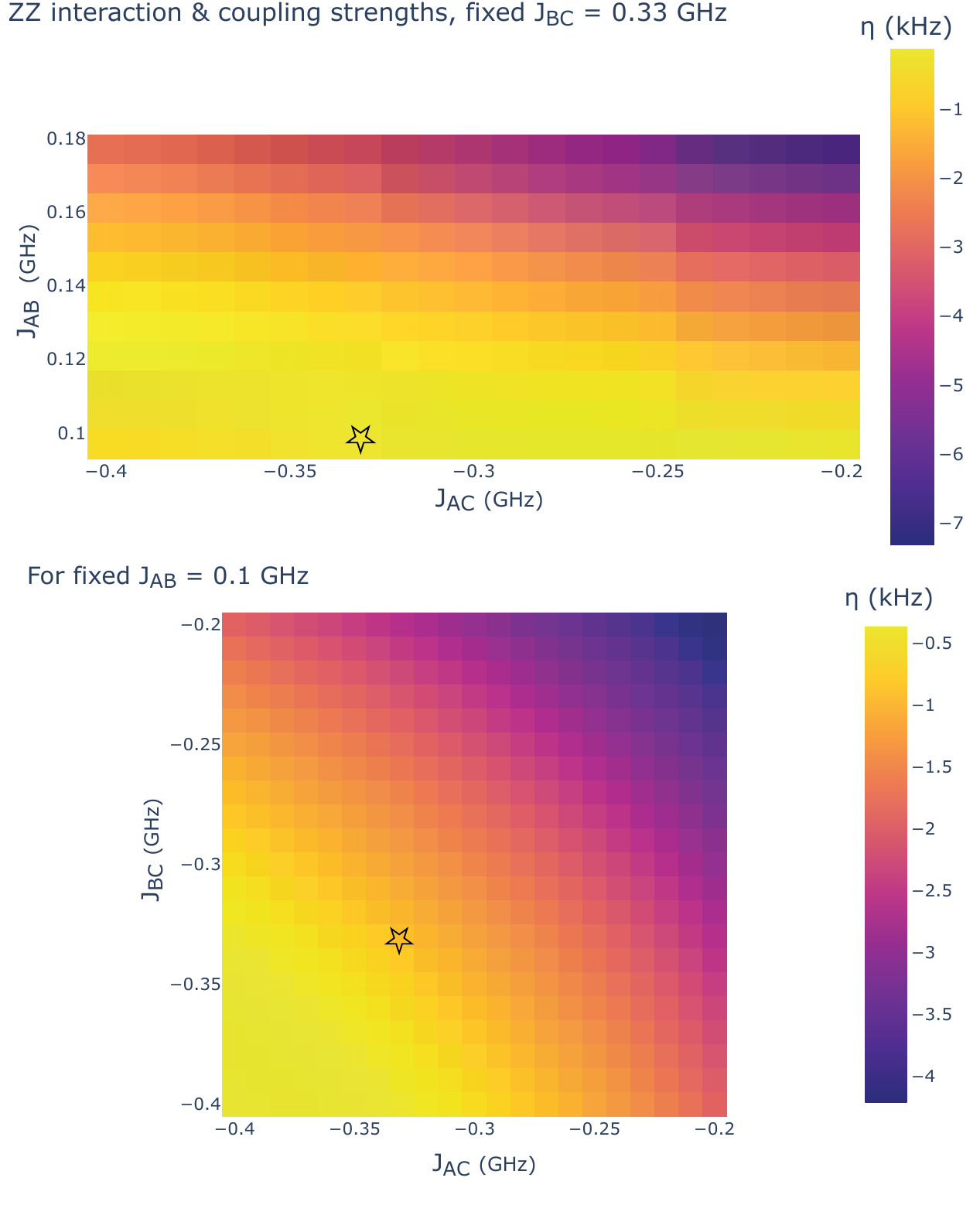}
    \caption{The ZZ interaction $\eta$, calculated by numerically diagonalizing the Hamiltonian, as a function of circuit coupling strengths. System parameters are the same as the main text. Top, sweeping $J_{B, c}$ and $J_{A, B}$, bottom, sweeping $J_{B, c}$ and $J_{A, c}$. Star symbols correspond to the values used in the main text. We find that the ZZ interaction strength stays below $\sim 2$ kHz for coupling strength changes on the order of $\sim 10\%$.}\label{fig:zz_sweep}
\end{figure}

\begin{figure}[ht]
  \centering
    \includegraphics[width=0.5\textwidth]{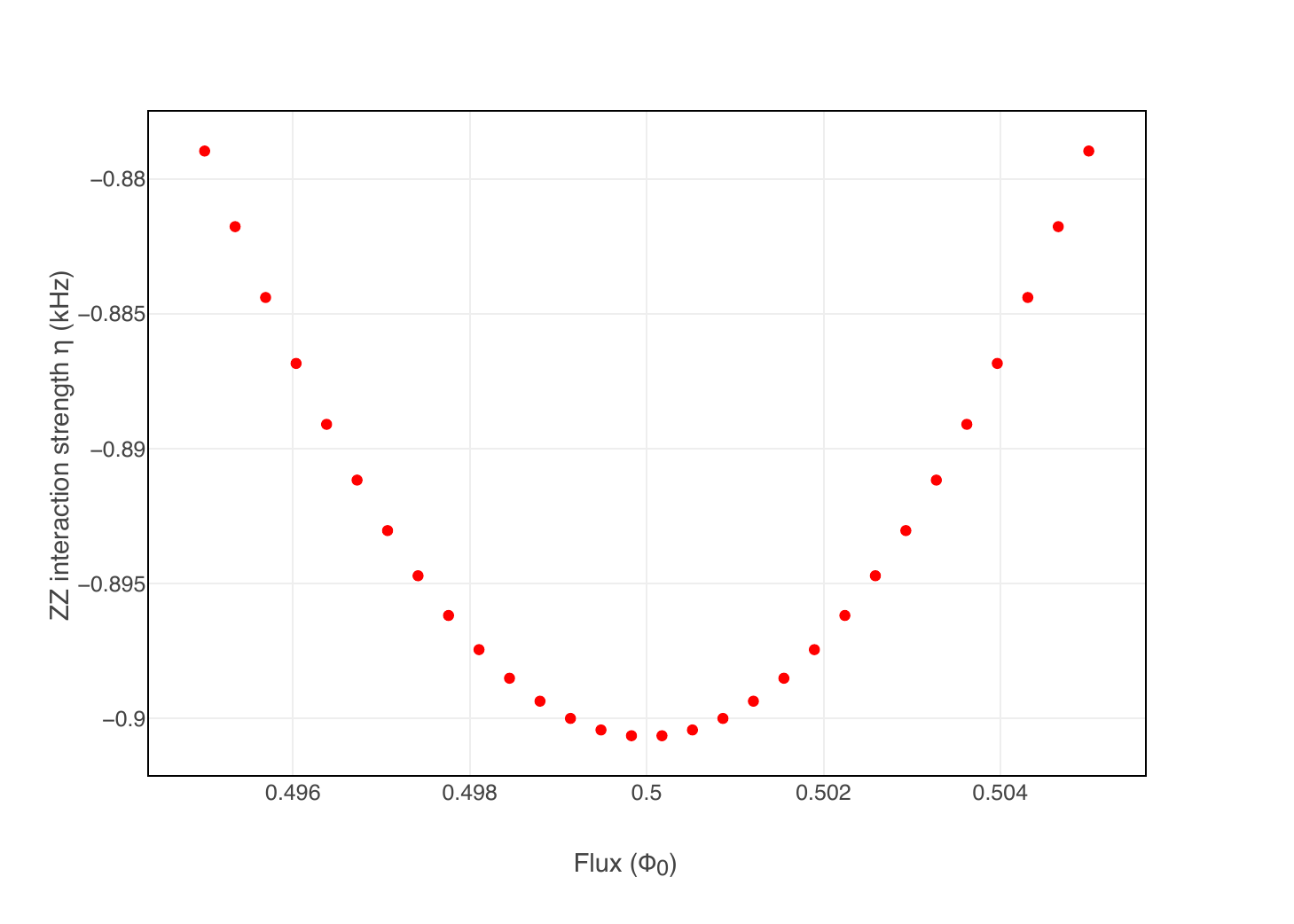}
  \caption{ZZ interaction strength $\eta$ as a function of external flux of fluxonium A. Circuit parameters are identical to those in the main text, except the flux value.  We find that the $\eta \sim 1$ kHz regime is well- maintained for generous deviations from the sweet spot of $10^{-2}\Phi_0$, with $\eta$ changing by about $3\%$ over this range.}\label{fig:zz_vs_flux}
\end{figure}

\begin{figure}[ht]
  \centering
    \includegraphics[width=0.5\textwidth]{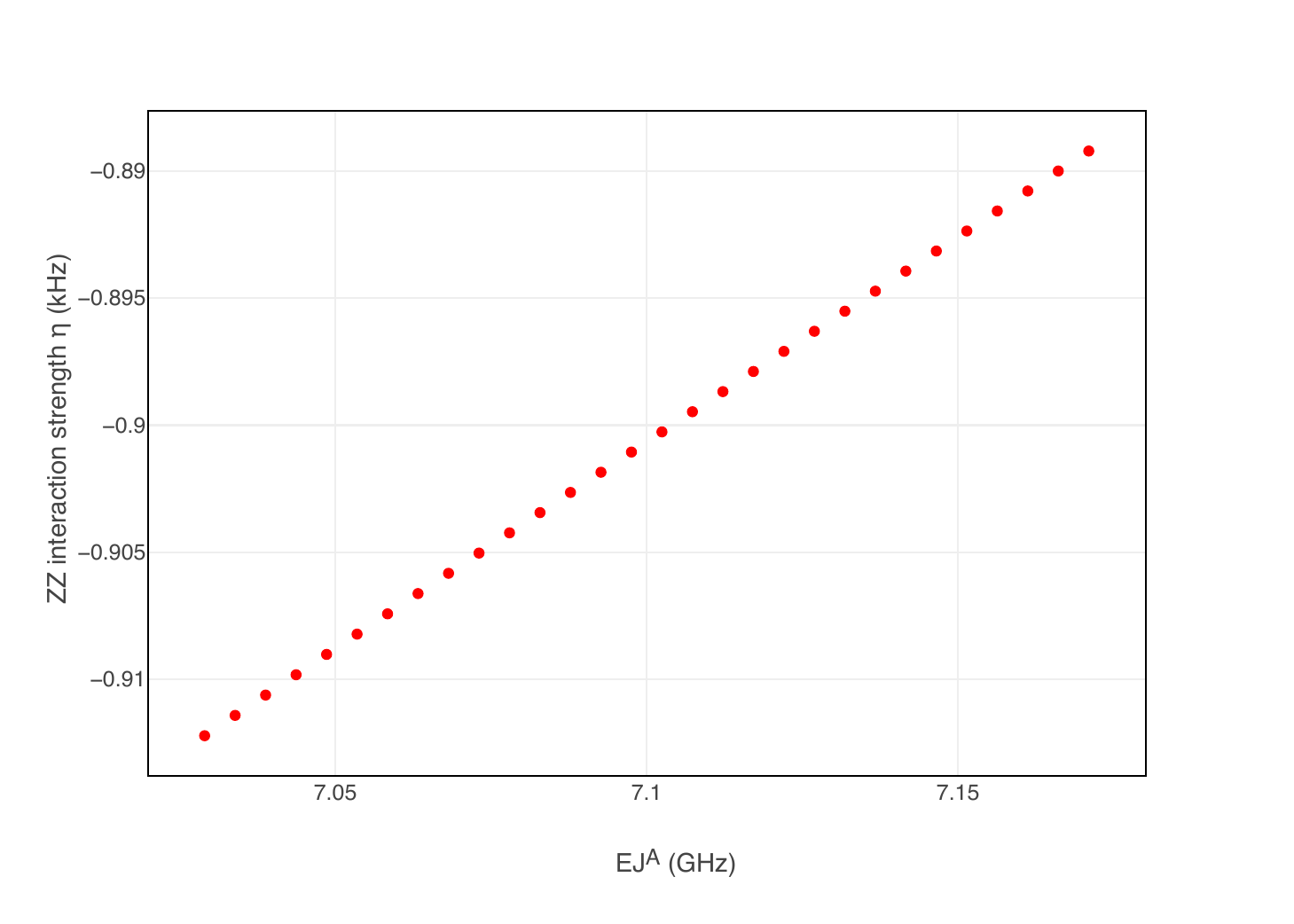}
  \caption{ZZ interaction strength $\eta$ as a function of the junction energy of fluxonium A. Circuit parameters are identical to those in the main text, except the junction energy.  We find that the $\eta \sim 1$ kHz regime is well-maintained for deviations from the targeted value of $1\%$, with $\eta$ changing by about $3\%$ over this range.}\label{fig:zz_vs_EJA}
\end{figure}

The ZZ interaction strength is most sensitive to $J_{A, B}$, where a change of $J_{A, B}$ of $10\%$ can lead to a change in $\eta$ of $\sim 1$ kHz. Converting to capacitance values, we find that the joint capacitance between fluxoniums $C_{A, B}$ must be within $\mathcal{O}(0.1)$ fF to maintain $|\eta|< 4$ kHz, with explicit values for each set of circuit parameters in Tables \ref{cap_matrix_gnd_inputs},
\ref{cap_matrix_diff_inputs}, and \ref{cap_matrix_diff_inputs_low_EC_2}. We find for all circuit parameter sets (for both the high low impedance resonators) that even at zero $C_{A, B}$, $|\eta|<4$ kHz -- i.e., the joint capacitance should be suppressed in circuit design as much as possible. Additionally plotting $\eta$ over a sweep of $J_{A, c}$ and $J_{B, c}$, we find that it is robust to changes in coupling strengths as shown in Fig. \ref{fig:zz_sweep}, particularly to changes in $J_{A, c}$ and $J_{B, c}$. For example, changes in $J_{A(B), c}$ of $\sim 10\%$ do not increase $\eta$ by more than $\sim 50\%$, maintaining the kHz regime (Fig. \ref{fig:zz_sweep}(b)).
We also report the $\eta$ value as a function of flux for our circuit parameters in the main text, see Fig. \ref{fig:zz_vs_flux}. We find that when the flux of fluxonium A changes by $10^{-2}\Phi_0$ from the sweet spot, $\eta$  changes by only $\sim 3\%$, demonstrating robustness of the low-$\eta$ regime to flux noise (in which deviations in flux to even smaller values of $\sim 10^{-3}\Phi_0$ are likely, as discussed in the main text).
Finally, we show $\eta$ under the presence of junction mistargeting, by sweeping $E_J^A$ and extracting $\eta$, numerically diagonalizing the Hamiltonian for each junction energy. We show in Fig. \ref{fig:zz_vs_EJA} find that when $E_J^A$ changes by $\sim 1\%$, $\eta$ changes by about $3\%$, i.e., the sufficiently-low $\eta$ is maintained in the presence of typical junction mistargeting.
\subsection{Perturbative analysis}\label{perturbative_anaysis}

We report the details of our perturbative analysis here, which leads to the observations listed above in section \ref{summary}.

\subsubsection{Matrix element hierarchy}\label{heirarchy}

Here, we list approximate values of the charge matrix elements in our circuit and the spectrum, using the circuit parameters in the main text. This informs which terms we can neglect in our perturbative analysis of the Hamiltonian, and further motivates the frequency allocations described in the main text. 

\begin{itemize}
        \item $n_{A, 01} \approx n_{B, 01} \approx 0.04$
        \item $n_{A, 12} \approx n_{B, 12} \approx 0.5$
        \item $n_{A, 03} \approx n_{B, 03}  \approx 0.5$. The third excied state of the fluxonium reaches the top of the cosine potential (see Fig. \ref{fig:si_wavefunctions}(a)), resulting in a smaller charge matrix element $n_{A(B), 03}$ compared to $n_{A(B), 12}$.
        \item $n_{A, 14} \approx n_{B, 14} \lesssim 0.02$. 
        \item $\omega_{A, 01} /2\pi \approx \omega_{B, 01} /2\pi \approx 100-300$ MHz.
        \item $\omega_{A, 12} / 2\pi \approx \omega_{B, 12} / 2\pi  \approx 4-9$ GHz, for $E_L \sim E_C/3$ to $E_L \sim E_C/6$. 
        \item $1 \lesssim n_{c, 01} \lesssim 3$ 
\end{itemize}

\subsubsection{Second order perturbation theory}
\paragraph{Second order shift on computational states}
The second order pathways are
\begin{widetext}
\begin{align}
\begin{split}
\label{2nd_order_101}
    E_{1, 0, 1}^{(2)} = J_{A, c}^2\left( \mathcal{O}(n_{A, 01}) + \color{red}\frac{ |\bra{2, 1, 1} n_A n_c \ket{1, 0, 1}|^2}{E_{1, 0, 1}^{(0)} - E_{2, 1, 1}^{(0)}}\color{black} + \frac{| \bra{4, 1,1} n_A n_c \ket{1, 0, 1}|^2 }{E_{1, 0, 1}^{(0)}-E_{4, 1, 1}}\right) \\
    + J_{B, c}^2\left( \mathcal{O}(n_{B, 01}) + \color{red}\frac{ |\bra{1, 1, 2} n_B n_c \ket{1, 0, 1}|^2}{E_{1, 0, 1}^{(0)} - E_{1, 1, 2}^{(0)}}\color{black} + \frac{| \bra{1, 1,4} n_B n_c \ket{1, 0, 1}|^2 }{E_{1, 0, 1}^{(0)}-E_{1, 1, 4}}\right)\\
    + J_{A, B}^2 \left(\frac{ |\bra{2, 0, 2} n_A n_B \ket{1, 0, 1}|^2}{E_{1, 0, 1}^{(0)} - E_{2, 0, 2}^{(0)}} + \frac{| \bra{4, 0,4} n_A n_B \ket{1, 0, 1}|^2 }{E_{1, 0, 1}^{(0)}-E_{4, 0, 4}} + \mathcal{O}(n_{A, 01} n_{B, 01} /  \omega_{01} ) + \mathcal{O}(n_{A, 01})  + \mathcal{O}(n_{B, 01})\right)
\end{split}
\end{align}
\end{widetext}
By about an order of magnitude, the largest terms are the \color{red}red \color{black} ones in equation \eqref{2nd_order_101} above involving a coupler excitation and a plasmon excitation, and they can be $\sim 15$ MHz. For the other computational states (see appendix), there will be identically large terms, i.e., $E_{1, 0, 0}^{(2)} \propto J_{A, c}^2  |\bra{2, 1, 0} n_A n_c \ket{1, 0, 0}|^2/(E_{1, 0, 0}^{(0)} - E_{2, 1, 0}^{(0)})$ and vice versa for $E_{0, 0, 1}$, as well as terms involving the $n_{03}$ of each fluxonium. For the second order component of the ZZ interaction $\eta^{(2)}$, all terms involving a coupler excitation cancel out, leaving
\begin{widetext}
\begin{align}\label{second_order_eta}
\begin{split}
-2\pi\eta^{(2)}/J_{A, B}^2 \sim  \frac{ |n_{A, 12} n_{B, 12}|^2}{\omega_{A, 12} + \omega_{B, 12}} - \frac{|n_{A, 03} n_{B, 12}|^2}{\omega_{A, 03} + \omega_{B, 12}} - \frac{|n_{A, 12} n_{B, 03}|^2}{\omega_{A, 12} + \omega_{B, 03}} +  \frac{|n_{A, 03} n_{B, 03}|^2}{\omega_{A, 03} + \omega_{B, 03}}
\end{split}
\end{align}
\end{widetext}
We additionally dropped terms order $\mathcal{O}(n_{A, 01}, n_{B, 14}, n_{A, 05}, n_{B, 05})$ or higher, since the associated matrix elements are small and the energy gaps are large. For our heirarchy, $\eta^{(2)}$ is small, with a generous upper bound of $(2\pi) J^2 |n_{12} | ^4 / (\omega_{A, 12}+ \omega_{B, 12}) \sim 0.5$ MHz. Also, it is clear that $\eta^{(2)} < 0$, as was stated in \cite{ding_high-fidelity_2023}, since $n_{03} < n_{12}$ and $\omega_{03} > \omega_{12}$.
\paragraph{Second order shift when the coupler is excited}
In the following we will again drop terms order $\mathcal{O}(n_{A, 01}, n_{B, 01}, n_{A, 14}, n_{B, 14}, n_{A, 05}, n_{B, 05})$ or higher. The second order contribution to the $E_{1, 1, 1}$ state is
\begin{widetext}
\begin{equation}\label{E_111}
\begin{aligned}
    E_{1, 1, 1}^{(2)} = J_{A, c}^2\left(\color{purple} \frac{ |\bra{2, 0, 1} n_A n_c \ket{1, 1, 1}|^2}{E_{1, 1, 1}^{(0)} - E_{2, 0, 1}^{(0)}}\color{black} + \color{red} \frac{ |\bra{2, 2, 1} n_A n_c \ket{1, 1, 1}|^2}{E_{1, 1, 1}^{(0)} - E_{2, 2, 1}^{(0)}}\color{black} \right) \\
    + J_{B, c}^2\left( \color{purple} \frac{ |\bra{1, 0, 2} n_B n_c \ket{1, 1, 1}|^2}{E_{1, 1, 1}^{(0)} - E_{1, 0, 2}^{(0)}}\color{black} + \color{red}\frac{ |\bra{1, 2, 2} n_B n_c \ket{1, 1, 1}|^2}{E_{1, 1, 1}^{(0)} - E_{1, 2, 2}^{(0)}}\color{black} \right) \\ 
    + J_{A,B}^2 \left(\frac{ |\bra{2, 1, 2} n_B n_A \ket{1, 1, 1}|^2}{E_{1, 1, 1}^{(0)} - E_{2, 1, 2}^{(0)}} \right)
      \end{aligned}
\end{equation}
\end{widetext}
Similarly as in the computational basis, the plasmon transition and a coupler excitation contribute non-trivially to the dispersive shift. However, the two analogous terms (\color{red}in red\color{black}) exhibit a bosonic enhancement of $2$, such that they are now order $\sim 30$ MHz. Particularly at lower coupler and plasmon frequencies, these could facilitate increased $\chi_{ij}$. More significantly, the dominating terms $\gtrsim 100$ MHz are in \color{purple}purple\color{black}, particularly when $\Delta_{A(B)} \sim J_{A(B), c}$ as described in the main text. For the coupler excitations conditional on other fluxonium states, we write $E_{0, 1, 1}^{(2)}$ as an example:
\begin{widetext}
\begin{equation}\label{E_011}
\begin{aligned}
    E_{0, 1, 1}^{(2)} = J_{A, c}^2\left(\color{purple} \frac{ |\bra{3, 0, 1} n_A n_c \ket{0, 1, 1}|^2}{E_{0, 1, 1}^{(0)} - E_{3, 0, 1}^{(0)}}\color{black} +   \color{red}\frac{ |\bra{3, 2, 1} n_A n_c \ket{0, 1, 1}|^2}{E_{0, 1, 1}^{(0)} - E_{3, 2, 1}^{(0)}}\color{black} \right) \\
    + J_{B, c}^2\left( \color{purple} \frac{ |\bra{0, 0, 2} n_B n_c \ket{0, 1, 1}|^2}{E_{0, 1, 1}^{(0)} - E_{0, 0, 2}^{(0)}}\color{black} + \color{red} \frac{ |\bra{0, 2, 2} n_B n_c \ket{0, 1, 1}|^2}{E_{0, 1, 1}^{(0)} - E_{0, 2, 2}^{(0)}}\color{black}\right) \\ 
    + J_{A,B}^2 \left( \frac{ |\bra{3, 1, 2} n_A n_B \ket{0, 1, 1}|^2}{E_{0, 1, 1}^{(0)} - E_{3, 1, 2}^{(0)}} \right)
    \end{aligned}
\end{equation}
\end{widetext}
Crucially, if $\omega_{A(B), 03} \sim \omega_{A(B), 12}$, the dominating terms (\color{purple}in purple\color{black}) can be equal to the $E_{1, 1, 1}^{(2)}$ contributions. These observations motivate our design tenets for large coupler excitation state-selectivity $\delta\chi_{ij}$: $J_{A(B)}n_{A(B), 12} n_{c, 01}/\Delta_{A(B)} \sim 1$, $\omega_{03} > \omega_{A(B), 12}$, and $\omega_{A(B), 12} - \omega_c \equiv 2\pi \Delta_{A(B)} > 0$. By symmetry, the same arguments follow for $E_{1, 1, 0}$ and $E_{0, 1, 0}$.
\begin{figure}[ht]
  \centering
    \includegraphics[width=0.5\textwidth]{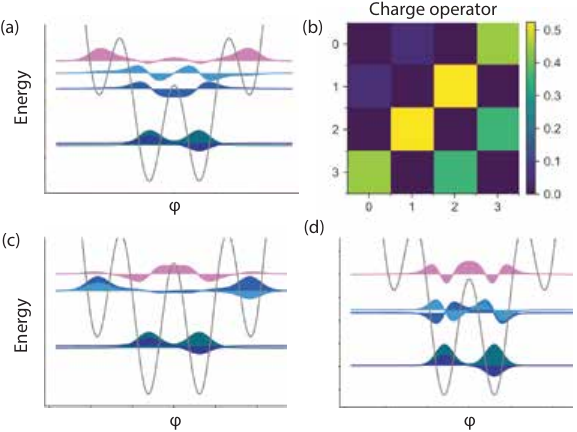}
  \caption{Target spectral qualities. Unless otherwise stated, $E_L = 0.3$ GHz, $E_C = 2$ GHz, and $E_J = 7 $ GHz. All plots generated using scqubits. (a) The example circuit parameter set in the main tect, $E_J = 7$ GHz, $E_C = 2$ GHz, $E_L = 0.3$ GHz. Note that the gap between states $\ket{2}$ and $\ket{3}$ is large, since they sit at the top of the cosine potential and aren't localized to the potential minima. This is desirable, since the $\ket{0}-\ket{3}$ transition causes a decrease in coupler $|\chi-\chi_{ij}|$ as $\propto 1/\omega_{23}$. (b) The charge matrix element of the fluxonium with parameters in plot(a). The $n_{01}$ matrix element remains small (here 0.04), while the $n_{03}$ element (0.48) starts to decrease compared to $n_{12}$ (0.56) as the $\ket{2}$ and $\ket{3}$ states experience the parabolic potential. (c) Insufficient $E_L = 0.15$ GHz. In this case, the $\ket{2}$ and $\ket{3}$ states are stuck in local minima of the cosine potential, since $E_L$ is small (parabola is shallow). The localization of those states reduces the $\omega_{23}$ gap which would cause a decrease in coupler selectivity as the dispersive shift on the coupler from the three-photon transition becomes comparable to that of the plasmon transition. (d) Insufficient $E_C = 0.8$ GHz. In this case, the $\ket{2}$ and $\ket{3}$ states are again stuck in local minima of the cosine potential, since $E_C$ is small (parabola is shallow). The localization of those states reduces the $\omega_{23}$ gap which would cause a decrease in coupler selectivity as the dispersive shift on the coupler from the three-photon transition becomes comparable to that of the plasmon transition.
  }\label{fig:si_wavefunctions}
\end{figure}
\paragraph{$\chi_{ij}$ values}\label{omega_23_section}
Ultimately, we need the \textit{difference} of these dispersive shifts on the coupler for various fluxonium states to be large to get frequency selectivity $\delta\chi_{ij}$ when driving the gate. Retaining only \color{purple}purple \color{black} terms in Eqs. \eqref{E_111} and \eqref{E_011}, we estimate $\chi_{ij}$ at second order $\chi_{ij}^{(2)}$ for all four fluxonium states: 
\begin{equation}\label{freq_selectivity}
    \begin{aligned}
        \chi^{(2)} = -\frac{J_{A, c}^2 n_{A, 12}^2  n_{c, 01}^2}{\Delta_A} - \frac{J_{B, c}^2 n_{B, 12}^2  n_{c, 01}^2}{\Delta_B} \\
        \chi_{01}^{(2)} = -\color{blue} \frac{2\pi J_{A, c}^2 n_{A, 03}^2 n_{c,01}^2}{2\pi\Delta_A + \omega_{A, 23} + \omega_{A, 01}} \color{black} -\frac{J_{B, c}^2 n_{B, 12}^2 n_{c, 01}^2 }{\Delta_B} \\
        \chi_{10}^{(2)} = -\color{blue} \frac{ 2\pi J_{B, c}^2 n_{B, 03}^2 n_{c,01}^2}{2\pi\Delta_B + \omega_{B, 23} + \omega_{B, 01}}\color{black} -\frac{J_{A, c}^2 n_{A, 12}^2 n_{c, 01}^2 }{\Delta_A} \\
        \chi_{00}^{(2)} = -\color{blue} \frac{2\pi J_{A, c}^2 n_{A, 03}^2 n_{c, 01}^2}{2\pi\Delta_A + \omega_{A, 23} + \omega_{A, 01}} - \frac{2\pi J_{B, c}^2 n_{B, 03}^2 n_{c, 01}^2}{2\pi\Delta_B + \omega_{B, 23} + \omega_{B, 01}}
    \end{aligned}
\end{equation}
For large $\delta \chi_{ij}$, we increase $\omega_{23}$ and decrease $n_{A(B), 03}$, to reduce the size of the terms in \color{blue}blue\color{black}. As $E_C$ decreases, $n_{03} \rightarrow n_{12}$ and $\omega_{23} \rightarrow \omega_{01} \rightarrow 0$. \textbf{This motivates a particular set of circuit parameter choices: choose a heavy fluxonium regime such that $n_{01}$  is small (to protect against charge noise), but also choose a sufficiently large $E_C/E_J \sim 0.1$ such that the $\ket{2}$ and $\ket{3}$ states are at the top of the cosine potential, not anchored within it.} This choice introduces a splitting $\omega_{23} \gg \omega_{01}$, as $\ket{2}$ and $\ket{3}$ experience the quadratic potential of the inductor as well as the cosine potential. It also decreases $n_{A(B), 03}$ for the same reason, further increasing $|\delta \chi_{ij}|$. 

Simultaneously, we choose a low $E_L \sim 0.3$ GHz, to protect against flux noise, but again a sufficiently large $E_L/E_C \approx 0.1$ such that the $\ket{2}$ and $\ket{3}$ states do not localize in the minima of the cosine potential at $\pm 2\pi$, retaining large $\omega_{A(B), 23}$. For examples of these circuit parameter choices, see Fig. \ref{fig:si_wavefunctions}. With these circuit parameter requirements, the coupler selectivity can be large, of order a few hundred MHz.
\subsubsection{Third order perturbation theory: deriving the ZZ interaction strength}

The result in \cite{ding_high-fidelity_2023} uses an interference effect between second, third, and fourth order terms in perturbation theory to cancel ZZ interactions. As found above, the second order ZZ interaction is at most $\sim 1$ MHz to 10 MHz, suggesting that the third and fourth order terms should also be similar in magnitude but opposite in sign.
\paragraph{Third order shift on computational states}
By similar calculations as above and assuming the same hierarchy, we identify the key terms which do not contain any factors of $\bra{0} n_{A, B} \ket{1}$ or $\bra{1} n_{A, B} \ket{4}$: 
\begin{itemize}
    \item $\ket{101}-\ket{211}-\ket{202}-\ket{101}$: gap $\omega_{A, 12}+ \omega_c$,  $\omega_{A, 12} + \omega_{B, 12}$
    \item $\ket{101}-\ket{112}-\ket{202}-\ket{101}$: gap $\omega_{B, 12}+ \omega_c$, $\omega_{A, 12} + \omega_{B, 12}$
    \item $\ket{101}-\ket{211}-\ket{112}-\ket{101}$: gaps $\omega_{A, 12} + \omega_c$, $\omega_{B, 12} + \omega_c$ (typically dominant)
\end{itemize}
All terms are positive, such that $E_{1, 0, 1}^{(3)} > 0$. For the other computational states, the strongest pathways and gaps are:
\begin{itemize}
    \item $\ket{001}-\ket{311}-\ket{012}-\ket{001}$: gaps $\omega_{A, 03} + \omega_c$, $\omega_c + \omega_{B, 12}$ (typically dominant)
    \item $\ket{001}-\ket{311}-\ket{302}-\ket{001}$: gaps $\omega_{A, 03} + \omega_c$, $\omega_{A, 03} + \omega_{B, 12}$.
    \item vice versa for the 100 computational state.
\end{itemize}
The strongest pathways for $E_{0, 0, 0}$ are: 
\begin{itemize}
    \item $\ket{000}-\ket{303}-\ket{013}-\ket{000}$: gaps $\omega_{A, 03} + \omega_{B, 03}$, and $\omega_{c} + \omega_{B, 03}$
    \item $\ket{000}-\ket{303}-\ket{310}-\ket{000}$: gaps $\omega_{A, 03} + \omega_{B, 03}$, and $\omega_{c} + \omega_{A, 03}$
\end{itemize}
\paragraph{ZZ cancellation condition (computational manifold)}
Neglecting the 0-3 excitations and fourth-order corrections, the third order ZZ interaction $\eta^{(3)}$ can be roughly approximated as 

\begin{figure}[H]
  \centering
    \includegraphics[width=0.25\textwidth]{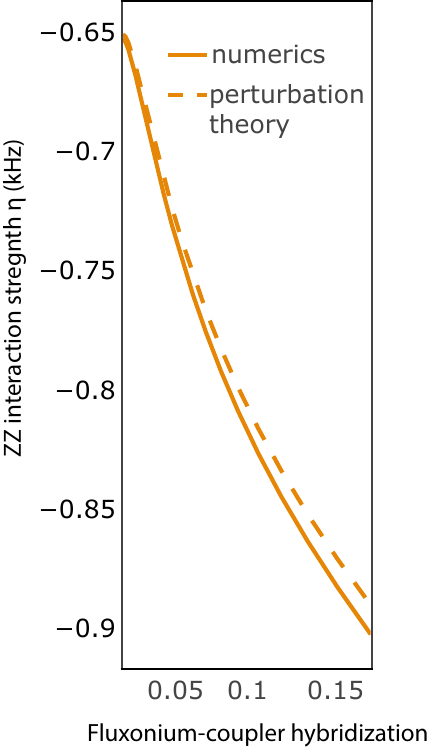}
  \caption{ZZ interaction $\eta$ as a function of coupler hybridization. Circuit parameters are identical to those in the main text, where we sweep the coupler frequency to tune the fluxonium-coupler hybridization, in analogy with Fig. 2 of the main text.  We observe agreement between our perturbative estimate to third order (dashed line) and the numerically exact result (solid line) within $\sim 5$\%.}\label{fig:sweep_zz_hybridization}
\end{figure}
\begin{widetext}
\begin{align}\label{cap_matrix_condition}
    -\eta^{(2)} = \eta^{(3)} \approx E_{1, 0, 1}^{(3)} 
    \approx  \\
    2 (2\pi)^2 J_{A, c} J_{B, c} |n_{c, 01}|^2 |n_{A, 12}|^2 |n_{B, 12}|^2 \left( \frac{1}{\omega_{A, 12} + \omega_{B, 12}} \left ( \frac{1}{\omega_{A, 12} + \omega_c} + \frac{1}{\omega_{B, 12} + \omega_c} \right) + \frac{1}{(\omega_{A, 12} + \omega_c)(\omega_{B, 12} + \omega_c)} \right)
\end{align}
\end{widetext}
The form $J_{A, B} \propto J_{c}^2$ where $J_{A, c} \sim J_{B, c}$ is consistent with the numerical findings in \cite{ding_high-fidelity_2023}, at eq. 2. Indeed, the third order shift $\eta^{(3)}$ turns out to be at most $\sim 1$ MHz, as needed for cancellation of $\eta^{(2)}$.  
No terms in eq \eqref{cap_matrix_condition} depend on $\Delta_{A(B)}$, suggesting that the energy gap between the plasmon and the coupler transition can be tuned to be small and we can still retain ZZ cancellation. We find that our perturbative calculation for $\eta \approx \eta^{(2)} + \eta^{(3)}$ well approximates the numerically exact result. See Fig. \ref{fig:sweep_zz_hybridization} below, sweeping the hybridization by changing the coupler frequency as done in the main text.
\paragraph{Third order shift when coupler is excited}
We consider third order perturbation theory on the states with a single coupler excitation.
The pathways that are significant for the shift $E_{1, 1, 1}^{(3)}$ include
\begin{itemize}\label{third_order_E_111_paths}
    \item $\ket{111}-\ket{201}-\ket{212}-\ket{111}$: gaps $2\pi \Delta_{A}$, $\omega_{A, 12} + \omega_{B, 12}$
    \item $\ket{111}-\ket{102}-\ket{212}-\ket{111}$: gaps $2\pi \Delta_{B}$, $\omega_{A, 12} + \omega_{B, 12}$
    \item $\ket{111}-\ket{201}-\ket{102}-\ket{111}$: gaps $2\pi \Delta_A$, $2\pi \Delta_B$ (typically dominant)
    \item $\ket{111}-\ket{221}-\ket{212}-\ket{111}$: gaps $\omega_{A, 12} + \omega_c$, $\omega_{A, 12} + \omega_{B, 12}$
    \item $\ket{111}-\ket{122}-\ket{212}-\ket{111}$: gaps $\omega_{B, 12} + \omega_c$, $\omega_{A, 12} + \omega_{B, 12}$
    \item $\ket{111}-\ket{122}-\ket{221}-\ket{111}$: gaps $\omega_{A, 12} + \omega_c$, $\omega_{B, 12} + \omega_c$.
\end{itemize}
For $\Delta_{A(B)} > 0$, all the third order terms are positive, decreasing $\delta \chi_{ij}$. The dominating term is $\propto 1/\Delta_A \Delta_B$, with strength of roughly $\sim 30$ MHz, decreasing $\delta \chi_{ij}$. 
This finding suggests that for maximizing the coupler selectivity $\delta \chi_{ij}$, while also minimizing the hybridization between the coupler and the plasmon transition, there is likely an optimal $J_{A, B} n_{c, 01} n_{A(B), 12}/\Delta_{A, B} \sim 1$, further confirming our design tenets.

\subsubsection{Coupler nonlinearity}

We want to study how the coupler nonlinearity is determined by the circuit parameters, as we found in our time-domain simulations that Stark shifts from detuned driving of the second coupler excitation contributes to the coherent error of the gate. Specifically, when we excite the $\ket{101}-\ket{111}$ transition, the $\ket{111}-\ket{121}$ experiences an off resonant drive, with detuning set by the coupler nonlinearity $\alpha$ when both fluxonium are in their $\ket{1}$ states:
\begin{equation}\label{def_K11}
    \alpha \equiv (E_{121}-E_{111})-(E_{111}-E_{101})
\end{equation}
The contributons to $\alpha$ at second and third order are zero, so we focus on the fourth order term below.

\paragraph{$\alpha$ to fourth order}
For the coupler nonlinearity, the fourth order terms are the most significant. In this section, we focus only on terms $\propto 1/\Delta^3$, such that each fluxonium can be treated as a two level system corresponding to the plasmon transition, and a Jaynes-Cummings coupling is approximated between the plasmon transitions of the fluxoniums and the resonator. Specifically, the Hamiltonian in our approximation is 
\begin{equation}\label{H_two_level_truncation}
\begin{aligned}
\mathcal{H}_{\text{trunc}}/h =\frac{ \omega_{A, 12}}{4\pi} \sigma_z^A + \frac{ \omega_{B, 12}}{4\pi} \sigma_z^B + \frac{\omega_c }{2\pi} c^\dagger c \\ + J_{A, c}n_{A, 12} n_{c, 01} \left( \sigma_-^A c^\dagger + h.c. \right) 
\\ + J_{B, c} n_{B, 12} n_{c, 01} \left( \sigma_-^B c^\dagger + h.c. \right) \\ + J_{A, B}n_{A, 12} n_{B, 12} (\sigma_-^B \sigma_+^A + h.c.)
\end{aligned}
\end{equation}

Considering the fourth order shift to $E_{1, 2, 1}$ in our Hilbert space truncation, there are only two fourth order paths: $\ket{121}-\ket{211}-\ket{112}-\ket{211}-\ket{121}$ and $\ket{121}-\ket{112}-\ket{211}-\ket{112}-\ket{121}$. Executing the fourth order perturbation theory, we arrive at
\begin{equation}
\begin{aligned}
E_{1, 2, 1}^{(4)} \approx n_{c, 01}^2 (...) + n_{c, 12}^4 \sum_{i \in \{A, B\}} \frac{J_{i, c}^2 n_{i, 12}^2}{\Delta_i} \sum_{j \in \{A, B\}} \frac{J_{j, c}^2 n_{j, 12}^2}{\Delta_j^2}
\end{aligned}
\end{equation}
We neglect the term $\propto n_{c, 01}^2$ since it'll cancel when subtracted with $2E_{1, 1, 1}^{(4)}$ in the coupler nonlinearity estimation. Indeed the $E_{1, 1, 1}^{(4)}$ terms are identical with one less excitation in the resonator such that the coupler nonlinearity at fourth order is 
\begin{equation}\label{fourth_order_K_11}
    \alpha^{(4)} \approx 2 n_{c, 01}^4 \sum_{i \in \{A, B\}} \frac{J_{i, c}^2 n_{i, 12}^2}{\Delta_i} \sum_{j \in \{A, B\}} \frac{J_{j, c}^2 n_{j, 12}^2}{\Delta_j^2}
\end{equation}
where we have retained only terms $\propto 1/\Delta^3$ and used $n_{c, 12}^4-2n_{c, 01}^4 = 2 n_{c, 01}^4$. As found numerically, the anharmonicity is positive for $\Delta_{A(B)} > 0$. However, the value is very sensitive to the couplings and circuit parameters since it depends on the coupling, resonator impedance, and plasmon matrix element to the fourth power. For example, using our typical parameters $J = 300$ MHz and $n_{A(B), 12} = 0.5$, and $\Delta_{A(B)} = 1$ GHz, we get a coupler anharmonicity of only +16 MHz. However if the plasmon transition charge matrix element increases to $n_{A(B), 12} = 0.6$, then the anharmonicity doubles.

Comparing to our numerical results, we find that the exact coupler nonlinearity is typically $\sim3$x smaller in magnitude than predicted by the fourth order perturbation theory. 
Therefore, we conclude that the coupler nonlinearity both depends on even higher orders in the perturbation theory, and is largely given by plasmon-coupler hybirdization. This motivates the approach in which strongly couple the plasmon-coupler subspace as described in the main text.

\section{Coherent error budget and DRAG implementation}\label{appendix_d_drag}

\subsection{Error budget without DRAG}

\begin{figure*}[ht]
\includegraphics[width=\textwidth]{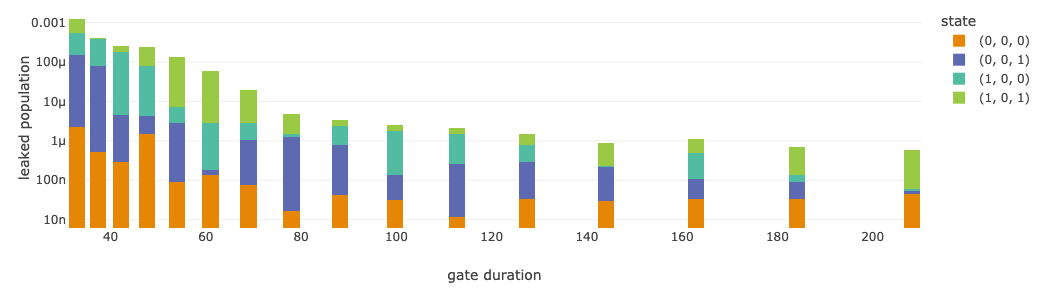}
\caption{Population leakage budget as a function of gate duration, given by state. Contributions are stacked vertically, such that for a given gate duration, the computational state with the largest contribution to leakage error is the one with the most area in the bar (taking into account the logscale). We observe that the leakage budget is dominated by leakage from the $\ket{1, 0, 1}$ state for nearly every gate duration (note the logscale). This suggests that population leakage into $\ket{1, 1, 1}$ or $\ket{1, 2, 1}$ from finite $\alpha$ dominates the error, motivating an application of the DRAG technique.}\label{fig:leakage_budget_no_drag}
\end{figure*}

As reported in the main text, the coherent error budget is dominated by population leakage. We perform unitary, time-domain simulations of our gate control, where the initial state is a computational eigenstate. Then, we extract the population that left the computational subspace, divide it by four to simulate leakage errors, and plot the results in Fig. \ref{fig:leakage_budget_no_drag} as a function of gate duration and computational state. We see that the error is dominated by leakage from the $\ket{1, 0, 1}$ state, suggesting that the finite $\alpha$ value limits the leakage budget and motivating our application of the Derivative Removal by Adiabatic Gate (DRAG) technique as discussed in the sections below.

\subsection{Implementation of DRAG}

We implement the Derivative Reduction by Adiabatic Gate technique to improve the coherent error of our gate. The total control is
\begin{equation}\label{drag_def}
    \Omega_\text{total}(t) = a\left(\Omega(t) \cos{\omega_d t} + d \frac{\dot{\Omega}(t) \sin{\omega_d t}}{\alpha}\right),
\end{equation}
where $\Omega(t) = 2 e^{-4 t^2/\tau^2} / \sqrt{\pi \tau} |\langle 1, 1, 1 | \hat{n}_c | 1, 0, 1\rangle|$ as in the main text, and $\omega_d$ is the drive frequency. 
\subsubsection{Calibration procedure}
\begin{figure*}[ht]
    \includegraphics[width=0.5\textwidth]{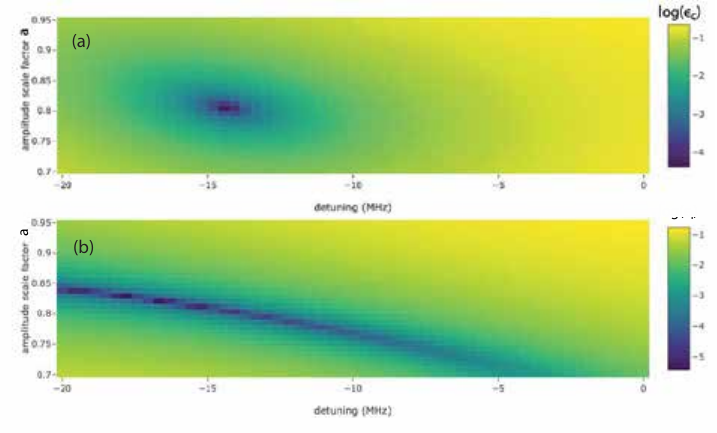}
    \centering
    \caption{DRAG calibration procedure. We sweep the amplitude scale $a$ on the y axis. The relative phase between the two tones is always fixed at $\pi/2$. Here, $d = 0.475$. On the x axis, we sweep the frequency $\omega_d$ of both tones, where the detuning is the deviation from the resonance frequency $\langle 111 |\mathcal{H} |111\rangle - \langle 101 |\mathcal{H} |101\rangle $. We observe a minimum in the coherent error (top). On the bottom plot, we observe the line in which the leakage error is reduced, corresponding to effective implementation of the DRAG technique.}\label{fig:drag_calibration}
\end{figure*}

We implement the DRAG technique by adding an in-quadrature tone proportional to the derivative of the Gaussian envelope. For a given gate duration, there are three parameters we optimize independently: (1) the magnitude $a$ of the original gaussian tone, (2) the magnitude $ad$ of the tone proportional to its derivative, and (3) the frequency $\omega_d$ of both tones. We first fix $d$, and sweep the amplitude $a$ of both tones, as well as their frequency $\omega_d$ in a two-dimensional grid. We find the point of the smallest coherent error, and then re-run the sweep with increased resolution. We repeat that whole process for different $d$ values. Finally, we choose the parameter set which minimizes the total coherent error, for all three parameters $a, d, \omega_d$. For an example calibration curve, we plot our 2D sweep over the drive amplitudes and frequencies in Fig. \ref{fig:drag_calibration}, showing the point of optimized coherent error and the line over which the leakage is reduced. In this case, the gate duration is 47.7 ns, and the ``drag scale" $d$ was 0.475. As shown in Figs. \ref{fig:drag_leakage_optimize_d} and \ref{fig:drag_leakage_optimize_d_2}, we perform this optimization procedure for many different $d$ values, and a range of gate durations $t_g$, extracting the optimal coherent error for each $t_g$. We still find the coherent error is entirely due to leakage, as we've once again effectively removed the phase errors by tuning $\omega_d$.

\subsubsection{Results}

\begin{figure*}
\subfloat[\label{fig:drag_leakage_optimize_d:a}]{%
  \includegraphics[width=\columnwidth]{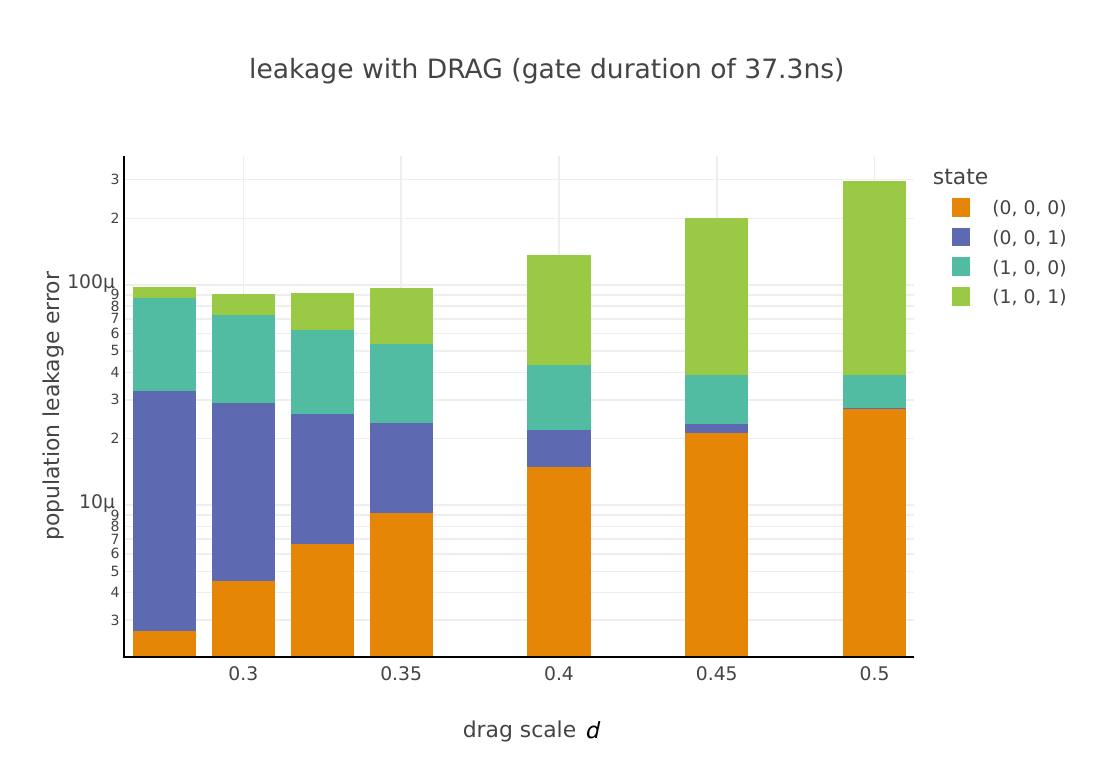}%
}
\hfill
\subfloat[\label{fig:drag_leakage_optimize_d:b}]{%
  \includegraphics[width=\columnwidth]{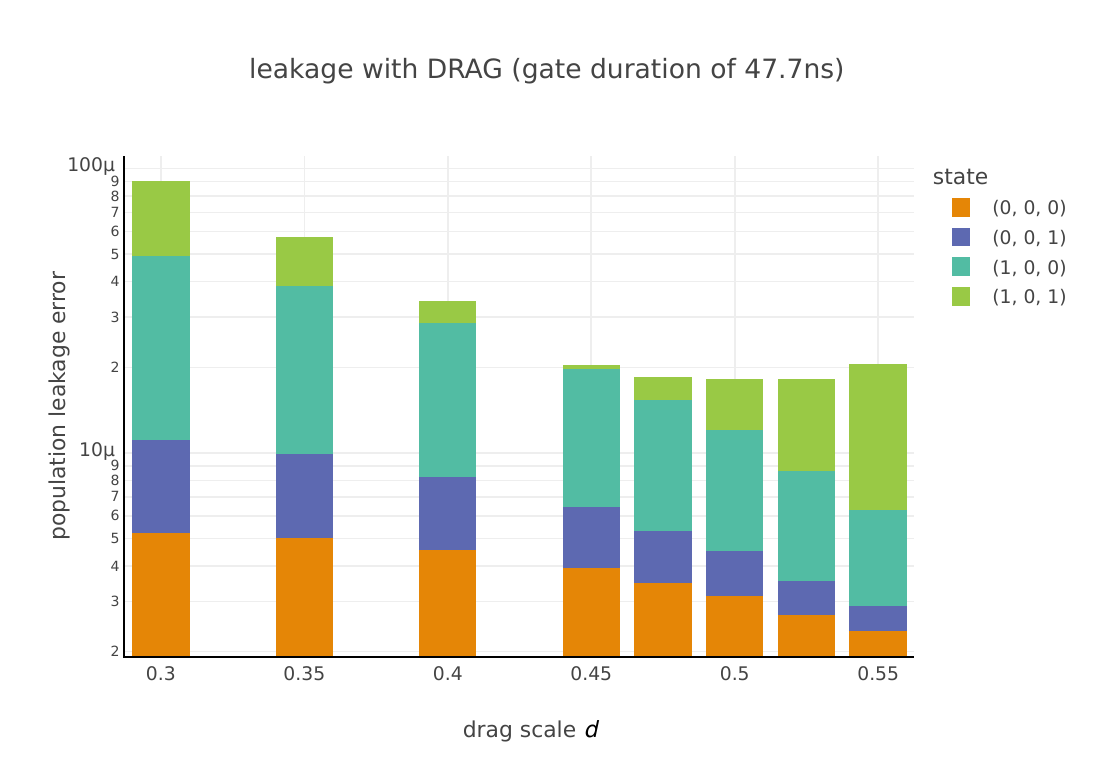}%
}
\caption{Leakage budget for short gate durations $37.3$ ns (top) and $47.7$ ns (bottom). We observe improvements in the $\ket{1, 0, 1}$ state leakage in particular at optimal $d$ values, corresponding to effective implementation in the DRAG technique. However, as discussed in Figs. \ref{fig:leakage_budget_DRAG_ratio_short_tg} and \ref{fig:leakage_budget_DRAG_ratio_long_tg} below, the overall coherent error does not significantly improve due to increased leakage of the other computational states.} \label{fig:drag_leakage_optimize_d}
\end{figure*}

\begin{figure*}
\subfloat[\label{fig:drag_leakage_optimize_d_2:a}]{%
  \includegraphics[width=\columnwidth]{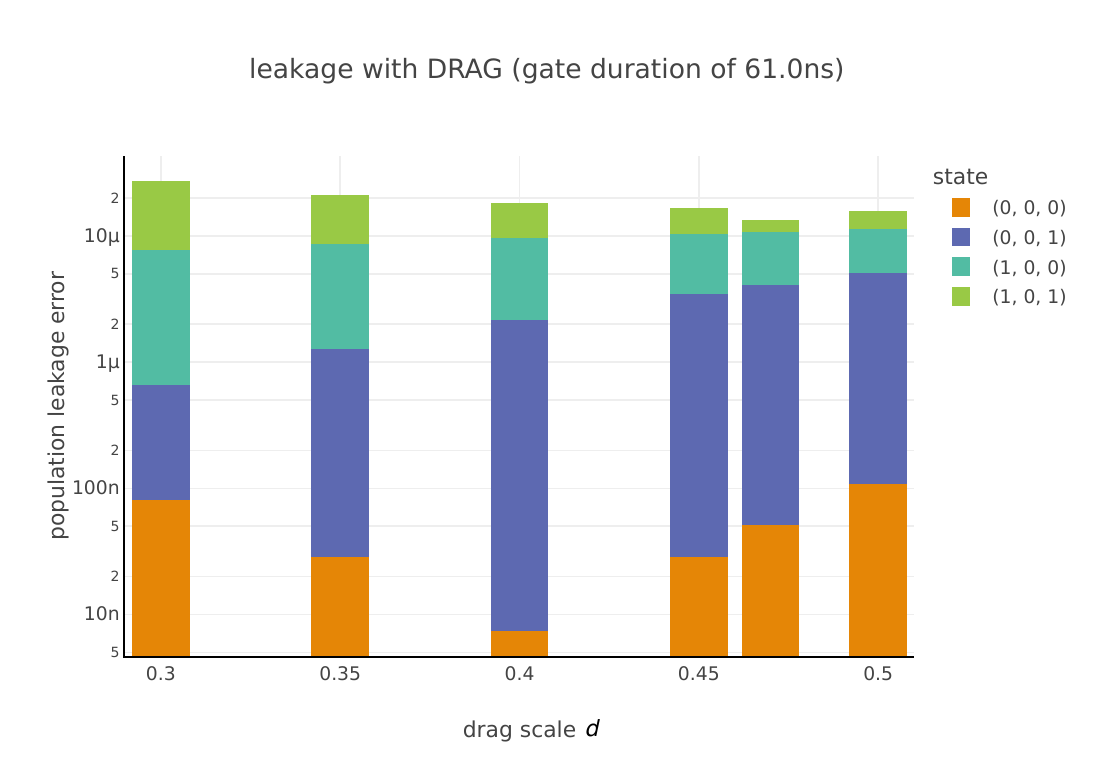}%
}\hfill
\subfloat[\label{fig:drag_leakage_optimize_d_2:b}]{%
  \includegraphics[width=\columnwidth]{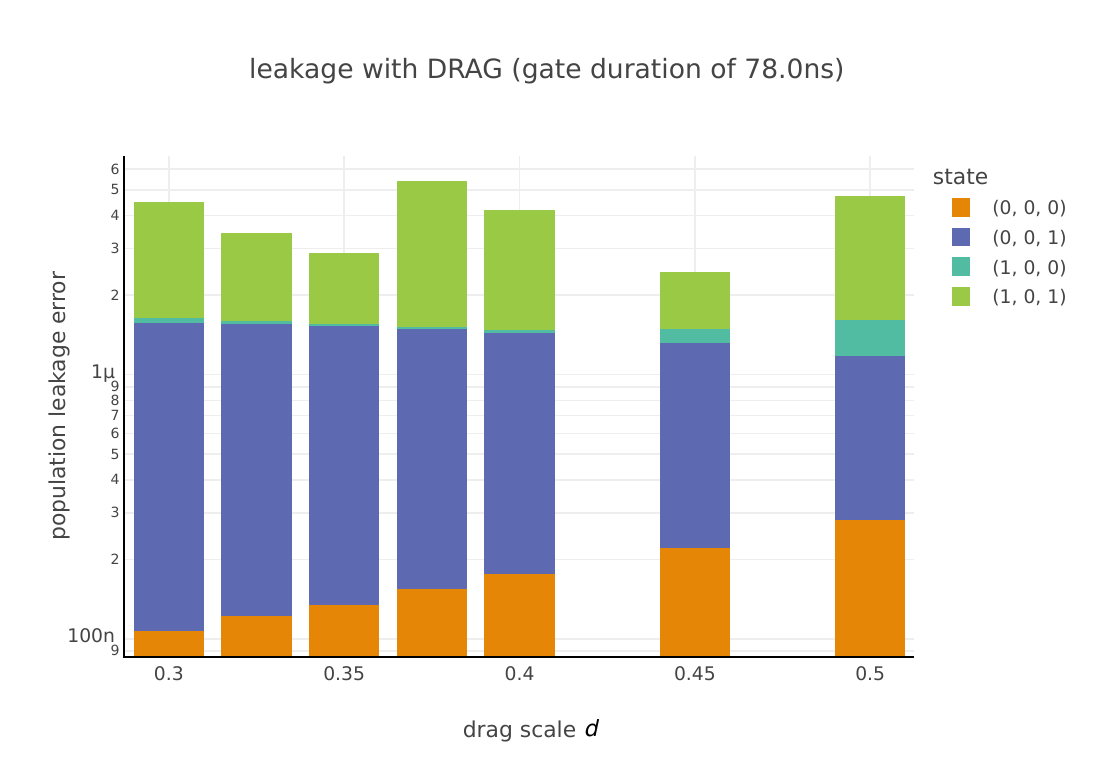}%
}
\caption{Leakage budget for short gate durations $61$ ns (top) and $78$ ns (bottom).} \label{fig:drag_leakage_optimize_d_2}
\end{figure*}

\begin{figure*}
\subfloat[\label{fig:leakage_budget_DRAG_ratio_short_tg:a}]{%
  \includegraphics[width=\columnwidth]{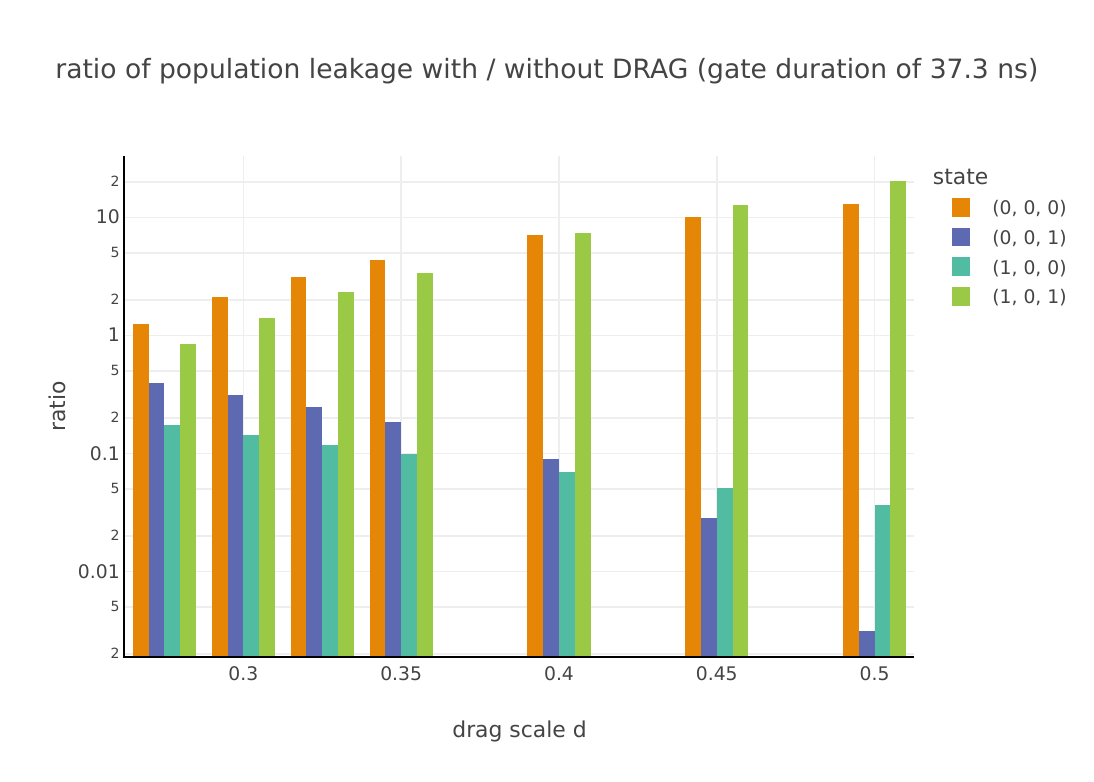}%
}\hfill
\subfloat[\label{fig:leakage_budget_DRAG_ratio_short_tg:b}]{%
  \includegraphics[width=\columnwidth]{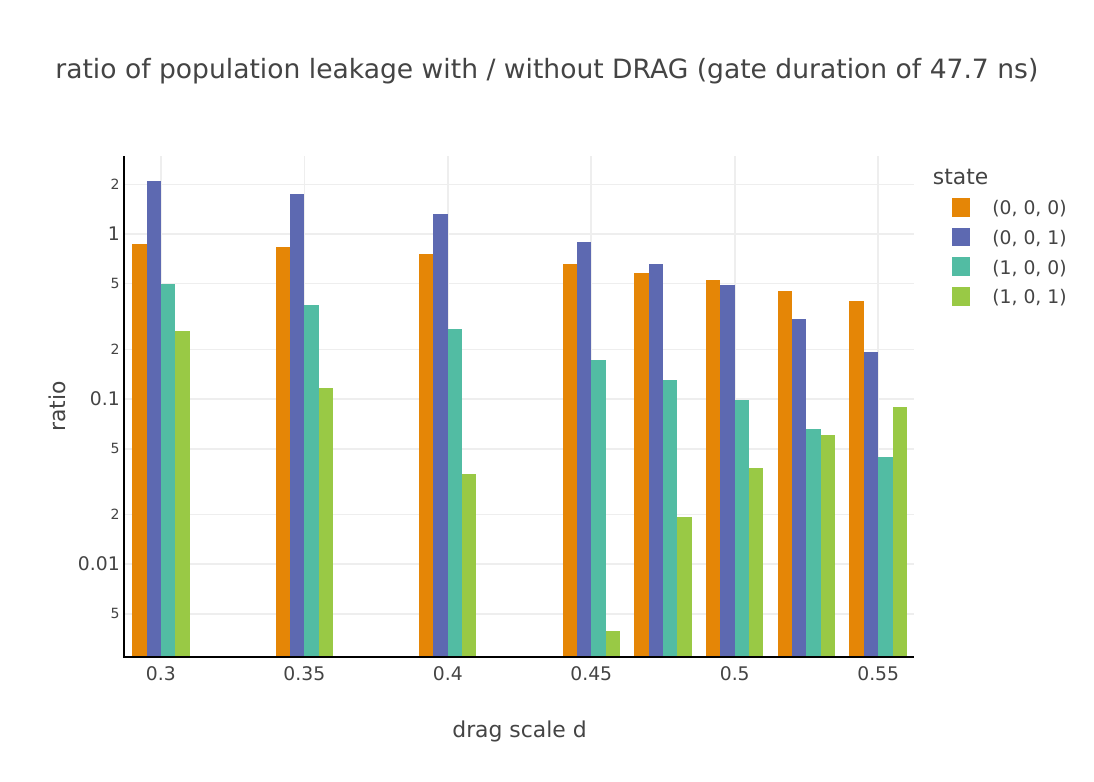}%
}
\caption{Comparison of leakage budget for $t_g = 37.3$  ns (top) and $47.7$ ns (bottom). } \label{fig:leakage_budget_DRAG_ratio_short_tg}
\end{figure*}

\begin{figure*}
\subfloat[\label{fig:leakage_budget_DRAG_ratio_long_tg:a}]{%
  \includegraphics[width=\columnwidth]{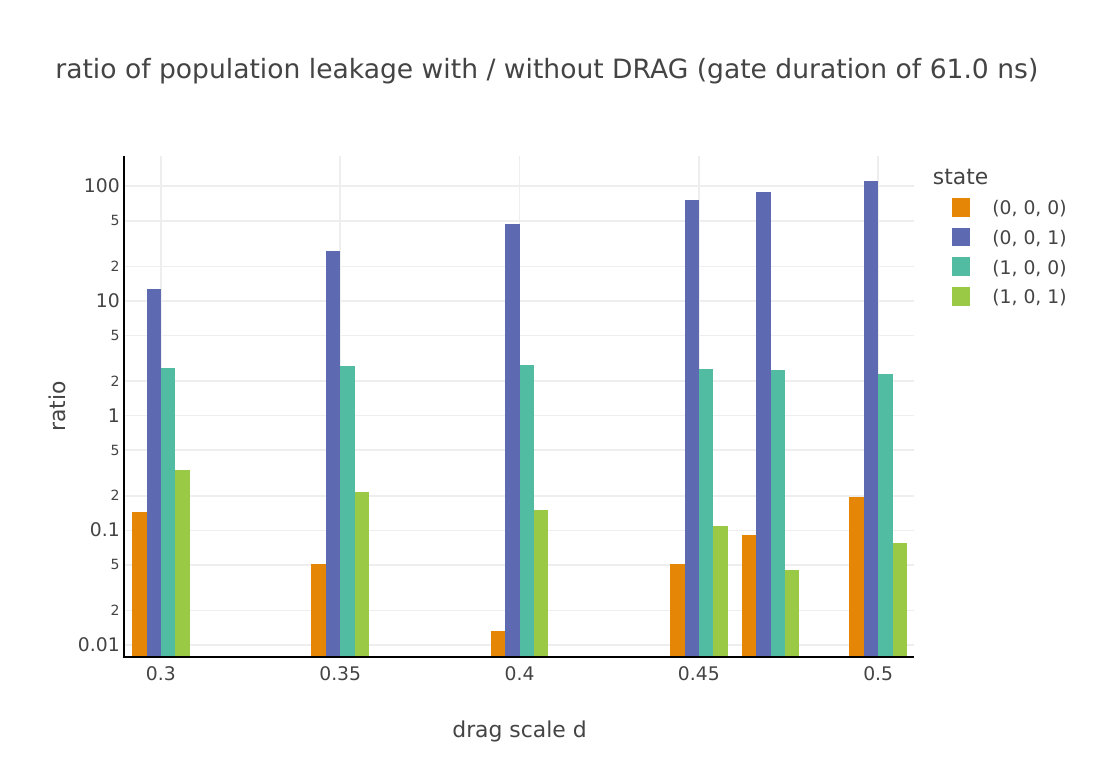}%
}\hfill
\subfloat[\label{fig:leakage_budget_DRAG_ratio_long_tg:b}]{%
  \includegraphics[width=\columnwidth]{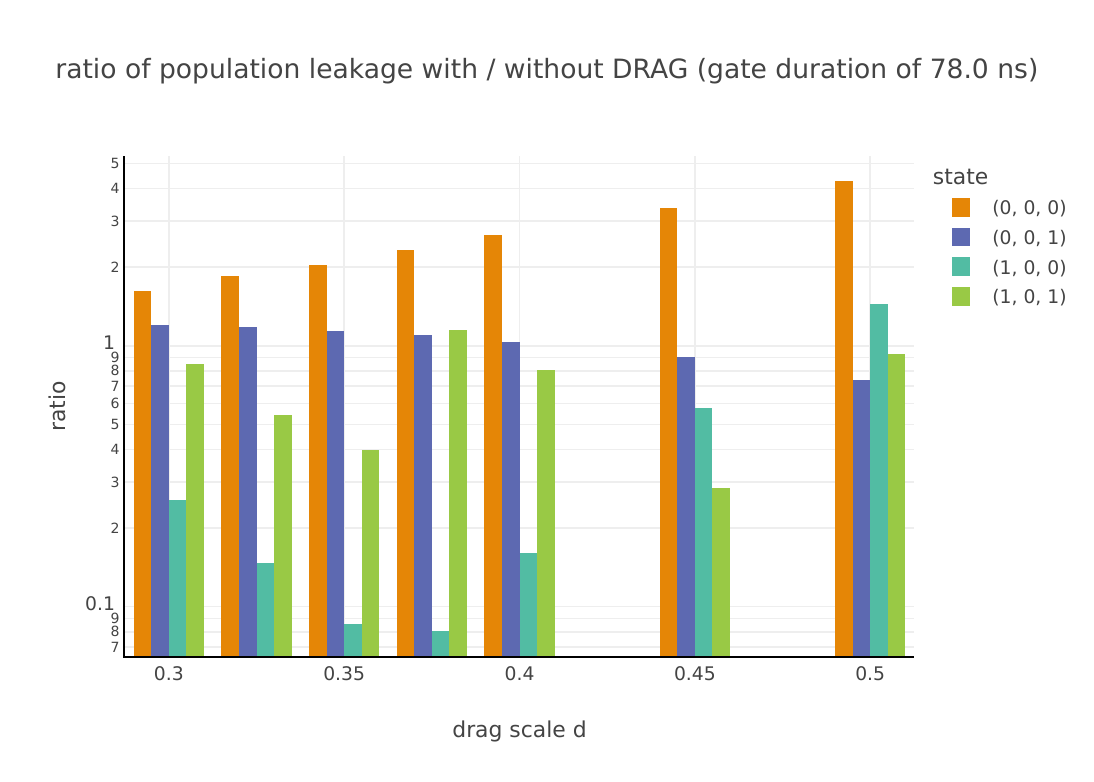}%
}
\caption{Comparison of leakage budget for $t_g = 61$  ns (top) and $78$ ns (bottom). } \label{fig:leakage_budget_DRAG_ratio_long_tg}
\end{figure*}

Here we show results for the population leakage under DRAG corrections for various $t_g$ values, sweeping $d$, and compare to the results without DRAG.  We observe an optimal $d$ value for each gate duration in Figs. \ref{fig:drag_leakage_optimize_d} and \ref{fig:drag_leakage_optimize_d_2}, resulting in the black points shown in Fig. 4(b) of the main text (coherent error with DRAG corrections).
We also compare to the leakage budget for each case without DRAG to determine the degree of improvement. As shown in Figs. \ref{fig:leakage_budget_DRAG_ratio_short_tg} and \ref{fig:leakage_budget_DRAG_ratio_long_tg}, we find that significant increase in leakage of the other computational states results in only modest improvement in coherent error by applying the DRAG technique. In principle, further pulse optimization can mitigate this to further improve the coherent error, outside the scope of this work.

\subsection{Coherent error when $\alpha < \delta \chi_{ij} $}\label{small_alpha}
In this section, we demonstrate that $\alpha \sim \delta \chi_{ij}$ is required to have low coherent errors ($\sim 10^{-6}$ within 100ns, $\sim 10^{-5}$ within 70ns). In particular, if $\alpha$ is too small, such that $\varphi$ is far from zero before drive frequency optimization, a large detuning $ \delta \omega$ is needed to set $\varphi \approx 0$ and remove the coherent phase errors. This large detuning in turn leads to increased leakage from multiple computational states.
\begin{table*}
\begin{tabular}{||c c c c c c c ||} 
 \hline
 Descriptor & $E_C^{A(B)}$ (GHz) & $Z_c$ ($\Omega$) & $J_{A(B), c}$ (GHz) & $J_{AB}$ (GHz) & $E_L^{A(B)}$ (GHz) & $E_J^{A(B)}$ (GHz) \\
 \hline\hline
 Small $\alpha$ & 2.0(2.0) & 51 & 0.24(0.24) & 0.14 & 0.25(0.3) & 7.3(7.2) \\ 
 \hline
 \end{tabular}
 \caption{\label{system_0_parameters}Hamiltonian parameters which yield smaller $\alpha = 57$ MHz compared to the main text, as discussed in section \ref{small_alpha}.}
\end{table*}

We illustrate this effect, by showing the coherent error budget for a set of circuit parameters in which $\alpha$ is smaller than the main text value. For the circuit parameters recorded in Table \ref{system_0_parameters}, the $\delta \chi_{ij}$ are about the same as the main text values, with $\delta \chi_{10} = 107$ MHz and $\delta \chi_{01} = 90$ MHz (similar to $\sim 80$ MHz and $100$ MHz in the main text). However, $\alpha = 57$ MHz (as opposed to $70$ MHz as in the main text), with reduced hybridization $h_A = 10\%$  and $h_B = 12\%$. We find that as expected, $\varphi$ is far from zero before drive optimization, with $\varphi_{ij}$ values shown in Fig. \ref{fig:system_0_spurious_phase}, following Fig. 3 of the main text. After tuning the drive frequency to (large) optimal detunings of $\delta \omega \sim (2\pi) 1$ MHz to $(2\pi) 3$ MHz, we extract the coherent errors, which are again entirely population leakage, as expected after the drive frequency optimization. We observe larger coherent errors compared to the main text, of e.g. $6 \times 10^{-6}$ in 100ns, which is dominated by the leakage of $\ket{1, 0, 1}$ (Fig. \ref{fig:system_0_population_leakage_ratio}). In principle, this contribution to the coherent error may be at least partially mitigated with DRAG techniques. However, the leakage of the other computational states is \textit{also} larger than the main text values, even despite the slightly larger $\delta \chi_{ij}$. This suggests that the increased detuning $\delta \omega$, required by the smaller $\alpha$, increases leakage of other non-computational transitions. In Fig. \ref{fig:system_0_population_leakage_ratio}, we plot the ratio of these larger leakage values to the population leakage for the main text case. We see that there is increased leakage with the smaller $\alpha$ for multiple computational states besides $\ket{1, 0, 1}$, implying that there is a limit to the detuning $\delta \omega$ and minimum $\alpha$ that is feasible to retain similar gate performance as the main text describes.
\begin{figure}[hb!]
    \centering
    \includegraphics[width=0.3\textwidth]{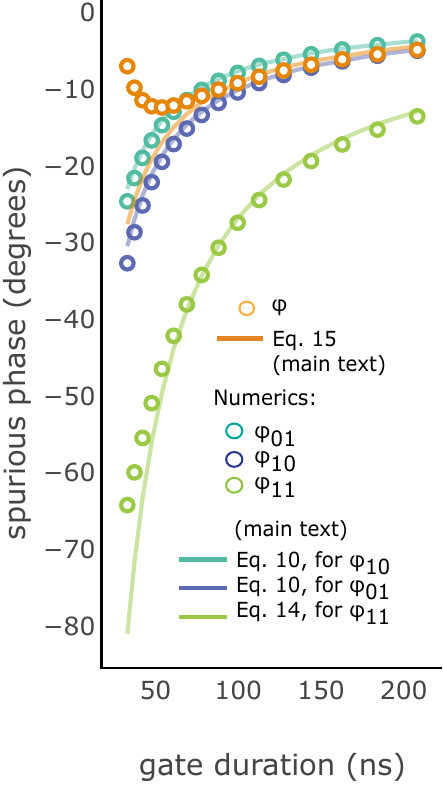}
    \caption{Spurious phases $\varphi_{ij}$ accumulated on the computational states as a function of gate duration, for Hamiltonian parameters given in Table \ref{system_0_parameters}. We find that the relative phase $\varphi$ is much larger than the main text values, as expected because $\alpha$ is small.}
    \label{fig:system_0_spurious_phase}
\end{figure}

\begin{figure}
    \centering
    \includegraphics[width=0.8\columnwidth]{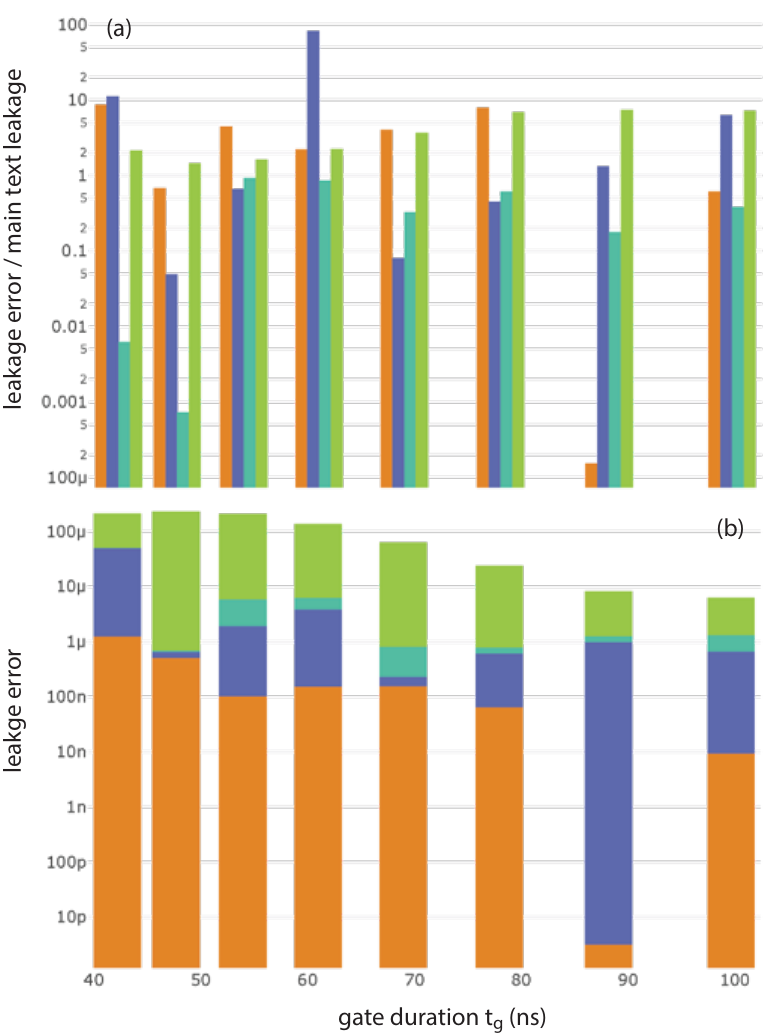}
    \caption{(a) Population leakage error for the Hamiltonian parameters given in Table \ref{system_0_parameters}, divided by leakage errors for the Hamiltonian parameters in the main text. We observe larger coherent errors, with $\epsilon_c \approx 10^{-5}$ at 100ns. (b) Population leakage error for Hamiltonian parameters given in Table \ref{system_0_parameters}.}
    \label{fig:system_0_population_leakage_ratio}
\end{figure}

\section{Incoherent error budget}\label{appendix_e_error_budget}

\subsection{Results}

\begin{figure*}
\subfloat[\label{fig:error_budget_short_tg:a}]{%
  \includegraphics[width=\columnwidth]{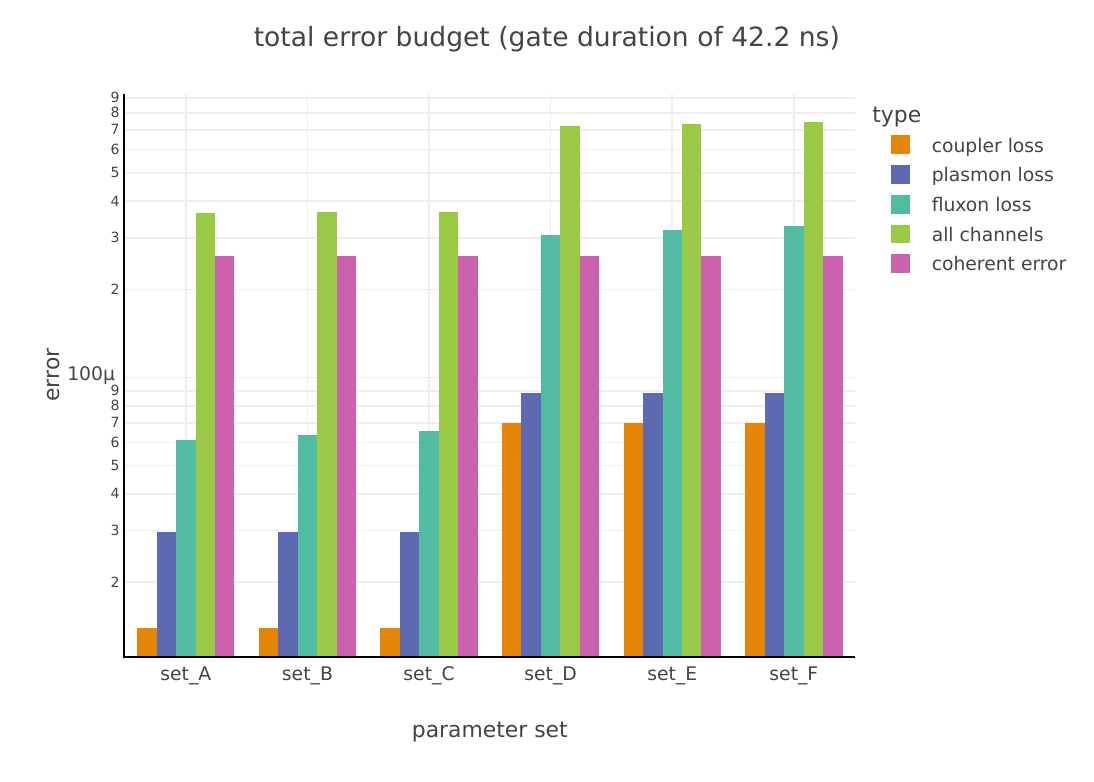}%
}\hfill
\subfloat[\label{fig:error_budget_short_tg:b}]{%
  \includegraphics[width=\columnwidth]{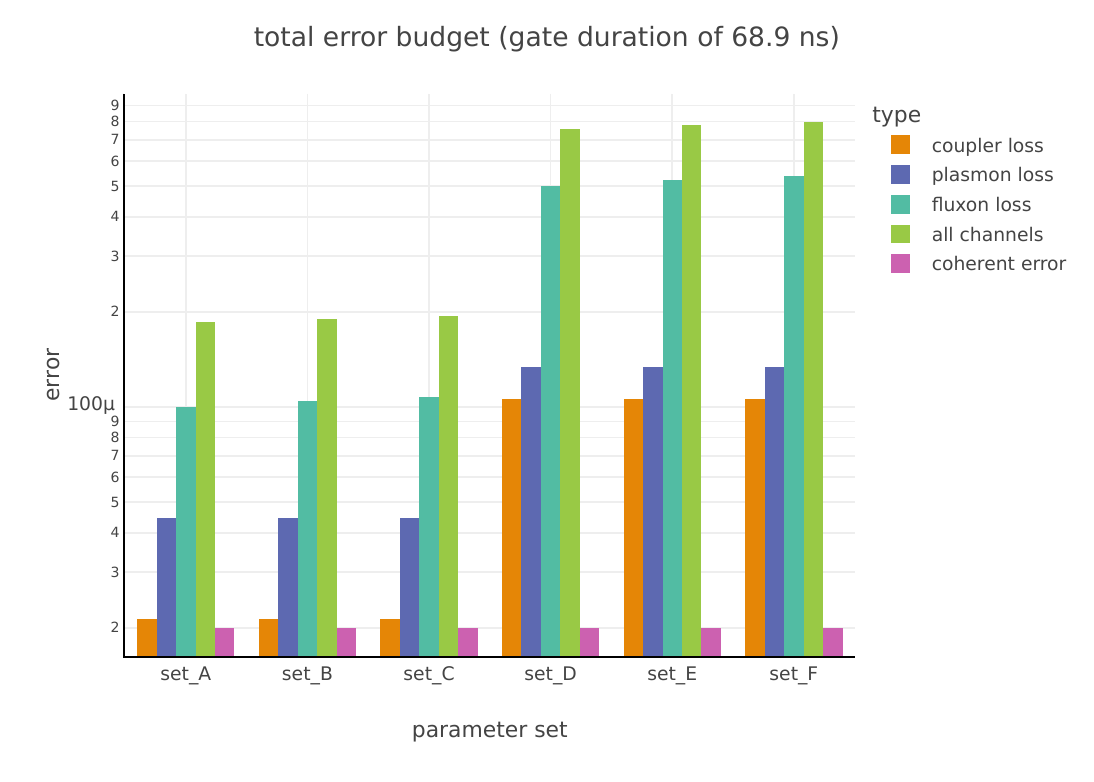}%
}
\caption{Error budget for gate duration of 42 ns (top) and 69 ns (bottom). As described in the main text, the error channels are: fluxon loss, fluxonium single-photon decay and heating, plasmon loss, inherited loss on the coupler from fluxonium 1-2 transition $T_1$;, coupler loss, and dissipation of the resonator.} \label{fig:error_budget_short_tg}
\end{figure*}

\begin{figure*}
\subfloat[\label{fig:error_budget_long_tg:a}]{%
  \includegraphics[width=\columnwidth]{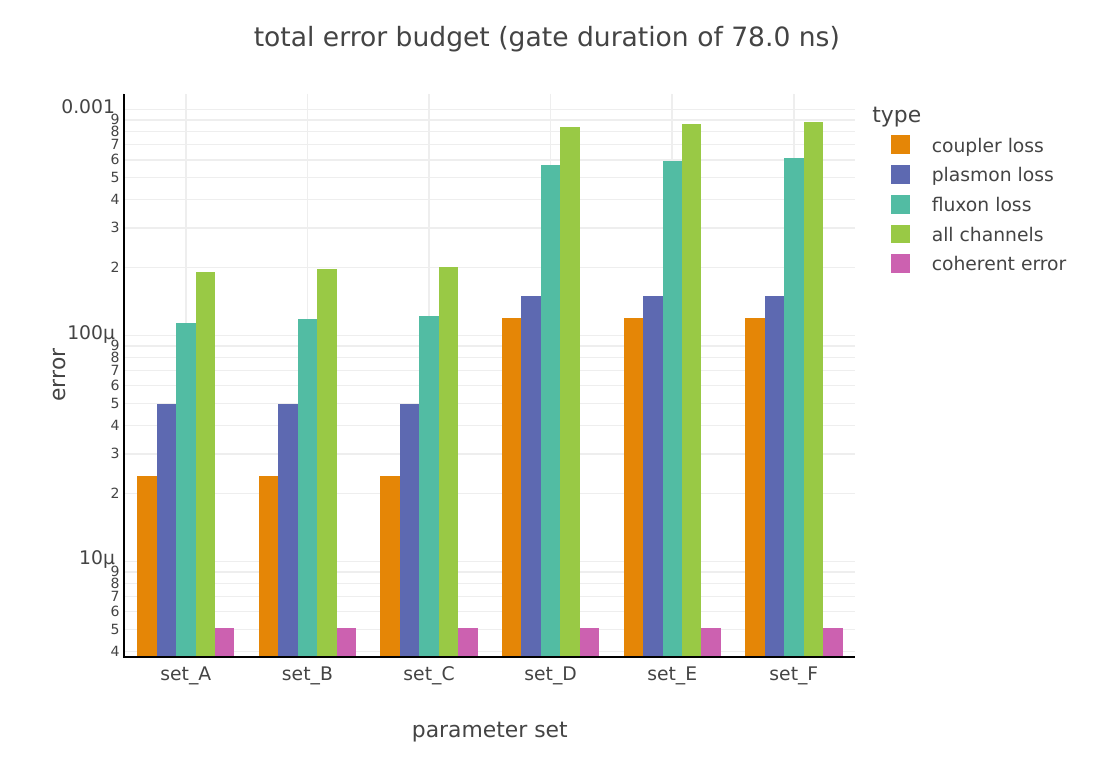}%
}\hfill
\subfloat[\label{fig:error_budget_long_tg:b}]{%
  \includegraphics[width=\columnwidth]{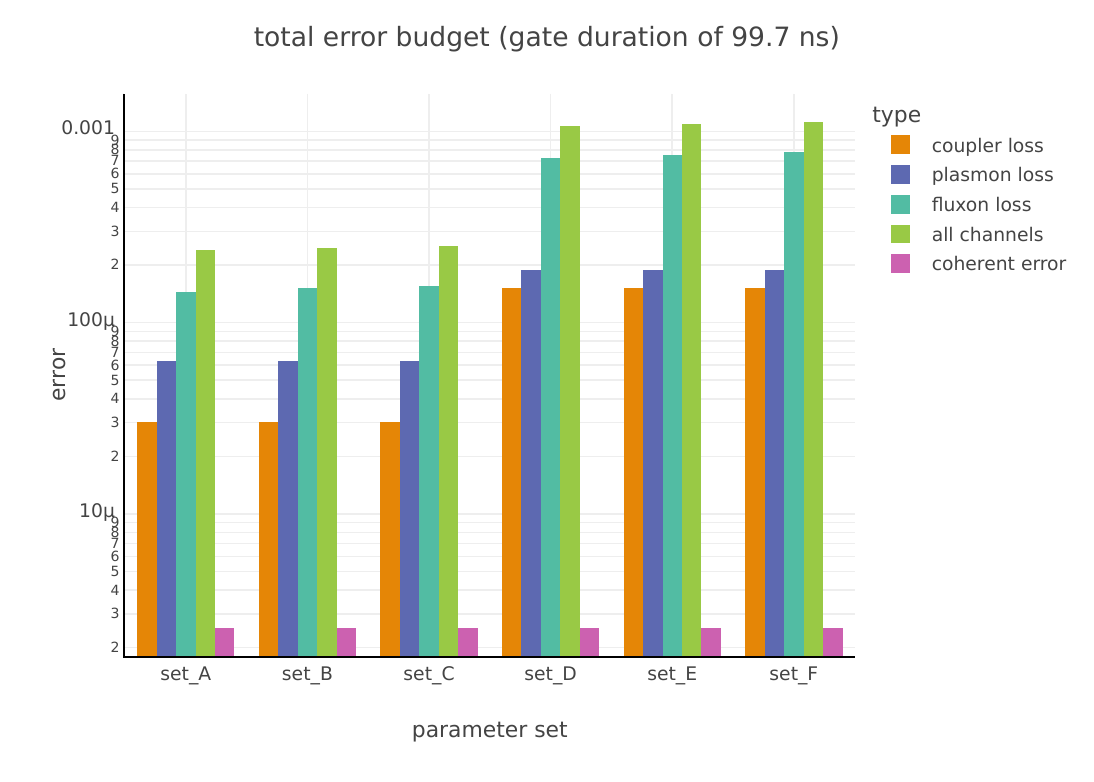}%
}
\caption{Error budget for gate duration of 78 ns (top) and 100 ns (bottom). Note that the error budget is limited by the fluxon loss, the longest dissipation rate in the system.} \label{fig:error_budget_long_tg}
\end{figure*}

We ran several master equation simulations, where each source of loss was the only one in the system in order to identify the incoherent error budget. Here we show the coherent error budget in Figs. \ref{fig:error_budget_short_tg} and \ref{fig:error_budget_long_tg}. We observe a trade-off of coherent and incoherent errors around the optimal gate duration. Note that the error budget is limited by the fluxonium's 0-1 transition (fluxon) loss, which is also the smallest dissipation rate for our parameter sets A-F reported in the main text. Therefore, improving single-qubit $T_1$ should lead immediately to improved gate fidelities with our scheme.

\subsection{Comparison to simple estimates}
\begin{figure*}
\subfloat[\label{fig:error_estimates_A_B:a}]{%
  \includegraphics[width=\columnwidth]{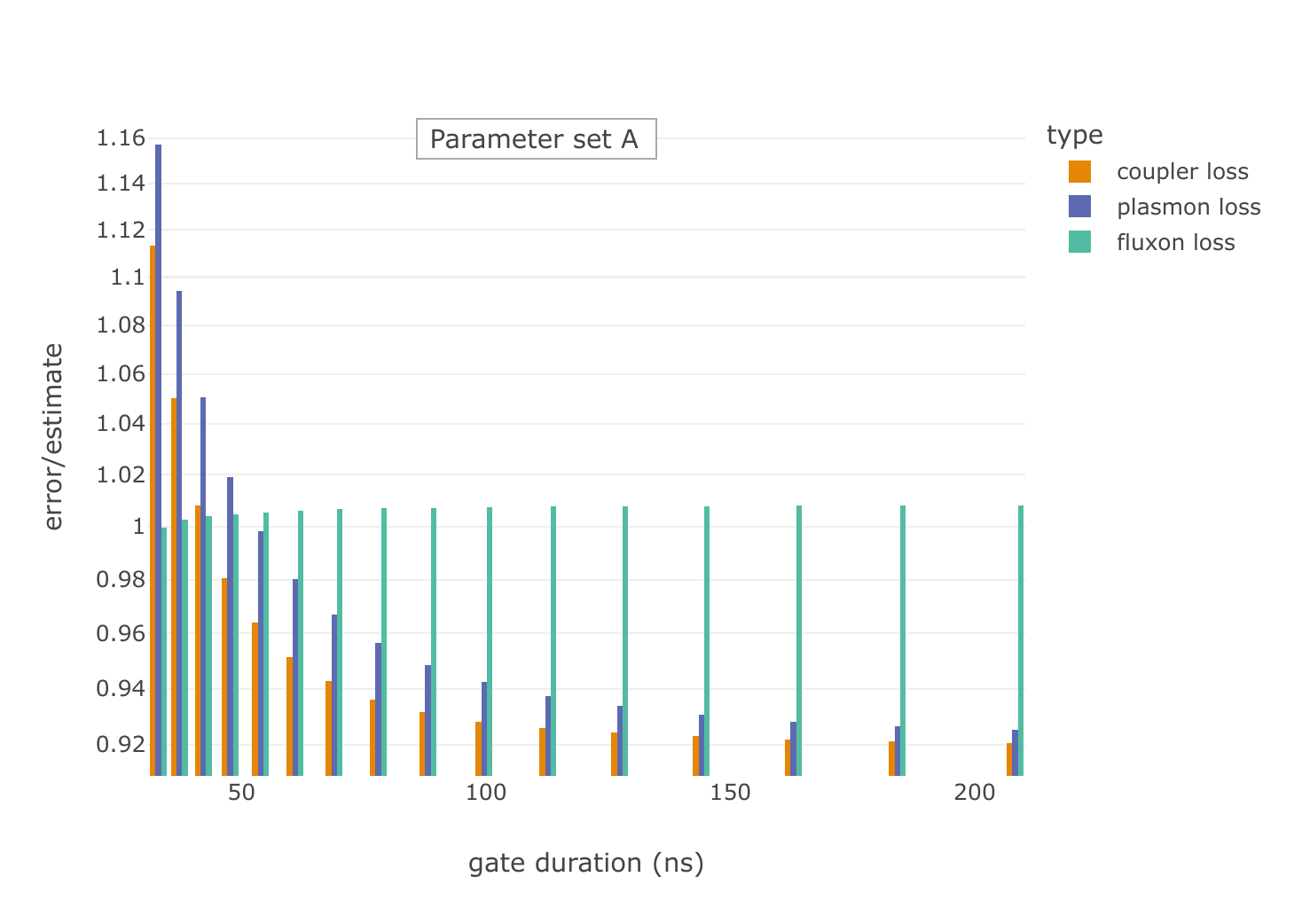}%
}\hfill
\subfloat[\label{fig:error_estimates_A_B:b}]{%
  \includegraphics[width=\columnwidth]{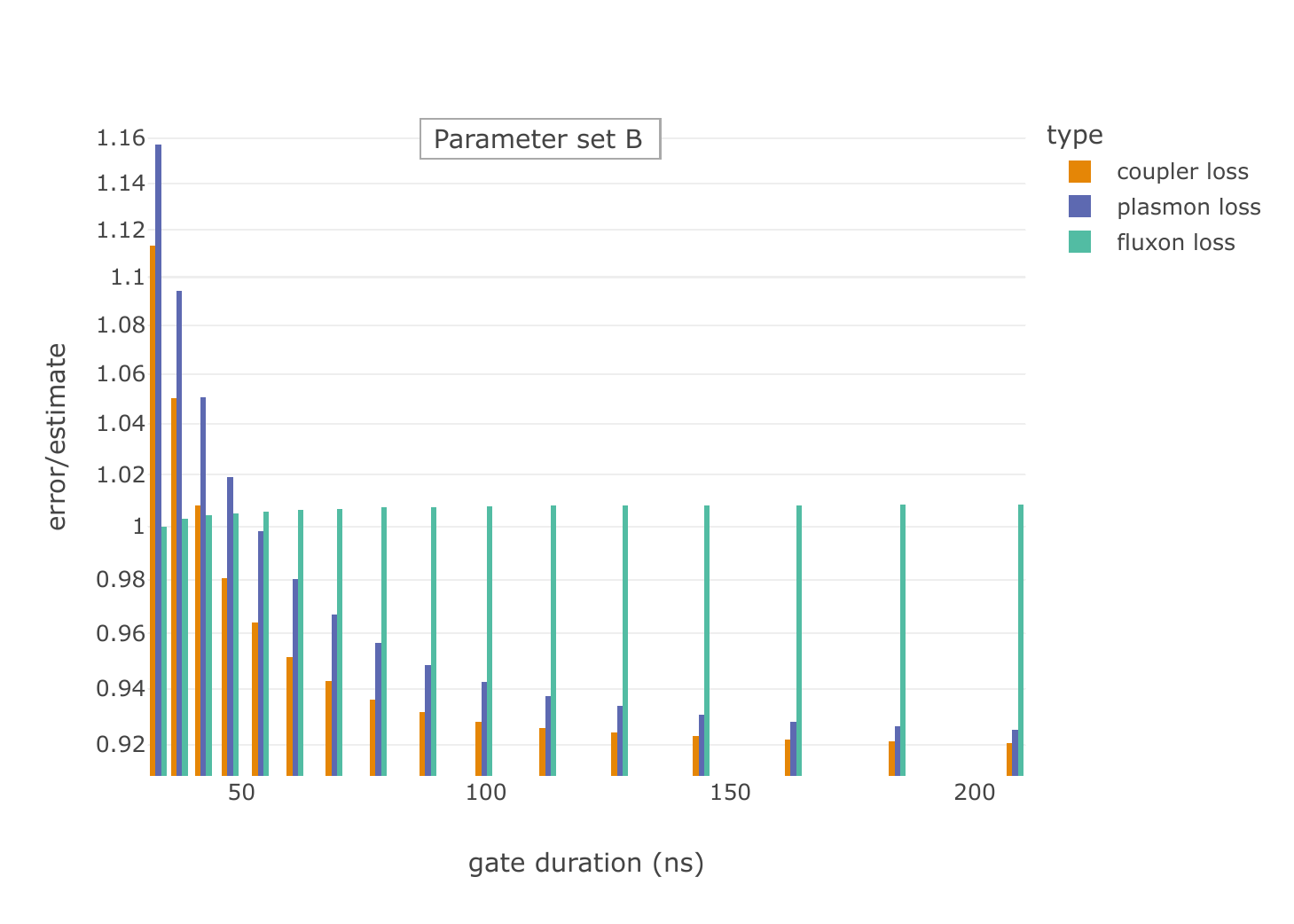}%
}
\caption{Error budget contributions compared to estimates in Eq.~\eqref{loss_estimates}, parameter sets A (top) and B (bottom). We observe agreement within 15\% for all gate durations.} \label{fig:error_estimates_A_B}
\end{figure*}

\begin{figure*}
\subfloat[\label{fig:error_estimates_C_D:a}]{%
  \includegraphics[width=\columnwidth]{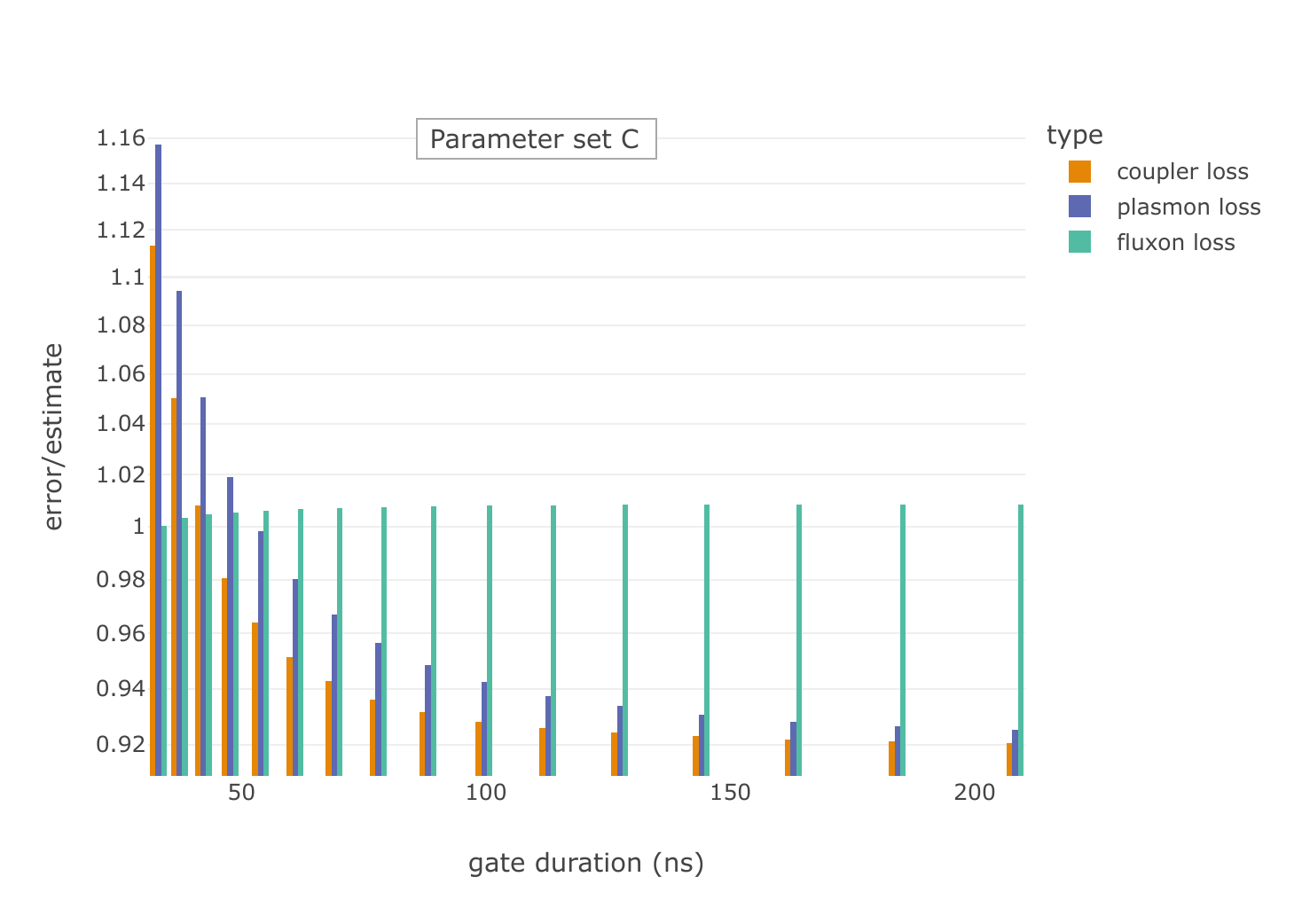}%
}\hfill
\subfloat[\label{fig:error_estimates_C_D:b}]{%
  \includegraphics[width=\columnwidth]{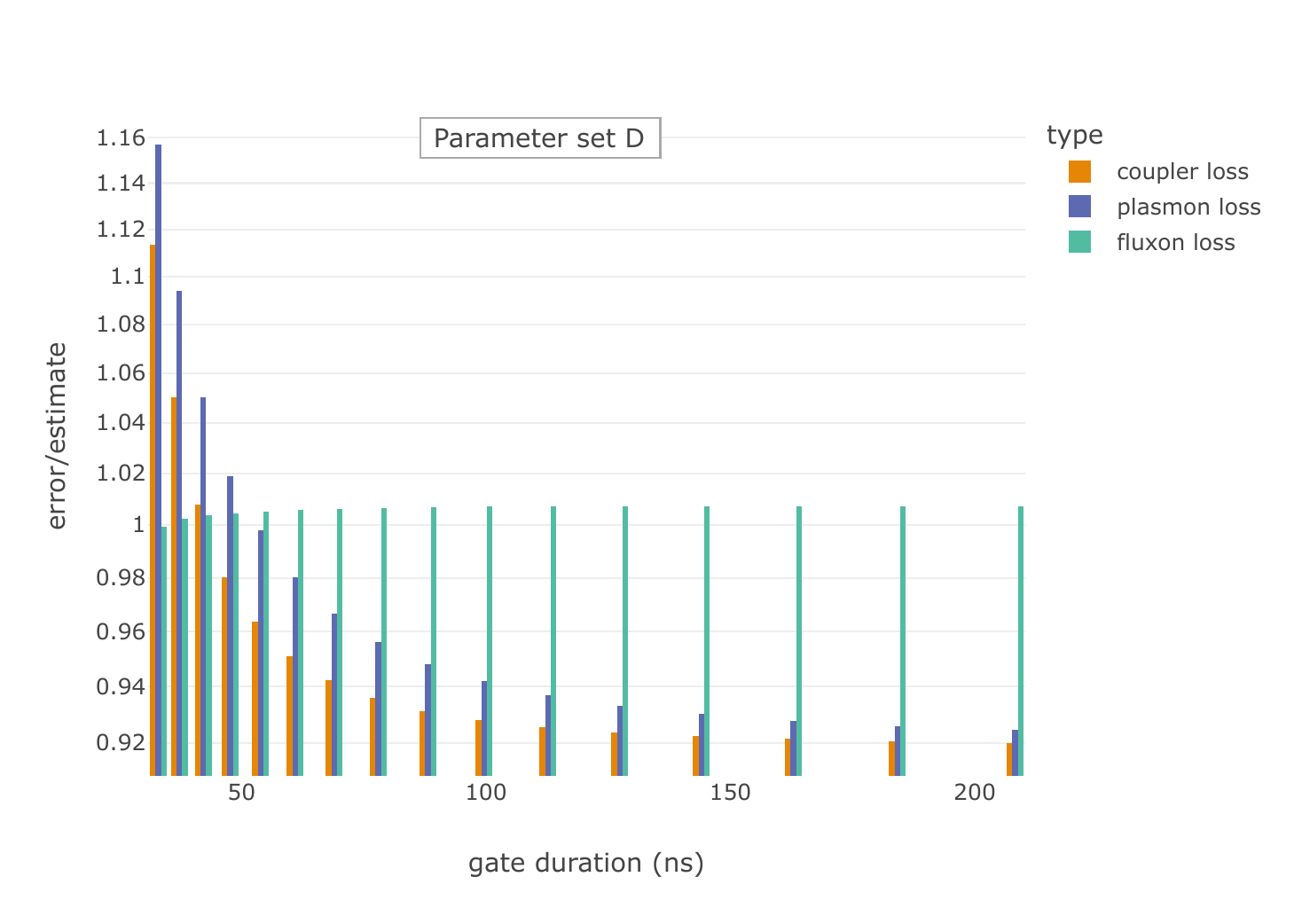}
}
\caption{Error budget contributions compared to estimates in Eq. \eqref{loss_estimates}, parameter sets C (top) and D (bottom).} \label{fig:error_estimates_C_D}
\end{figure*}

\begin{figure*}
\subfloat[\label{fig:error_estimates_E_F:a}]{%
  \includegraphics[width=\columnwidth]{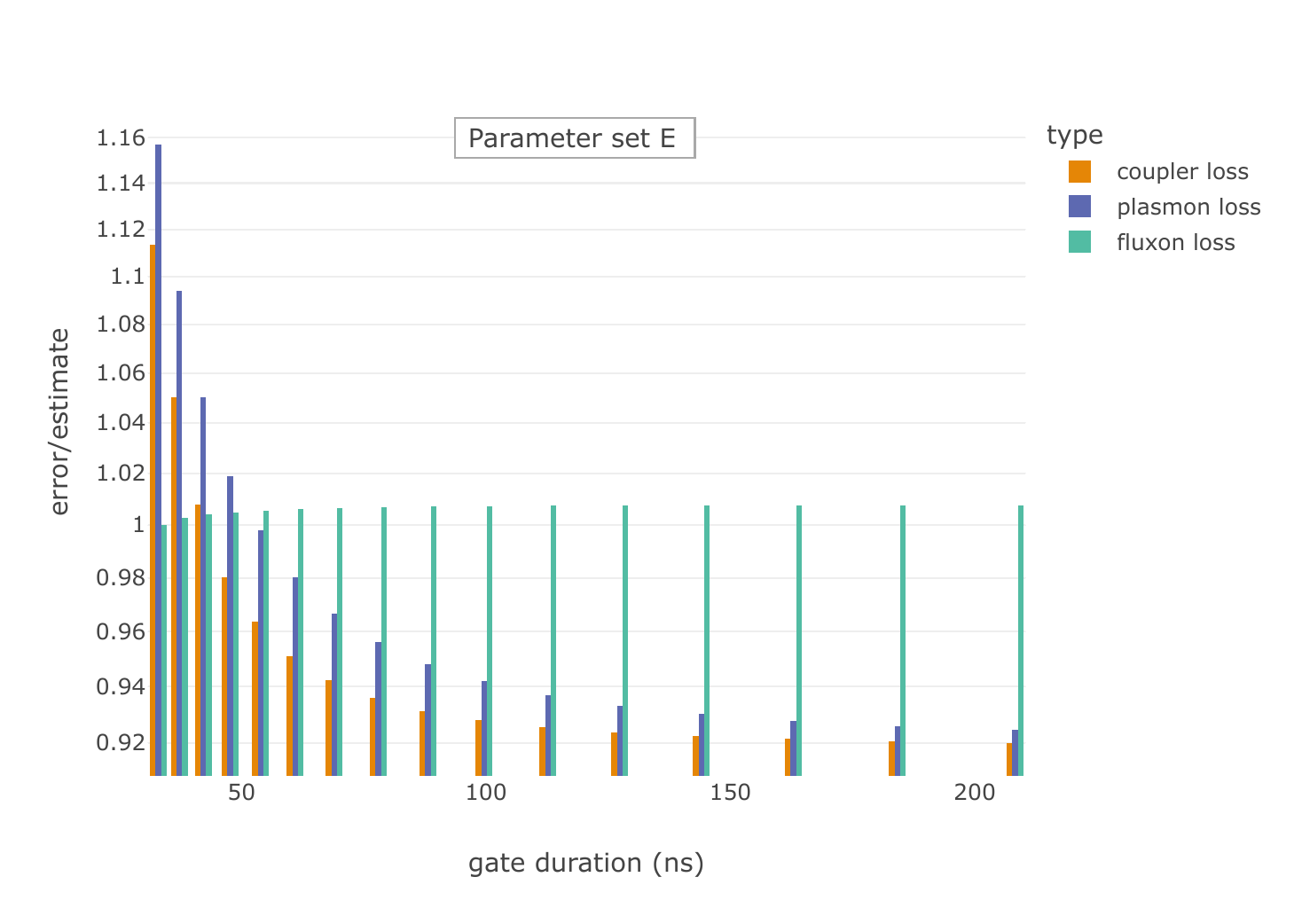}%
}\hfill
\subfloat[\label{fig:error_estimates_E_F:b}]{%
  \includegraphics[width=\columnwidth]{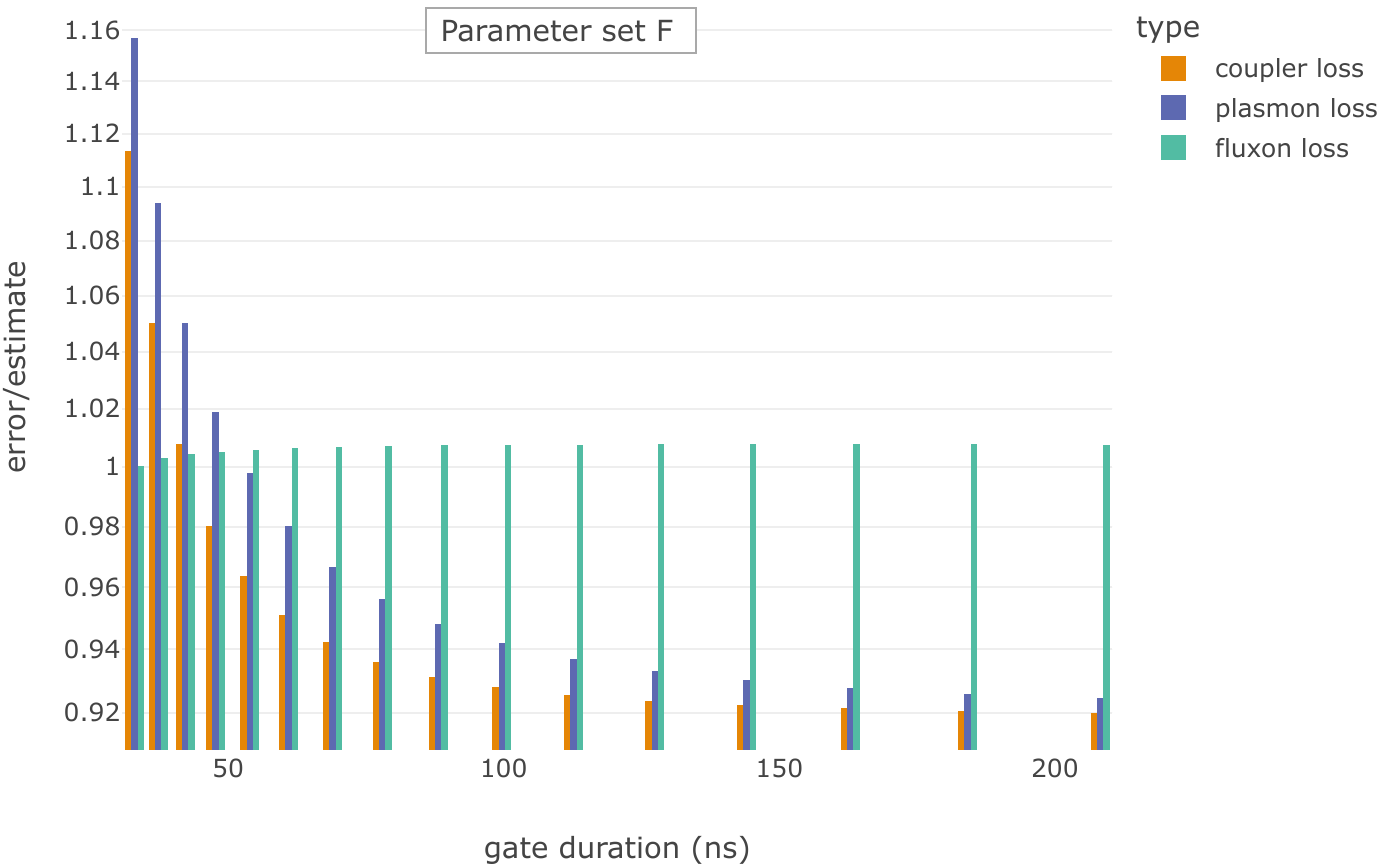}
}
\caption{Error budget contributions compared to estimates in Eq. \eqref{loss_estimates}, parameter sets E (top) and F (bottom).} \label{fig:error_estimates_E_F}
\end{figure*}

We recover approximations within 15\% of the incoherent error, using simple estimates for each channel's contributions as a function of the loss parameters. In particular the loss is described as 
\begin{widetext}
\begin{equation}\label{loss_estimates}
\begin{aligned}
    \epsilon_{\text{fluxon}} \approx \frac{4}{5} t_g \left(\frac{\kappa_{01}^A}{2}(2 n_{th}^A + 1) + \frac{\kappa_{01}^B}{2}(2 n_{th}^B + 1)\right) \\
    \epsilon_{\text{coupler}} \approx \tau |\langle 1, 0, 1| \hat{n}_c | 1, 1, 1\rangle |^2 \kappa_c / 8 \\
    \epsilon_{\text{plasmon}} \approx \left( | \langle 1, 0, 1| \hat{n}_A | 1, 1, 1\rangle |^2  \kappa_{12}^A  + | \langle 1, 0, 1| \hat{n}_B | 1, 1, 1\rangle |^2  \kappa_{12}^B \right)\tau / 8 
\end{aligned}
\end{equation}
\end{widetext}
where as described in the main text, $t_g = 2.2 \tau$ is the full gate time and $\tau/\sqrt{2}$ is the full width half maximum of the Gaussian pulse. In Figs. \ref{fig:error_estimates_A_B}, \ref{fig:error_estimates_C_D} and \ref{fig:error_estimates_E_F}, we show the ratio of each error contribution compared to the associted estimate in \eqref{loss_estimates}. We observe agreement within 15\% for all six parameter sets A-F, and for all gate durations $t_g$.
\subsection{Simulations on smaller Hilbert space dimension}
\begin{figure*}
\subfloat[\label{fig::a}]{%
  \includegraphics[width=0.8\columnwidth]{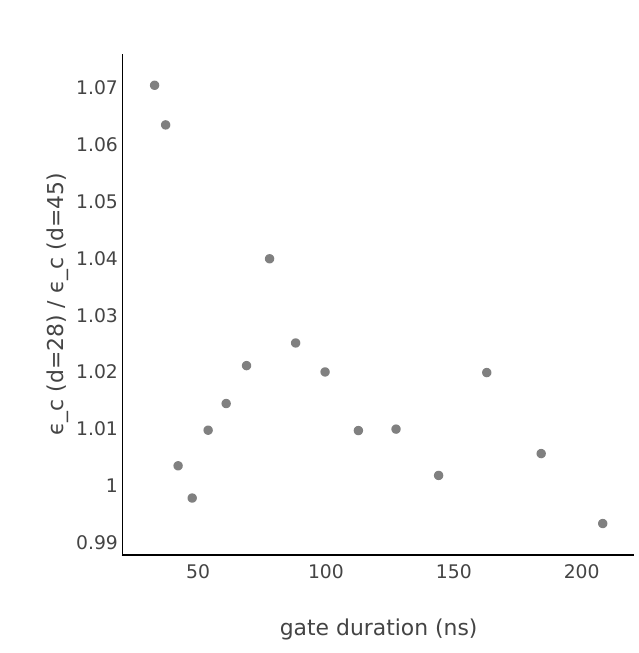}%
}\hfill
\subfloat[\label{fig::b}]{%
  \includegraphics[width=0.8\columnwidth]{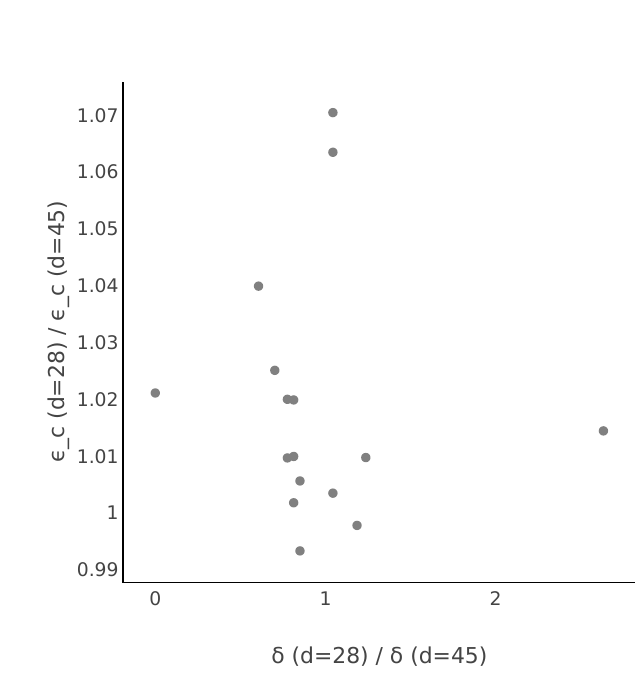}
}
\caption{Coherent error for the larger simulation case, compared to the truncated error as required for master equation simulations. Top: we observe relative changes in coherent error within 7\% for all gate durations simulated. Bottom: there is a change in optimized drive frequencies between the two cases. This is likely due to small changes in Stark shifts from higher-lying fluxonium eigenstates, such as the shift on fluxonium $\ket{1}$ states due to the $\ket{1}-\ket{4}$ transition, which is included in the $d=45$ simulation but not in the $d=28$ simulation. The points at zero and 2.5 correspond to ~100 kHz changes in optimal detuning, where the $d=45$ optimal detuning was close to zero, such that the relative change in optimal detuning is large.}\label{fig:change_dim_ec}
\end{figure*}

For tractable master equation simulations, we need to reduce the simulation dimensions from the coherent error case. In particular, we diagonalize the Hamiltonian with 1,208 dimensions, then truncate the Hilbert space to $d = 45$ (coherent error simulations) compared to $d = 28$ (loss simulations). Here we show that the change in coherent error from reducing the simulation dimension is minimal, within a 7\% change for all gate durations simulated (see Fig. \ref{fig:change_dim_ec}).
\section{Methods}\label{appendix_f_methods}
\subsection{Hilbert space dimension}
\begin{figure}\label{H_space_sweep}
\centering
\includegraphics[width=\columnwidth]{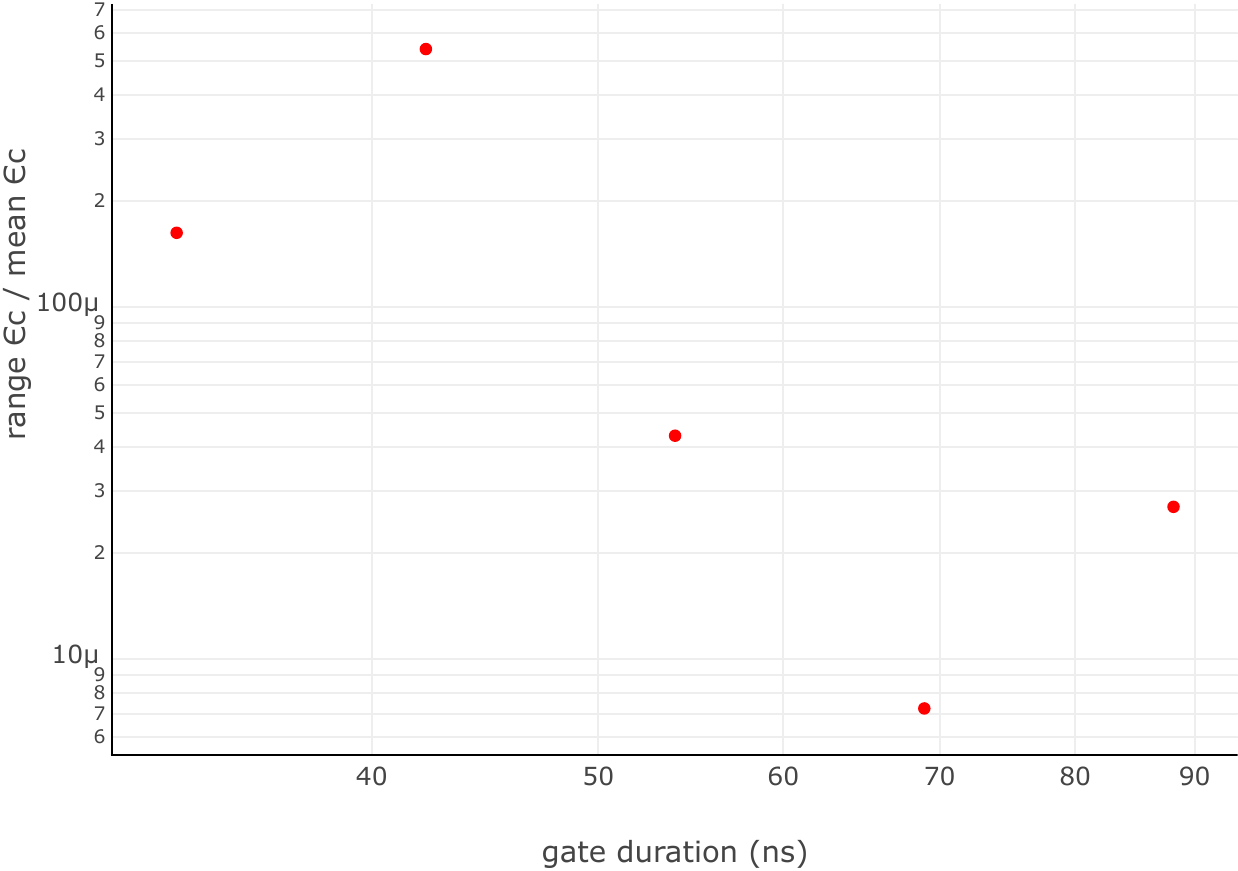}\caption{Sweeping the Hilbert space dimension of the simulation from 40 to 50. We extract the range of values and divide by the mean to quantify deviations due to the finite simulation size. In the main text, our Hilbert space dimension is 45.}\label{fig:H_space_sweep}
\end{figure}

Our simulations constitute a fairly intensive computational task. As described in the main text, our simulations are performed in the lab frame, without a rotating wave approximation, and using the full cosine potential of the fluxoniums' Hamiltonian. Our circuit elements are also strongly nonlinear and strongly coupled such that a large Hilbert space size is required for accuracy. We run our simulations on AWS EC2 for speed. Here, we show results from performing our time domain simulations without loss to extract the coherent error, sweeping the Hilbert space dimension from 40 to 50 for various gate durations $t_g$, without changing any other simulation parameters (e.g., drive parameters are fixed). We extract the range of the coherent error observed across the Hilbert space dimension sweep, and compare to the mean, finding a change of about 1 part in $10^{4}$ of the coherent error across the gate durations $t_g$ we tested. For these results see Fig. \ref{fig:H_space_sweep}.
\subsection{Derivation of $\epsilon_{\varphi}$}
We consider a unitary that introduces a phase on each computational state,
\begin{equation}
    U_\varphi = \text{diag}(e^{i \varphi_{00} }, e^{i \varphi_{10} }, e^{i \varphi_{01} }, e^{i(\varphi_{11} + \pi)}).
\end{equation}
Considering an initial product state $\psi_0 = |+\text{x} \rangle \otimes |+\text{x}\rangle$, application of a perfect CZ gate would maximally entangle the two qubits to $\psi = \frac{1}{2}(|00\rangle + |10\rangle + |01\rangle - |11\rangle$, with a reduced density matrix on each qubit of the identity. However, for arbitrary phases, after application of $U_\varphi$, the reduced density matrix tracing out one of the qubits would have an off diagonal element of $\rho_{01} = e^{i(\varphi_{00} - \varphi_{01})} + e^{i(\varphi_{10} - \varphi_{11} - \pi)}$. For the reduced density matrix on each qubit to be identity, i.e., maximize the entanglement we require
\begin{equation}\label{varphi}
\varphi_{00} - \varphi_{10} - \varphi_{01} + \varphi_{11} = 0
\end{equation}
Any deviation from this condition such that if $\varphi_{00} - \varphi_{10} - \varphi_{01} + \varphi_{11} = \varphi \neq 0$, the entangling power of the gate will be reduced which cannot be fixed with single qubit rotations. Adding single qubit $\sigma_z$ rotations $U_{A} = \text{exp}(-i (\varphi_{10}-\varphi_{00}) \sigma_{z, A} / 2)$ and $U_{B} = \text{exp}(-i (\varphi_{01}-\varphi_{00}) \sigma_{z, B} / 2)$ along with a global phase of $(\varphi_{10} + \varphi_{01})/2$ simplifies the evolution so that we can write it as $U_\varphi U_A U_B = \text{diag}(1, 1, 1, -\text{exp}(i(\varphi_{11} + \varphi_{00} - \varphi_{10}-\varphi_{01})) \equiv \text{diag}(1, 1, 1, -e^{i\varphi})$. The coherent error $\epsilon_\varphi$ following the standard definition for gate fidelity would then be 
\begin{widetext}
\begin{equation}\label{epsilon_varphi}
\begin{split}
\epsilon_\varphi = 1- \frac{1}{20}\Big(\text{Tr}\left((U_\varphi U_A U_B)^\dagger (U_\varphi U_A U_B)\right) + |\text{Tr}\left((U_\varphi U_A U_B)^\dagger U_{\text{CZ}}\right)|^2\Big) = 1-\frac{1}{20}(4 + 10 + 6 \cos{\varphi}) \approx 3 \varphi^2 / 20
\end{split}
\end{equation}
\end{widetext}
where $U_{\text{CZ}}$ is the desired ideal evolution. Note that the same result is found in the Appendix of \cite{ding_high-fidelity_2023}.

%

\clearpage
\end{document}